\documentclass[letterpaper, 10 pt, conference]{ieeeconf}%
\usepackage{graphicx}
\usepackage{epsfig}
\usepackage{mathptmx}
\usepackage{times}
\usepackage{amsmath}
\usepackage{thmtools,thm-restate}
\usepackage{amssymb}
\usepackage[section]{placeins}
\usepackage{placeins}
\usepackage{amsmath}
\usepackage{multirow}
\usepackage{graphicx}
\usepackage[normalem]{ulem}
\usepackage{hyperref}
\usepackage{subcaption}
\usepackage{algorithm,tabularx}
\usepackage{algpseudocode}
\usepackage{amsfonts}
\usepackage{cases}
\usepackage{romannum}
\usepackage{textcomp}
\usepackage{xcolor}%
\providecommand{\U}[1]{\protect\rule{.1in}{.1in}}
\providecommand{\U}[1]{\protect\rule{.1in}{.1in}}
\IEEEoverridecommandlockouts
\newcommand{\R}{\mathbb{R}}

\newtheorem{assumption}{Assumption}
\newtheorem{theorem}{Theorem}
\newtheorem{corollary}{Corollary}
\newtheorem{lemma}{Lemma}

\newtheorem{remark}{Remark}
\useunder{\uline}{\ul}{}
\makeatletter
\newcommand{\multiline}[1]{  \begin{tabularx}{\dimexpr\linewidth-\ALG@thistlm}[t]{@{}X@{}}
#1
\end{tabularx}
}
\makeatother
\usepackage{cite}
\begin{document}

\title{{\LARGE \textbf{
Event-Driven Receding Horizon Control for Distributed Estimation in Network Systems
}}
}
\author{Shirantha Welikala and Christos G. Cassandras \thanks{$^{\star}$Supported in part by NSF under grants ECCS-1931600, DMS-1664644, CNS-1645681, by AFOSR under grant FA9550-19-1-0158, by ARPA-E's NEXTCAR program under grant DE-AR0000796 and by the MathWorks.} \thanks{The authors are with the Division of Systems Engineering and Center for Information and Systems Engineering, Boston University, Brookline, MA 02446, \texttt{{\small \{shiran27,cgc\}@bu.edu}}.}}
\maketitle

\begin{abstract}
We consider the problem of estimating the states of a distributed network of nodes (targets) through a team of cooperating agents (sensors) persistently visiting the nodes so that an overall measure of estimation error covariance evaluated over a finite period is minimized. We formulate this as a multi-agent persistent monitoring problem where the goal is to control each agent's trajectory defined as a sequence of target visits and the corresponding dwell times spent making observations at each visited target. A distributed on-line agent controller is developed where each agent solves a sequence of receding horizon control problems (RHCPs) in an event-driven manner. A novel objective function is proposed for these RHCPs so as to optimize the effectiveness of this distributed estimation process and its unimodality property is established under some assumptions. Moreover, a machine learning solution is proposed to improve the computational efficiency of this distributed estimation process by exploiting the history of each agent's trajectory. Finally, extensive numerical results are provided indicating significant improvements compared to other state-of-the-art agent controllers.
\end{abstract}

\thispagestyle{empty} \pagestyle{empty}


\section{Introduction}
This paper considers the problem of controlling a team of mobile agents (\emph{sensors}) deployed to monitor a finite set of points of interest (\emph{targets}) in a mission space. Each target state follows independent (from other targets) stochastic dynamics and the goal of the agent team is to estimate the target states so that an overall measure of estimation error covariance evaluated over a finite period is minimized.

As introduced in \cite{Welikala2020P7}, this problem is different from conventional \emph{distributed estimation} problems \cite{He2020} because: 
(i) we use mobile, rather than stationary, sensors,
(ii) we aim to estimate a distributed set of target states rather than a common global state and  
(iii) we focus on developing an optimal distributed control strategy for the mobile sensor trajectories rather than an optimal fusion framework for the distributed sensor measurements. 
However, the proposed approach can still be seen as one of optimal data fusion, but one that estimates a potentially large number of target states using a typically small number of sensors.  
In fact, this shortage of sensors is what motivates the exploitation of each sensor's mobility. 
We also highlight that this is the same motivation that expanded the study of conventional optimal \emph{coverage} problems \cite{Zhong2011} (where stationary agents are used to monitor a given mission space) to optimal \emph{persistent monitoring} problems \cite{Zhou2019} (where mobile agents are used). Considering this analogy and many other similarities, we cast this optimal estimation problem as an optimal \emph{persistent monitoring} problem.

In the literature, persistent monitoring problems have been widely studied and find many applications such as in sensing \cite{Trevathan2018}, data collection \cite{Smith2011}, surveillance \cite{Leahy2016} and energy management \cite{Mathew2015}. Many variants of persistent monitoring problems have been considered in the literature under different forms of (i) target state dynamics \cite{Zhou2019,Rezazadeh2019}, (ii) global objectives \cite{Pinto2021,Yu2015,Welikala2020J2,Hari2019}, (iii) agent motion dynamics \cite{Welikala2020J4,Wang2017} and (iv) mission spaces \cite{Zhou2019,Maini2018,Zhou2018,Song2014}.

A closely related persistent monitoring problem is studied in \cite{Zhou2019} where the target state dynamics are assumed to be deterministic (i.e., each target state itself is a measure of uncertainty with no explicit stochasticity) and the agent team is tasked with minimizing the target state (uncertainty) values via sensing targets. To find the optimal agent trajectories in this problem setting, \cite{Zhou2019} proposes a network (graph) abstraction for the target-agent system along with a gradient-based distributed on-line parametric control solution. The subsequent work in \cite{Welikala2020J2} appends to this solution a centralized off-line stage to find an effective set of periodic agent trajectories as an initial condition. For the same persistent monitoring problem, our recent work in \cite{Welikala2020J4} takes an alternative approach and develops a distributed on-line solution based on Event-Driven Receding Horizon Control (RHC) \cite{Li2006}. This RHC solution has many attractive features, such as being gradient-free, parameter-free, initialization-free, computationally cheap and adaptive to various forms of state and system perturbations.

In contrast to \cite{Zhou2019,Welikala2020J2,Welikala2020J4}, the persistent monitoring problems considered in this paper and \cite{Pinto2021,Park2019,Lan2013} are more challenging as they assume that each target state follows independent stochastic dynamics and task the agent team to persistently estimate the set of target states so that an overall measure of error covariance associated with target state estimates is minimized. 
However, despite such differences in the problem setup, this class of persistent monitoring problems can be addressed by adopting the key concepts used in \cite{Welikala2020J2,Welikala2020J4}. 
For example, \cite{Pinto2021} formulates a minimax problem over an infinite horizon and proposes a centralized off-line periodic solution inspired by \cite{Welikala2020J2}.
Similarly, this paper considers a mean overall estimation error covariance objective evaluated over a finite horizon and develops a distributed on-line RHC solution (not constrained to be periodic) inspired by \cite{Welikala2020J4}.

While \cite{Pinto2021} and \cite{Lan2013} adopt the persistent monitoring setting to address the underlying estimation task, they only consider single-agent scenarios. Therefore, they require additional clustering and assignment stages to handle multi-agent scenarios (analogous to \cite{Welikala2020J2}). Moreover, both \cite{Pinto2021} and \cite{Lan2013} consider infinite horizon objective functions and develop periodic solutions in a centralized off-line stage. 
In contrast, \cite{Park2019} and this work (both of which also adopt the persistent monitoring setting to address the underlying estimation task) use finite horizon objective functions and develop distributed solutions well-suited for multi-agent scenarios. 
However, the solution proposed in \cite{Park2019} is computationally expensive, off-line and time-driven. To address these limitations, this work restricts the target state dynamics to a one-dimensional space (as in \cite{Zhou2019,Welikala2020J2,Welikala2020J4}) and develops a computationally efficient, on-line and event-driven persistent monitoring solution.

The contributions of this paper are as follows. 
First, it is shown that each agent's trajectory is fully defined by the sequence of \emph{control decisions} it makes at specific discrete event times. 
Second, a \emph{receding horizon control problem} (RHCP) is formulated for an agent to solve at any one of these event times so as to determine the immediate set of optimal control decisions to execute within a \emph{planning horizon}. 
Hence, the event-driven nature of this control approach significantly reduces the computational complexity due to its flexibility in the frequency of control updates. 
A novel element in this RHCP is that, unlike conventional RHC where the planning horizon is an exogenously selected global parameter \cite{Li2006,Khazaeni2018,Chen2020,Ma2020}, it can simultaneously determine the \emph{optimal} planning horizon length along with the optimal control decisions locally and asynchronously by each agent.
Note that, similar to all RHC solutions \cite{Li2006,Khazaeni2018,Chen2020}, the determined optimal control decisions are subsequently executed only over a shorter \emph{action horizon} defined by the next event that the agent observes, thus defining an event-driven process. 
Third, a novel RHCP objective function form (as opposed to the one used in \cite{Welikala2020J4}) is proposed to maximize the utilization of each agent's sensing capabilities over its planning horizon. 
Properties of this RHCP objective function form are studied, leading to establish its unimodality under certain conditions. 
This unimodality property is crucial as it ensures that each agent can independently solve their RHCPs globally and computationally efficiently using a simple gradient descent algorithm.
In addition, a machine learning solution is developed to improve the computational efficiency of the proposed RHC-based agent controllers (i.e., of the overall distributed estimation process) by exploiting the history of each agent's optimal controls. 
Finally, the performance of the proposed RHC-based agent controllers is investigated in terms of providing accurate target state estimates and enabling target state controls compared to other state-of-the-art agent controllers.

Finally, we summarize some fundamental differences of this work compared to the closely related work \cite{Welikala2020P7,Zhou2019,Pinto2021,Welikala2020J2,Welikala2020J4,Welikala2021Ax6,Pinto2020}. 
The work in \cite{Zhou2019,Welikala2020J2,Welikala2020J4,Welikala2021Ax6} considers a class of persistent monitoring problems where the agents are tasked with regulating deterministic piece-wise linear target state trajectories. In particular, \cite{Zhou2019,Welikala2020J2} propose parametric control solutions that require parameter tuning stages and \cite{Welikala2020J4,Welikala2021Ax6} propose distributed on-line solutions based on RHC. Compared to \cite{Zhou2019,Welikala2020J2,Welikala2020J4,Welikala2021Ax6}, the persistent monitoring problems considered in this paper and \cite{Pinto2020,Pinto2021,Welikala2020P7} are entirely different as they task the agents with estimating stochastic linear target state trajectories. In particular, \cite{Pinto2020} proposes a parametric control solution, \cite{Pinto2021} proposes a centralized off-line periodic solution, and this paper and \cite{Welikala2020P7} propose distributed on-line solutions based on RHC. Compared to \cite{Welikala2020P7}, here we provide: 1) the details of all the involved RHCPs, 2) a more mild and intuitive assumption, 3) several new theoretical results and all the proofs, 5) a machine learning solution to improve the computational efficiency and 6) several new numerical results.

The paper is organized as follows. The problem formulation is presented in Section \ref{Sec:ProblemFormulation} and a few preliminary theoretical results are discussed in Section \ref{Sec:PreliminaryResults}. Section \ref{Sec:RHCPFormulation} and \ref{Sec:RHCPSolution} present the RHCP formulation and its solution, respectively. The subsequent Section \ref{Sec:RHCLSolution} describes how machine learning can be used to assist in solving RHCPs efficiently. The performance of the proposed RHC method is demonstrated using simulation results in Section \ref{Sec:SimulationResults}. Finally, concluding remarks and future work are provided in Section \ref{Sec:Conclusion}.

\section{Problem Formulation}
\label{Sec:ProblemFormulation}
We consider $M$ stationary nodes (targets) in the set $\mathcal{V} = \{1,2,\ldots,M\}$ and $N$ mobile agents (sensors) in the set $\mathcal{A} = \{1,2,\ldots,N\}$ located in an $l$-dimensional mission space. The location of a target $i\in\mathcal{V}$ is fixed at $Y_i\in\R^l$ and that of an agent $a\in\mathcal{A}$ at time $t$ is denoted by $s_a(t)\in \R^l$.

\paragraph{\textbf{Graph Topology}}
A directed graph topology $\mathcal{G}=(\mathcal{V},\mathcal{E})$ is embedded into the mission space so that the \emph{targets} are represented by the graph \emph{vertices} $\mathcal{V}$ and the inter-target \emph{trajectory segments} available for agents to travel between targets are represented by the graph \emph{edges} $\mathcal{E}\subseteq \{(i,j):i,j\in\mathcal{V}\}$. 
These trajectory segments are allowed to take arbitrary shapes in $\R^l$ to account for possible constraints in the mission space and agent dynamics. We use $\rho_{ij}$ to represent the \emph{travel time} that an agent spends on a trajectory segment $(i,j)\in\mathcal{E}$ to reach target $j$ from target $i$. 

\paragraph{\textbf{Target Dynamics}}
Each target $i\in\mathcal{V}$ has an associated state $\phi_i(t)\in\R^n$ that follows the dynamics
\begin{equation}
    \label{Eq:TargetStateDynamics}
    {\dot{\phi}}_i(t) = A_i{\phi}_i(t) + B_i\upsilon_i(t) + {w}_i(t),
\end{equation}
where $\{w_i(t)\}_{i\in\mathcal{V}}$ are mutually independent, zero mean, white, Gaussian distributed processes with $E[{w}_i(t){w}^\prime_i(t)]=Q_i$. Similar to \cite{Pinto2021,Park2019,Lan2013}, in this work, the main focus is on persistently maintaining an accurate set of estimates of these target states using the fleet of mobile agents as target state sensors. Hence, in this setting, agents are not responsible for controlling target states and thus we assume that each target $i$ independently selects (or is aware of) its control input $\upsilon_i(t)$.

When an agent visits a target $i\in\mathcal{V}$, it takes the measurements $z_i(t)\in\R^m$ which follow a linear observation model
\begin{equation}
    \label{Eq:TargetObservationModel}
    {z}_{i}(t)=H_i{\phi}_i(t)+{v}_{i}(t),
\end{equation}
where $\{v_i(t)\}_{i\in\mathcal{V}}$ are mutually independent, zero mean, white, Gaussian distributed processes with $E[{v}_i(t){v}^\prime_i(t)]=R_i$.

The matrices in \eqref{Eq:TargetStateDynamics} and \eqref{Eq:TargetObservationModel}: $A_i,\,B_i,\,Q_i,\,H_i$ and $R_i$ are time invariant and known at target $i\in\mathcal{V}$. Moreover, the matrices $Q_i$ and $R_i$ are positive definite.

\paragraph{\textbf{Kalman-Bucy Filter}}
Considering the models \eqref{Eq:TargetStateDynamics} and \eqref{Eq:TargetObservationModel}, the maximum likelihood estimator $\hat{\phi}_i(t)$ of the target state $\phi_i(t)$ is a Kalman-Bucy filter \cite{Athans1967} evaluated at target $i$\,:
\begin{equation*}\label{Eq:StateEstimatorDynamics}
    \dot{\hat{{\phi}}}_i(t) = A_i\hat{{\phi}}_i(t)+B_i\upsilon_i(t)+\eta_i(t)\Omega_i(t){H}_i'{R}_i^{-1}\left({{z}}_i(t)-{H}_i\hat{{\phi}}_i(t)\right),
\end{equation*}
where $\Omega_i(t)$ is the estimation \emph{error covariance} (i.e., $\Omega_i(t) = E(e_i(t)e_i^{\prime}(t))$ with $e_i(t) = \phi_i(t)-\hat{\phi}_i(t)$) given by the matrix Riccati equation \cite{Nazarzadeh1998}:
\begin{equation}\label{Eq:ErrorCovarianceDynamics}
    \dot{\Omega}_i(t) = A_i\Omega_i(t)+ \Omega_i(t)A_i'+Q_i-\eta_i(t)\Omega_i(t){G}_i\Omega_i(t), 
\end{equation}
with $G_i=H_i^{\prime}R_i^{-1}H_i$ and 
\begin{equation}\label{Eq:AgentPresenceAtATarget}
    \eta_i(t) \triangleq \textbf{1}\{\mbox{Target $i$ is observed by an agent at time $t$}\},
\end{equation}
($\textbf{1}\{\cdot\}$ is the usual indicator function).

Based on \eqref{Eq:ErrorCovarianceDynamics}, when a target $i\in\mathcal{V}$ is being sensed by an agent (i.e., when $\eta_i(t)=1$), the covariance $\Omega_i(t)$ decreases and, as a result, the state estimate $\hat{\phi}_i$ becomes more accurate. The opposite occurs when a target is not being observed by an agent. It is also worth pointing out that the dynamics of the covariance $\Omega_i(t)$ \eqref{Eq:ErrorCovarianceDynamics} are independent of the target state control $\upsilon_i(t)$ in \eqref{Eq:TargetStateDynamics}, due to the principle of separation \cite{Friedland2012}.

\paragraph{\textbf{Agent Model}}
According to the adopted graph topology, we assume that once an agent $a\in\mathcal{A}$ spends the required travel time to reach a target $i\in\mathcal{V}$, its location $s_{a}(t)$ falls within a certain range from the target location $Y_i$ which enables establishing a constant sensing ability (i.e., $H_i$, $R_i$ are fixed) of the target state $\phi_i(t)$. Without loss of generality, we denote the number of agents present at target $i\in\mathcal{V}$ at time $t$ as 
\begin{equation}\label{Eq:NumOfAgentsAtTarget}
 N_i(t) \triangleq \sum_{a\in\mathcal{A}}\textbf{1}\{s_a(t)=Y_i\}.   
\end{equation}

To prevent resource (agent sensing) wastage and simplify the analysis, similar to \cite{Welikala2020J2,Welikala2020J4}, we next introduce a control constraint that prevents \emph{simultaneous target sharing} by multiple agents at each target $i\in\mathcal{V}$ as 
\begin{equation}\label{Eq:NoTargetSharing}
    N_i(t) \in \{0,1\},\ \forall t\geq 0.
\end{equation}
Evidently, this constraint only applies if $N\geq2$. Moreover, under \eqref{Eq:NoTargetSharing}, according to the definitions \eqref{Eq:AgentPresenceAtATarget} and \eqref{Eq:NumOfAgentsAtTarget}, $\eta_i(t)=N_i(t)$. Also, note that due to the use of a fixed set of travel times $\{\rho_{ij}:(i,j)\in\mathcal{E}\}$, the analysis in this paper is independent of the agent motion dynamic model (similar to \cite{Zhou2019,Welikala2020J2,Welikala2020J4}). However, analogous to the work in \cite{Welikala2021Ax6}, with some modifications, the proposed solution in this paper can be adapted to accommodate specific agent dynamic models.

\paragraph{\textbf{Global Objective}}
The goal is to design controllers for the agents to minimize the finite horizon global objective
\begin{equation}\label{Eq:GlobalObjective}
    J_T \triangleq \frac{1}{T}\int_0^T\sum_{i\in\mathcal{V}}\mbox{tr}(\Omega_i(t)) dt.
\end{equation}
This choice of global objective \eqref{Eq:GlobalObjective} was inherited from related prior work \cite{Pinto2020}. Based on its form, clearly it motivates finding agent trajectories that minimize target state estimation error covariance values across the target network over the specified finite horizon. Therefore, \eqref{Eq:GlobalObjective} is a natural and intuitive choice for the global objective. Moreover, as we will see in the sequel, \eqref{Eq:GlobalObjective} is simple enough to be conveniently decomposed, and thus, it also motivates the development of distributed on-line optimal agent control strategies.

Nevertheless, we also propose an alternative global objective 
\begin{equation}\label{Eq:GlobalObjectiveHat}
    \hat{J}_T  \triangleq -
\frac{\int_0^T\sum_{i\in\mathcal{V}}\eta_i(t)\mbox{tr}(\Omega_i(t))dt}
{\int_0^T\sum_{i\in \mathcal{V}}\mbox{tr}(\Omega_i(t))dt}.
\end{equation}
First, note that the denominator of \eqref{Eq:GlobalObjectiveHat} is proportional to \eqref{Eq:GlobalObjective}. Hence, optimizing (minimizing) \eqref{Eq:GlobalObjectiveHat} will lead to agent trajectories that minimize \eqref{Eq:GlobalObjective} (notice the negative sign in \eqref{Eq:GlobalObjectiveHat}). 
Second, note that the numerator of \eqref{Eq:GlobalObjectiveHat} represents a quantity that reflects the overall agent \emph{sensing effort} as $\eta_i(t)=1$ only when an agent senses the target $i$. 
Therefore, this alternative objective \eqref{Eq:GlobalObjectiveHat}, omitting the negative sign, can be seen as the efficiency of overall agent sensing effort, i.e., the fraction of resources (agents) used for sensing (as opposed to both sensing and traveling).
Hence, it is clear that optimizing \eqref{Eq:GlobalObjectiveHat} requires agents to optimally allocate their sensing resources across the target network over the finite horizon $[0,T]$. Moreover, it is worth pointing out that compared to \eqref{Eq:GlobalObjective}, \eqref{Eq:GlobalObjectiveHat} is conveniently normalized, thus its value has a direct intuitive interpretation.

\begin{figure}[!h]
     \centering
     \begin{subfigure}[b]{0.48\columnwidth}
         \centering
         \includegraphics[width=\textwidth]{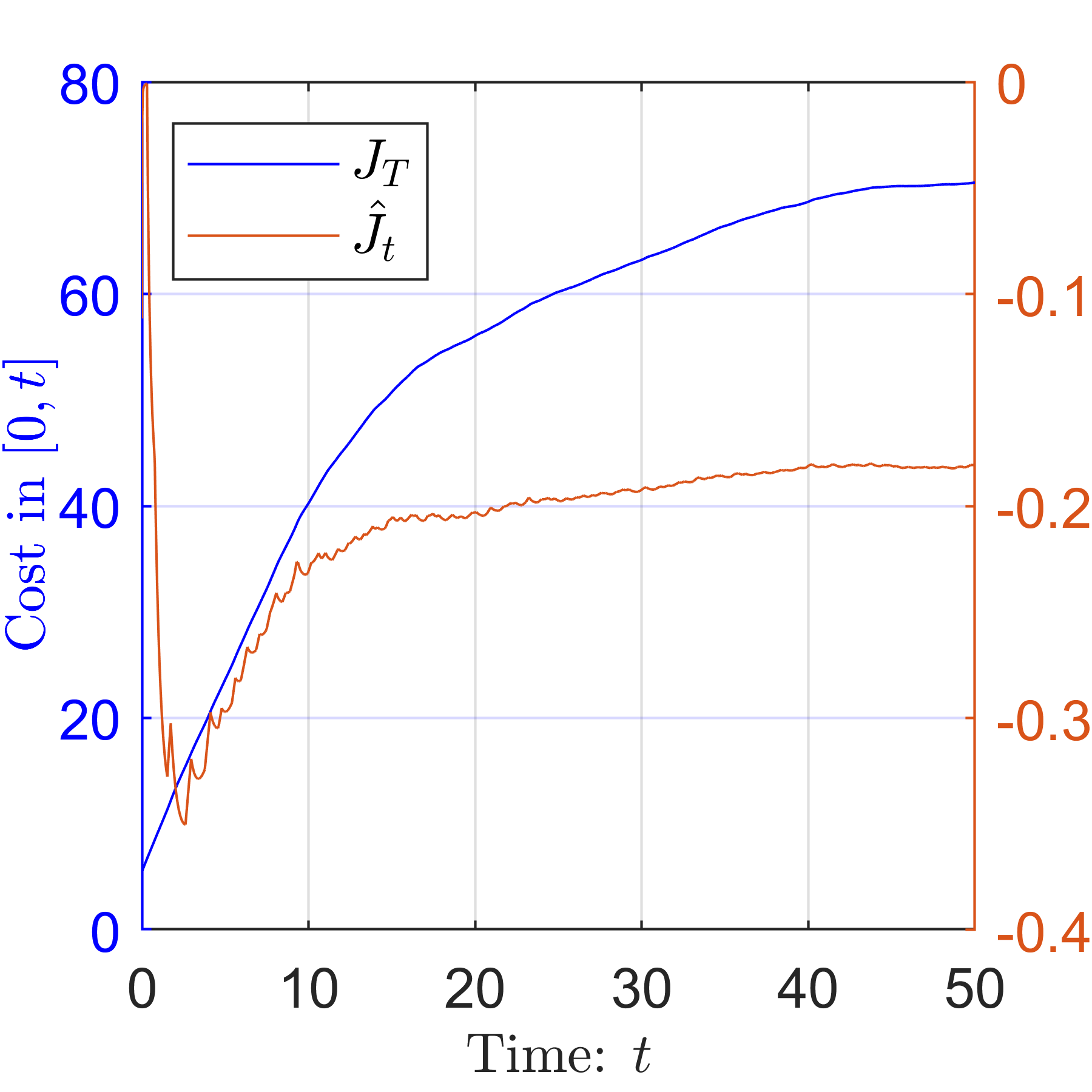}
         \caption{Accumulated cost}
         \label{SubFig:1}
     \end{subfigure}
     \begin{subfigure}[b]{0.48\columnwidth}
         \centering
         \includegraphics[width=\textwidth]{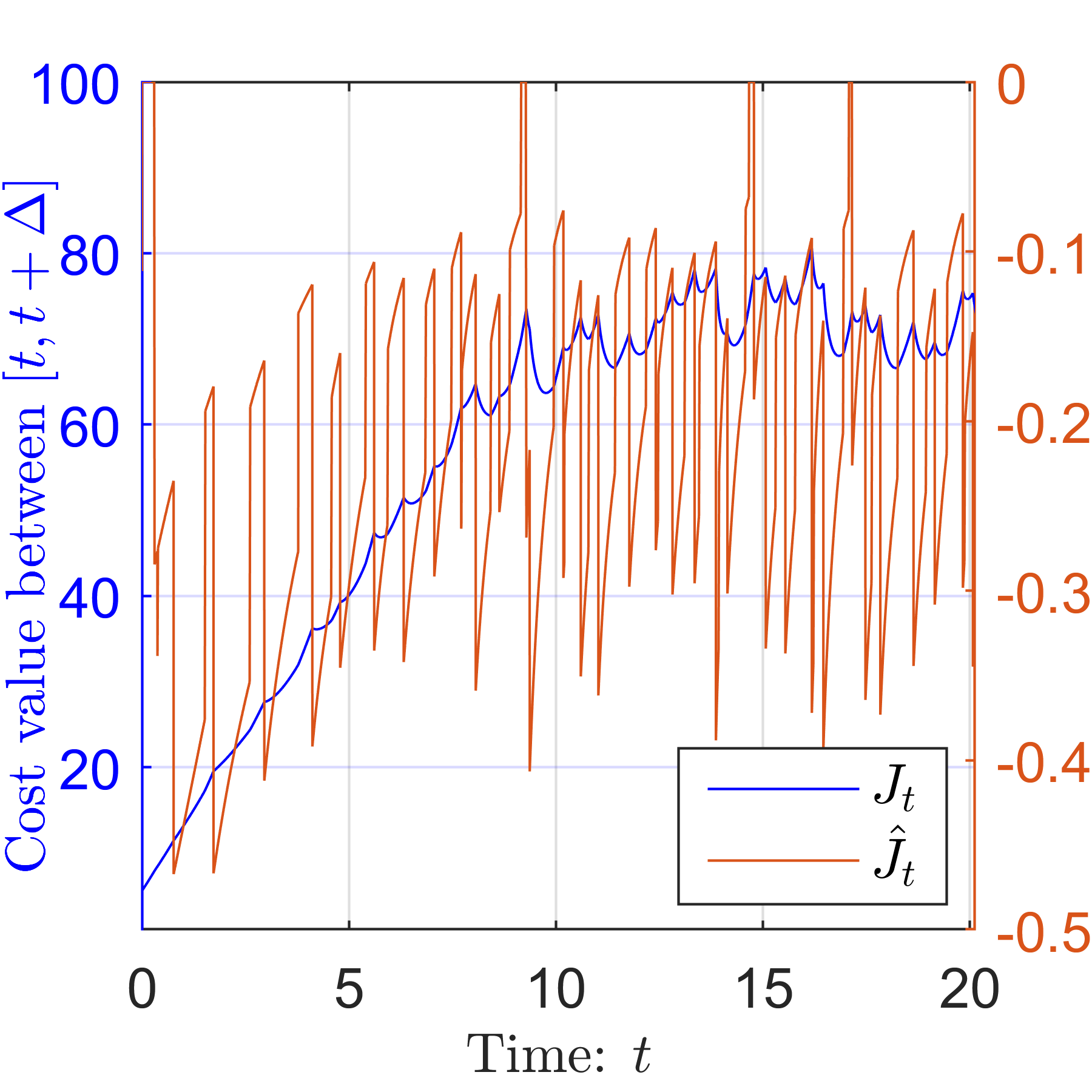}
         \caption{Instantaneous cost}
         \label{SubFig:2}
     \end{subfigure}
     \caption{
     {
     Comparison between $J_t$ in \eqref{Eq:GlobalObjective} and $\hat{J}_t$ in \eqref{Eq:GlobalObjectiveHat}
     }
     }
     \label{Fig:ObjectiveFunctionComparison}
\end{figure}

A final observation is that numerical experiments have shown that both $J_t$ \eqref{Eq:GlobalObjective} and $\hat{J}_t$ \eqref{Eq:GlobalObjectiveHat} profiles (under the same agent controls over $t\in[0,T]$) behave in the same manner after a brief transient phase (e.g., see Figs. \ref{Fig:ObjectiveFunctionComparison}(a) and \ref{Fig:ObjectiveComparison}). Therefore, we can see that both \eqref{Eq:GlobalObjective} and \eqref{Eq:GlobalObjectiveHat} are equally appropriate global objective function choices for this optimal estimation problem.  

As mentioned earlier, we view this optimal estimation problem as a \emph{persistent monitoring on networks} (PMN) problem. Moreover, we assume that the initial condition of this PMN problem setup, defined by $\phi_i(0)$, $\hat{\phi}_i(0)$ and $\Omega_i(0)$, $\forall i\in\mathcal{V}$, is known at respective targets. 

\paragraph{\textbf{Agent Control}}
Based on the graph topology $\mathcal{G} = (\mathcal{V},\mathcal{E})$, we define the \emph{neighbor set} and the \emph{neighborhood} of a target $i\in\mathcal{V}$ as 
$
    \mathcal{N}_{i}\triangleq\{j:(i,j)\in\mathcal{E}\} \ \mbox{ and }\ 
    \bar{\mathcal{N}}_{i} \triangleq \mathcal{N}_{i}\cup\{i\},
$
respectively. The basic agent controls are as follows. Whenever an agent $a\in\mathcal{A}$ is ready to leave a target $i\in\mathcal{V}$, it selects a \emph{next-visit} target $j\in\mathcal{N}_{i}$. Thereafter, the agent travels over $(i,j)\in\mathcal{E}$ to arrive at target $j$ upon spending $\rho_{ij}$ amount of time. Subsequently, it selects a \emph{dwell-time} $u_{j}\in\mathbb{R}_{\geq0}$ to spend at target $j$ (which contributes to decreasing $\Omega_{j}(t)$, making $\hat{\phi}_j(t)\rightarrow \phi_j(t)$), and then makes another next-visit decision.

Therefore, the overall control exerted by an agent consists of a sequence of \emph{dwell-times} $u_{i}\in\mathbb{R}_{\geq0}$ and \emph{next-visit} targets $j\in\mathcal{N}_{i}$. Our goal is to determine $(u_{i}(t),j(t))$ for any agent residing at a target $i\in\mathcal{V}$ at any time $t\in[0,T]$ which are optimal in the sense of minimizing the global objective \eqref{Eq:GlobalObjective}. 

This PMN problem is much more complicated than the well-known NP-hard traveling salesman problem (TSP) based on the metrics: 
(i) the number of decision variables, and 
(ii) the nature of feasible decision spaces.
In particular, note that in PMN: 
(i) we need to determine dwell-times at each visited target, and account for   
(ii) the involved target dynamics, 
(iii) the presence of multiple agents, and 
(iv) the freedom to make multiple visits to targets. 

The same reasons make it computationally intractable to apply dynamic programming techniques so as to obtain the optimal controls, even for a relatively simple PMN problem.

\paragraph{\textbf{Receding Horizon Control}}
In order to address this hard dynamic optimization problem, this paper proposes an \emph{Event-Driven Receding Horizon Controller} (RHC) at each agent $a\in\mathcal{A}$. Even though the key idea behind RHC derives from Model Predictive Control (MPC), it exploits the event-driven nature of the considered problem to significantly reduce the complexity by effectively decreasing the frequency of control updates. As introduced and extended later on in \cite{Li2006} and \cite{Khazaeni2018,Chen2020,Welikala2020J4}, respectively, the RHC method involves solving an optimization problem of the form \eqref{Eq:GlobalObjective} limited to a finite \emph{planning horizon}, whenever an event of interest to the (agent) controller is observed. The determined optimal controls are then executed over an \emph{action horizon} defined by the occurrence of the next such event. This event-driven process is continued iteratively. 

Pertaining to the PMN problem considered in this paper, the RHC, when invoked at time $t$ for an agent residing at target $i\in\mathcal{V}$, aims to determine 
(i) the immediate dwell-time $u_i$ at target $i$, (ii) the next-visit target $j\in\mathcal{N}_{i}$ and (iii) the next dwell-time $u_j$ at target $j$. These \emph{control decisions} are jointly denoted by $U_{i}(t) \triangleq [u_i(t),\,j(t),\,u_j(t)]$ and its optimal value is obtained by solving an optimization problem of the form 
\begin{equation}\label{Eq:RHCProblem0}
U_{i}^{\ast}(t) = \underset{U_{i}(t)\in\mathbb{U}(t)}{\arg\min}\ \left[J_{H}(X_{i}(t),U_{i}(t);\,H)+\hat{J}_{H}(X_{i}(t+H))\right],  %
\end{equation}
where $X_{i}(t)$ is the current local state and $\mathbb{U}(t)$ is the feasible control set at $t$ (exact definition is provided later). The term $J_{H}(X_{i}(t),U_{i}(t);H)$ represents the immediate cost over the planning horizon $[t,t+H]$ and $\hat{J}_{H}(X_{i}(t+H))$ stands for an estimate of the future cost based on the state at $t+H$.

In contrast to standard methods where the planning horizon length $H$ is exogenously selected, here we adopt the \emph{variable horizon} approach proposed in \cite{Welikala2020J4} where the planning horizon length is treated as an upper-bounded function of control decisions $w(U_{i}(t)) \leq H$. In other words, we constrain the planning horizon to be $[t,t+w(U_i(t))] \subseteq [t, t+H]$ where $[t, t+H]$ is now treated as a predefined \emph{fixed planning horizon}.

Hence, this approach incorporates the selection of planning horizon length $w(U_{i}(t))$ into the optimization problem \eqref{Eq:RHCProblem0}, which now can be re-stated as 
\begin{equation}
\label{Eq:RHCProblem}
\begin{alignedat}{4}
& U_{i}^{\ast}(t) = 
& \underset{\makebox[2cm]{\footnotesize $U_{i}(t)\in\mathbb{U}(t)$}}{\arg\min} & \ && J_{H}(X_{i}(t),U_{i}(t);\,w(U_{i}(t))) & \\
&  & \makebox[2cm]{subject to} &  && w(U_{i}(t))\leq H. &
\end{alignedat}
\end{equation}
Note that the term $\hat{J}_{H}(X_{i}(t+H))$ in \eqref{Eq:RHCProblem0} has been omitted in \eqref{Eq:RHCProblem}. This is to make the proposed RHC method \emph{distributed} so that each agent can separately solve \eqref{Eq:RHCProblem} using only its local state information.

However, since we now optimize the planning horizon length $w(U_i(t))$, this compensates for the intrinsic inaccuracies resulting from the said omission of $\hat{J}_{H}(X_{i}(t+H))$. This claim is justified in the following remark. 

\begin{remark}
Previously in \eqref{Eq:RHCProblem0}, the involved $\hat{J}_{H}(X_{i}(t+H))$ term motivated an agent to optimize its state at the end of the planning horizon (i.e., the \emph{state}: $X_{i}(t+H)$) via optimizing the control decisions $U_i(t)$. However, under this new setting in \eqref{Eq:RHCProblem}, in lieu of the $\hat{J}_{H}(X_{i}(t+H))$ term, the involved $w(U_i(t))$ term motivates an agent to optimize its end of the planning horizon (i.e., the \emph{time}: $t+w(U_i(t))$) via optimizing the control decisions $U_i(t)$. Therefore, the optimization of the planning horizon can be seen as a compensation for the omission of the future cost. 
\end{remark}

Finally, we also point out that the said omission of the future cost term in \eqref{Eq:RHCProblem} calls for a systematic design of the objective function $J_{H}(X_{i}(t),U_{i}(t);\,w(U_{i}(t)))$ in \eqref{Eq:RHCProblem}. 
Upon establishing some preliminary results in Section \ref{Sec:PreliminaryResults} we will return to discussing this topic in Section \ref{Sec:RHCPFormulation}.

\section{Preliminary Results}
\label{Sec:PreliminaryResults}

Based on \eqref{Eq:ErrorCovarianceDynamics}, the error covariance $\Omega_i(t)$ of any target $i\in\mathcal{V}$ is continuous and piece-wise differentiable. Specifically, $\dot{\Omega}_{i}(t)$ jumps only when one of the following two (strictly local) \emph{events} occurs:
(\romannum{1}) an agent arrival at target $i$, or 
(\romannum{2}) an agent departure from target $i$. 
These two events occur alternatively and respectively trigger two different \emph{modes} of subsequent $\Omega_i(t)$ behaviors, named \emph{active} and \emph{inactive} modes, described by using $\eta_i(t)=1$ and $\eta_i(t) = 0$ in \eqref{Eq:ErrorCovarianceDynamics}. In the following discussion, we use $\eta_i(t)\in\{0,1\}$ to represent the mode of target $i\in\mathcal{V}$ and $I_n$ as the identity matrix in $\R^n$.

\begin{lemma}\label{Lm:ErrorCovarianceProfileActive}
If a target $i\in\mathcal{V}$ is in the mode $\eta_i(t)\in\{0,1\}$ during a time period $[t_0,t_1]$, its error covariance $\Omega_i(t)$ for any time $t \in [t_0,t_1]$ is given by  
\begin{equation}\label{Eq:ErrorCovarianceProfileActive}
    \Omega_i(t) = C_i(t)D_i^{-1}(t),
\end{equation}
where
\footnotesize 
$
\begin{bmatrix}
    C_i(t)\\D_i(t) 
    \end{bmatrix} 
    = e^{\Psi_i(t-t_0)} 
    \begin{bmatrix}
    \Omega_i(t_0)\\ I_n 
    \end{bmatrix}
$
\normalsize
 with 
\footnotesize 
$ \Psi_i = 
\begin{bmatrix}
A_i & Q_i \\ \eta_i(t)G_i & -A_i^T 
\end{bmatrix}   
$\normalsize.
\end{lemma}

\emph{Proof: }
Omitting the argument $t$ for notational convenience and using the substitution $\Omega_i = C_iD_i^{-1}$ in \eqref{Eq:ErrorCovarianceDynamics} gives
$$
\dot{C}_i - C_iD_i^{-1} \dot{D}_i = 
(A_iC_i + Q_iD_i) - C_iD_i^{-1}(\eta_iG_iC_i -A_i^TD_i).
$$
Notice that both sides of the above equation takes an affine linear form with respect to $-C_iD_i^{-1}$. Therefore, equating the coefficients of $1$ and $-C_iD_i^{-1}$ terms above gives
$$
\begin{bmatrix}
\dot{C}_i(t) \\ \dot{D}_i(t)
\end{bmatrix} 
=
\begin{bmatrix}
A_i & Q_i\\
\eta_i(t) G_i & -A_i^T
\end{bmatrix}
\begin{bmatrix}
C_i(t)\\
D_i(t)
\end{bmatrix}.
$$
Recall that $\eta_i(t) = 1$ if the target $i$ is active and $\eta_i(t) = 0$ otherwise. Finally, setting the initial conditions: $C_i(t_0) = \Omega_i(t_0)$ and $D_i(t_0) = I_n$, the above linear differential equation can be solved to obtain the result in \eqref{Eq:ErrorCovarianceProfileActive}.
\hfill $\blacksquare$

Using the above lemma, a simpler expression for $\Omega_i(t)$ than \eqref{Eq:ErrorCovarianceProfileActive} can be derived if target $i$ is in the inactive mode: $\eta_i(t)=0$.

\begin{corollary}\label{Co:ErrorCovarianceProfileInactive}
If a target $i\in\mathcal{V}$ is inactive during $t\in[t_0,t_1]$, the corresponding $\Omega_i(t)$ for any time $t \in [t_0,t_1]$ is given by
\begin{align}\label{Eq:ErrorCovarianceProfileInactive}
   \Omega_i(t) = \Phi_i(t)
   \Big[ 
   \Omega_i(t_0)+ (t-t_0) \int_0^1 \psi_i(t,x) dx
   \Big]\Phi_i^T(t), 
\end{align}
where $\Phi_i(t) = e^{A_i(t-t_0)}$ and $\psi_i(t,x) = (\Phi_i(t))^{-x}Q_i(\Phi_i^T(t))^{-x}$.
\end{corollary}

\emph{Proof: }According to Lemma \ref{Lm:ErrorCovarianceProfileActive}, when $\eta_i(t)=0$, $\Omega_i(t)$ is given by \eqref{Eq:ErrorCovarianceProfileActive} where $\Psi_i$ is a block triangular matrix. Therefore, $e^{\Psi_i(t-t_0)}$ can be written (using \cite[p.~1]{Dieci2001}) as
$$
e^{\Psi_i(t-t_0)} = 
\begin{bmatrix}
\Phi_i(t) & (t-t_0)\int_0^1 (\Phi_i^T(t))^{1-x} Q_i (\Phi_i^T(t))^{-x}dx\\
0 & (\Phi_i^T(t))^{-1}
\end{bmatrix}.
$$
Applying this in \eqref{Eq:ErrorCovarianceProfileActive} gives $D_i(t) = \Phi_i^{-1}(t)$ and 
$$
C_i(t) = \Phi_i(t)\Omega_i(t_0) + (t-t_0)\int_0^1 (\Phi_i^T(t))^{1-x} Q_i (\Phi_i^T(t))^{-x}dx.
$$
Finally, $\Omega_i(t) = C_i(t)D_i^{-1}(t)$ gives the expression in \eqref{Eq:ErrorCovarianceProfileInactive}.
\hfill $\blacksquare$

From Lemma \ref{Lm:ErrorCovarianceProfileActive} and Corollary \ref{Co:ErrorCovarianceProfileInactive}, it is clear that the exact form of the ``tr$(\Omega_i(t))$'' expression required for the global objective \eqref{Eq:GlobalObjective} cannot be written more compactly - unless the matrices $A_i$ and $\Psi_i$ have some additional properties.

\paragraph{\textbf{One-Dimensional PMN Problem}}
In the remainder of this paper, similar to \cite{Zhou2019,Welikala2020J2,Welikala2020J4}, we constrain ourselves to one-dimensional target state dynamics and agent observation models by setting  \begin{equation}\label{Eq:OneDimensionConstraint}
    n = m = 1,
\end{equation}
in \eqref{Eq:TargetStateDynamics} and \eqref{Eq:TargetObservationModel}. 
This is a reasonable assumption given that the goal of this work is to derive necessary theoretical results to apply the RHC method and then to explore its feasibility for the considered particular PMN problem setup. In future work, we expect to generalize these theoretical results and the RHC solution to higher-dimensional target state models. Therefore, we henceforth consider the fixed target parameters $A_i,\,Q_i,\,H_i,\,R_i,\,G_i$ and the time-varying quantities $\phi_i(t),\,z_i(t),\,\hat{\phi}_i(t),\,\Omega_i(t),\,\forall i\in\mathcal{V}$ as scalars.

\paragraph{\textbf{Local Contribution}}
The contribution to the global objective $J_T$ in \eqref{Eq:GlobalObjective} by a target $i\in\mathcal{V}$ during a time period $[t_0,t_1]$ is defined as $\frac{1}{T}J_i(t_0,t_1)$ where  
\begin{equation}\label{Eq:LocalContribution}
    J_i(t_0,t_1) \triangleq \int_{t_0}^{t_1}\mbox{tr}(\Omega_i(t))dt = 
    \int_{t_0}^{t_1}\Omega_i(t)dt. 
\end{equation}
We further define the corresponding active and inactive portions of the above \emph{local contribution} term $J_i(t_0,t_1)$ in \eqref{Eq:LocalContribution} respectively as $J_i^A(t_0,t_1)$ and $J_i^I(t_0,t_1)$ where 
\begin{equation}\label{Eq:LocalContributionComponents}
\begin{aligned}
    J_i^A(t_0,t_1) &\triangleq \int_{t_0}^{t_1}\eta_i(t)\Omega_i(t)dt\ \mbox{ and }\ \\
    J_i^I(t_0,t_1) &\triangleq \int_{t_0}^{t_1}(1-\eta_i(t))\Omega_i(t)dt. 
\end{aligned}
\end{equation}
Notice that, by definition, $J_i(t_0,t_1) = J_i^A(t_0,t_1) + J_i^I(t_0,t_1)$.

\begin{lemma}\label{Lm:ErrorCovarianceActive}
If a target $i\in\mathcal{V}$ is \emph{active} during  $t\in[t_0,t_1]$, the corresponding $\Omega_i(t)$ for any time $t \in [t_0,t_1]$ is given by
\begin{equation}\label{Eq:ErrorCovarianceActive}
    \Omega_i(t) = 
    \frac{c_{i1} + c_{i2}e^{-\lambda_i(t-t_0)}}
    {v_{i1}c_{i1} + v_{i2}c_{i2}e^{-\lambda_i(t-t_0)}},
\end{equation}
where 
$\lambda_i = 2\sqrt{A_i^2 + Q_iG_i}$, 
$v_{i1} = \frac{1}{Q_i}(-A_i + \sqrt{A_i^2 + Q_iG_i})$, 
$v_{i2} = \frac{1}{Q_i}(-A_i - \sqrt{A_i^2 + Q_iG_i})$, 
$c_{i1} = v_{i2}\Omega_i(t_0)-1$ and 
$c_{i2} = -v_{i1}\Omega_i(t_0)+1$. The corresponding local contribution $J_i(t_0,t_0+w)$ in \eqref{Eq:LocalContribution} (where $w = (t-t_0)$) is given by $J_i(t_0,t_0+w) = J_i^A(t_0,t_0+w)$ where
\begin{equation}\label{Eq:ContributionActive}
    J_i^A(t_0,t_0+w) = \frac{1}{G_i}
    \log\left(\frac{v_{i1}c_{i1} + v_{i2}c_{i2}e^{-\lambda_i w}}{v_{i2}-v_{i1}}\right)
    + \frac{1}{v_{i1}} w.
\end{equation}
\end{lemma}

\emph{Proof: }We first use Lemma \ref{Lm:ErrorCovarianceProfileActive} to derive \eqref{Eq:ErrorCovarianceActive}. Note that under \eqref{Eq:OneDimensionConstraint}, $\Psi_i$ in \eqref{Eq:ErrorCovarianceProfileActive} is such that $\Psi_i\in\R^{2\times2}$ and it can be simplified using $\eta_i(t)=1$, since, by assumption, the target $i$ is active. The eigenvalues of $\Psi_i$ are $\pm \lambda_i/2$ and the corresponding generalized eigenvector matrix is 
\footnotesize 
$\begin{bmatrix}
1 & 1 \\
v_{i1} & v_{i2}
\end{bmatrix}$\normalsize. Therefore, the matrix exponent $e^{\Psi_i w}$ required in \eqref{Eq:ErrorCovarianceProfileActive} can be evaluated as
$$
e^{\Psi_i w} =  
\begin{bmatrix}
1 & 1 \\ v_{i1} & v_{i2}
\end{bmatrix}
\begin{bmatrix}
e^{\frac{\lambda_i w}{2}} & 0 \\ 0 & e^{-\frac{\lambda_i w}{2}}
\end{bmatrix}
\begin{bmatrix}
1 & 1 \\ v_{i1} & v_{i2}
\end{bmatrix}^{-1}.
$$
Applying this result in \eqref{Eq:ErrorCovarianceProfileActive} gives
$$
\begin{bmatrix}
C_i(t) \\ D_i(t) 
\end{bmatrix}
 = e^{\frac{\lambda_i w}{2}}
\begin{bmatrix}
v_{i2} - v_{i1}e^{-\lambda_i w} & 
-1 + e^{-\lambda_i w} \\
v_{i1}v_{i2}(1 - e^{-\lambda_i w}) &
-v_{i1}+v_{i2}e^{\lambda_i w}
\end{bmatrix}
\begin{bmatrix}
\Omega_i(t_0) \\ 1
\end{bmatrix}.
$$
Since $\Omega_i(t) = C_i(t)/D_i(t)$ (from \eqref{Eq:ErrorCovarianceProfileActive}), we now can use the above result to obtain \eqref{Eq:ErrorCovarianceActive}. 

Finally, using \eqref{Eq:LocalContributionComponents},  \eqref{Eq:ContributionActive} can be obtained by analytically evaluating the integral of $\Omega_i(t)$ over the period $[t_0,t_0+w]$:  
$$
\quad \quad \ \ 
J_i^A(t_0,t_0+w) = \int_{t_0}^{t_0+w}  
\frac{c_{i1} + c_{i2}e^{-\lambda_i(\tau-t_0)}}
{v_{i1}c_{i1} + v_{i2}c_{i2}e^{-\lambda_i(\tau-t_0)}} d\tau.
\quad \quad \blacksquare
$$

\begin{lemma}\label{Lm:ErrorCovarianceInactive}
If a target $i\in\mathcal{V}$ is \emph{inactive} during  $t\in[t_0,t_1]$, the corresponding $\Omega_i(t)$ for any time $t \in [t_0,t_1]$ is given by
\begin{equation}\label{Eq:ErrorCovarianceInactive}
    \Omega_i(t) = \left(\Omega_i(t_0)+\frac{Q_i}{2A_i}\right)e^{2A_i(t-t_0)} - \frac{Q_i}{2A_i}, 
\end{equation}
and the corresponding local contribution $J_i(t_0,t_0+w)$ (where $w = (t-t_0)$) is given by $J_i(t_0,t_0+w) = J_i^I(t_0,t_0+w)$ where
\begin{equation}\label{Eq:ContributionInactive}
    J_i^I(t_0,t_0+w) = \frac{1}{2A_i}
    \left(\Omega_i(t_0)+\frac{Q_i}{2A_i}\right)(e^{2A_iw}-1) - \frac{Q_i}{2A_i}w.
\end{equation}
\end{lemma}

\emph{Proof: }These two results \eqref{Eq:ErrorCovarianceInactive} and \eqref{Eq:ContributionInactive} can be proved by using $\Psi_i$ in \eqref{Eq:ErrorCovarianceProfileActive} as $\Psi_i = [A_i \ Q_i; 0\ -A_i^T]$ and following the same steps used in the proof of Lemma \ref{Lm:ErrorCovarianceActive}.
\hfill $\blacksquare$

\section{RHC problem (RHCP) formulation}
\label{Sec:RHCPFormulation}

\begin{figure}[!h]
\centering
\includegraphics[width=\columnwidth]{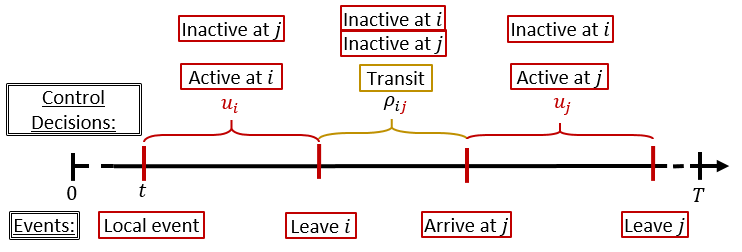} \caption{Event timeline and control decisions under RHC.}%
\label{Fig:OneVisitTimeline}%
\end{figure}

Consider an agent $a\in\mathcal{A}$ residing on a target $i\in\mathcal{V}$ at some time $t\in [0,T]$. Recall that control $U_i(t)$ in \eqref{Eq:RHCProblem} consists of the dwell-time $u_{i}$ at the current target $i$, the next-visit target $j\in\mathcal{N}_{i}$, and the dwell-time $u_{j}$ at the next-visit target $j$ (see Fig. \ref{Fig:OneVisitTimeline}). Therefore, agent $a$ has to optimally select the three control decisions (control vector) $U_{i}(t) = [u_{i}(t),j(t),u_{j}(t)]$.

\paragraph{\textbf{The RHCP}} 

Let us denote the real-valued component of the control vector $U_{i}(t)$ in \eqref{Eq:RHCProblem} as $U_{ij}\triangleq [u_{i},u_{j}]$ (omitting time arguments for notational simplicity). The discrete component of $U_i(t)$ is simply the next-visit target $j\in\mathcal{N}_i$. 
In this setting, we define the planning horizon length $w(U_i(t))$ in  \eqref{Eq:RHCProblem} as 
\begin{equation}
w = w(U_{ij}) \triangleq \vert U_{ij} \vert + \rho_{ij} = u_i +\rho_{ij} + u_j \label{Eq:VariableHorizon}%
\end{equation} 
($|\cdot|$ denotes the cardinality operator or the 1-norm depending on the argument)
so that it covers the control decisions and corresponding controllable events that can occur in the immediate future pertaining to the current neighborhood $\bar{\mathcal{N}}_i$ as shown in Fig. \ref{Fig:OneVisitTimeline}. 
The current local state $X_{i}(t)$ in \eqref{Eq:RHCProblem} is taken as $X_{i}(t)=\{\Omega_{j}(t):j\in\bar{\mathcal{N}}_{i}\}$. Then, the optimal controls are obtained by solving \eqref{Eq:RHCProblem}, which can be re-stated as the following set of optimization problems, henceforth called the RHC Problem (RHCP): 
\begin{align}
&\begin{alignedat}{4}\label{Eq:RHCGenSolStep1}
& U_{ij}^{\ast} = & \underset{\makebox[2cm]{\footnotesize $U_{ij}\in\mathbb{U}$}}{\arg\min} & \quad && J_{H}(X_{i}(t),U_{ij};\,w(U_{ij}));\ \forall j\in\mathcal{N}_{i},\ & \\
& & \makebox[2cm]{subject to} &  && w(U_{ij}) \leq H & \\
\end{alignedat}\\
&\begin{alignedat}{4}\label{Eq:RHCGenSolStep2}
&\ \ j^{\ast} = & \underset{\makebox[2cm]{\footnotesize $j\in\mathcal{N}_{i}$}}{\arg\min} & \quad && J_{H}(X_{i}(t),U_{ij}^{\ast};\,w(U_{ij}^*)). &
\end{alignedat}
\end{align}
Before getting into details, note that \eqref{Eq:RHCGenSolStep1} involves solving $|\mathcal{N}_{i}|$ optimization problems, one for each neighbor $j\in\mathcal{N}_{i}$. Then, \eqref{Eq:RHCGenSolStep2} determines $j^{\ast}$ through a simple numerical comparison. Therefore, the optimal control vector $U_i^*(t)$ of \eqref{Eq:RHCProblem} is the composition: $U_i^*(t) = \{U_{ij^{\ast}}^{\ast},\,j^{\ast}\}$.

The RHCP objective function $J_{H}(\cdot)$ is chosen in terms of the \emph{local objective function} of target $i$, which is denoted by $\bar{J}_i(t_0,t_1)$ over any interval $[t_0,t_1]\subseteq[0,T]$ (the exact definition of $\bar{J}_i(t_0,t_1)$ is provided later on in \eqref{Eq:LocalObjectiveFunction}). In particular, we define the RHCP objective function as the local objective function of target $i$ evaluated over the planning horizon $[t,t+w]$:
\begin{equation}
\begin{aligned} 
J_H(X_i(t),U_{ij};H) = \bar{J}_i(t,t+w),
\end{aligned}\label{Eq:RHCNewChoices}%
\end{equation}
and the RHCP feasible control space as $\mathbb{U} = \{U:U \in \mathbb{R}^2,\ U \geq 0,\ \vert U\vert  + \rho_{ij} \leq H\}$ (including the constraint $w(U_{ij})\leq H$)

\paragraph{\textbf{Planning Horizon}} 
In conventional RHC methods, the RHCP objective function is evaluated over a fixed planning horizon, e.g., $[t,t+H]\subseteq[0,T]$, where $H$ is selected exogenously. This leads to control solutions that are dependent on the choice of the used fixed planning horizon length $H$. When developing on-line control methods, having such a dependence on a predefined parameter is undesirable, as it prevents the controller from having the opportunity to fine-tune $H$ and re-evaluate its controls. 

However, through \eqref{Eq:VariableHorizon} and \eqref{Eq:RHCNewChoices} above, we have made the RHCP solution (i.e., \eqref{Eq:RHCGenSolStep1} and \eqref{Eq:RHCGenSolStep2}) free of the parameter $H$ (i.e., fixed planning horizon length) by only using $H$ as an upper-bound to the actual planning horizon length $w(U_{ij})$ in \eqref{Eq:VariableHorizon} and selecting $H$ to be sufficiently large (e.g., $H = T-t$). Moreover, since the planning horizon length $w(U_{ij})$ is control-dependent, this RHCP formulation simultaneously determines the optimal planning horizon length $w^{\ast}=|U_{ij^{\ast}}^{\ast}|+\rho_{ij^{\ast}}$ in terms of the optimal control vector $U_{i}^{\ast}(t)=\{U_{ij^{\ast}}^{\ast},\,j^{\ast}\}$.

In all, we use two planning related horizon concepts in this paper: 1) the fixed planning horizon $[t,t+H]$, and 2) the planning horizon $[t,t+w(U_{ij})]$. They are related such that $[t,t+w(U_{ij})]\subseteq [t,t+H]$ where $H$ is a predefined sufficiently large constant such as $H = T-t$ and $w(U_{ij})$ is optimized on-line through \eqref{Eq:RHCGenSolStep1}-\eqref{Eq:RHCGenSolStep2} to be $w(U_{ij^*}^*)$.

\paragraph{\textbf{Local Objective}}
As mentioned earlier, the local objective function of a target $i\in\mathcal{V}$ over a period $[t_{0},t_{1})\subseteq [0,T]$ is denoted by $\bar{J}_i(t_0,t_1)$. The purpose of $\bar{J}_i$ is to be used in \eqref{Eq:RHCNewChoices} as the RHCP objective by each agent that visits target $i$ for the selection of its controls $U_i = [u_i,j,u_j]$. 

The local versions of the global objective \eqref{Eq:GlobalObjective} (based on ``local contribution'' functions \eqref{Eq:LocalContribution}): $J_i$ and $\sum_{j\in\mathcal{N}_i}J_j$ are two conventional candidates for the local objective function $\bar{J}_i$ \cite{Welikala2020J4}. 
However, note that: (i) such candidate forms can be written as summations of the contribution terms in \eqref{Eq:ContributionActive} and \eqref{Eq:ContributionInactive}, (ii) both \eqref{Eq:ContributionActive} and \eqref{Eq:ContributionInactive} increase monotonically with its argument $w$ and (iii) in this case, $w=u_i+\rho_{ij}+u_j$ \eqref{Eq:VariableHorizon}. Therefore, when minimizing  $\bar{J}_i$, both of its candidate forms ($J_i$ and $\sum_{j\in\mathcal{N}_i}J_j$) yield the controls $u_i^*=0,\,u_j^*=0$ in an attempt to minimize $w$ (making $w^*=\rho_{ij}$). This would imply that no agent ever dwells at any target. 

Hence, instead of using a local version of the global objective \eqref{Eq:GlobalObjective}, we propose to use a local version of the alternative global objective \eqref{Eq:GlobalObjectiveHat} as the local objective function: 
\begin{equation}
\bar{J}_{i}(t_{0},t_{1}) \triangleq -
\frac{\sum_{j\in\bar{\mathcal{N}}_{i}}J_{j}^A(t_{0},t_{1})}
{\sum_{j\in\bar{\mathcal{N}}_{i}}J_{j}(t_{0},t_{1})}.\label{Eq:LocalObjectiveFunction}%
\end{equation}
This choice of local objective represents the \emph{normalized active contribution} (i.e., the contribution during agent visits in  \eqref{Eq:LocalContributionComponents}) of the targets in the neighborhood $\bar{\mathcal{N}}_i$ over the interval $[t_0,t_1)$. Due to this particular form \eqref{Eq:LocalObjectiveFunction}, when it is used as the RHCP objective function \eqref{Eq:RHCNewChoices}, the agent (residing at target $i$) will have to optimally allocate its sensing capabilities (resources) over the target $i$ and the next-visit target $j\in\mathcal{N}_i$ (i.e., have to optimally select the controls $u_i$ and $u_j$). Moreover, in the sequel, we will show that this local objective function is \emph{unimodal} in most cases of interest.

Since we have already shown that \eqref{Eq:GlobalObjectiveHat} and \eqref{Eq:GlobalObjective} perform in an equivalent manner (see Fig. \ref{Fig:ObjectiveFunctionComparison}(a)), we can also conclude that agents minimizing a local version of \eqref{Eq:GlobalObjectiveHat} (i.e., \eqref{Eq:LocalObjectiveFunction}) can in fact lead to minimizing \eqref{Eq:GlobalObjective}. Moreover, when the instantaneous values of $J_t$ \eqref{Eq:GlobalObjective} and $\hat{J}_t$ \eqref{Eq:GlobalObjectiveHat} (i.e., evaluated over a very small period $[t,t+\Delta]$) are compared for $t\in[0,T-\Delta]$, we have seen that $\hat{J}_t$ is more sensitive to the variations of the system (while remaining within a small interval) compared to $J_t$ (e.g., see Fig \ref{Fig:ObjectiveFunctionComparison}(b)). These qualities imply the feasibility of  \eqref{Eq:LocalObjectiveFunction} as a local objective function for the use of agents to decide their controls (in a distributed manner) so as to optimize the global objective \eqref{Eq:GlobalObjective}.
However, to date, we have not provided a formal proof to the statement that agents minimizing the local objective function \eqref{Eq:LocalObjectiveFunction} will lead to a minimization of the global objective function \eqref{Eq:GlobalObjective}; this is a subject of future research. 

Finally, note that the local objective function $\bar{J}_i(t_0,t_1)$ in \eqref{Eq:LocalObjectiveFunction} can be obtained locally at target $i$ by evaluating the required $J_j^A(t_0,t_1)$ and $J_j^I(t_0,t_1)$ terms for all $j\in\bar{\mathcal{N}}_i$ using Lemmas \ref{Lm:ErrorCovarianceActive} and \ref{Lm:ErrorCovarianceInactive}, respectively (recall $J_j(t_0,t_1) = J_j^A(t_0,t_1)+J_j^I(t_0,t_1)$).

\paragraph{\textbf{Event-Driven Action Horizon}}
Similar to all receding horizon controllers, an optimal receding horizon control solution computed over a planning horizon $[t,t+w^*] \subseteq [t,t+H]$ is generally executed only over a shorter \emph{action horizon} $[t,t+h] \subseteq [t,t+w^*]$.
In this event-driven persistent monitoring setting, the value of $h$ is determined by the first event that takes place after the time instant when the RHCP was last solved. Therefore, in the proposed RHC approach, the control is updated whenever asynchronous events occur. This prevents unnecessary steps to re-solve the RHCP (i.e., \eqref{Eq:RHCGenSolStep1}-\eqref{Eq:RHCGenSolStep2}) unlike time-driven receding horizon control.

In general, the determination of the action horizon $h$ may be \emph{controllable} or \emph{uncontrolled}. The latter case occurs as a result of random or external events in the system (if such events are allowed), while the former corresponds to the occurrence of any one event resulting from an agent finishing the execution of a RHCP solution determined at an earlier time. We next define two \emph{controllable} events associated with an agent when it resides at target $i$. Both of these events define the action horizon $h$ based on the RHCP solution $U_i^*(t) = [u_i^*(t), j^*(t), u_j^*(t)]$ obtained by the agent at time $t\in [0,\,T]$:

\textbf{1. Event }$[h\rightarrow u_{i}^{\ast}]$\textbf{: }
This event occurs at time $t+u_{i}^{\ast}(t)$ and indicates the termination of the active time at target $i$. By definition, this coincides with an agent departure event from target $i$.

\textbf{2. Event }$[h\rightarrow\rho_{ij^{\ast}}]$\textbf{: }
This event occurs at time $t+\rho_{ij^{\ast}(t)}$ and is only feasible after an event $[h\rightarrow u_{i}^{\ast}]$ has occurred (including the possibility that $u_{i}^{\ast}(t)=0$). Clearly, this coincides with an agent arrival event at target $j^{\ast}(t)$.

Among these two types of events, only one is feasible at any one time. However, it is also possible for a different event to occur after $t$, before one of these two events occurs. Such an event is either external, random (if our model allows for such events) or is controllable but associated with a different target than $i$. In particular, let us define two additional events that may occur at any neighboring target $j\in\mathcal{N}_{i}$ and affect the agent residing at target $i$. These events aim to ensure the control constraint \eqref{Eq:NoTargetSharing} (to prevent simultaneous target sharing) and apply only to \emph{multi-agent} persistent monitoring problems.

At time $t$, if a target $j\in\mathcal{V}$ already has a residing agent or if an agent is en route to visit it from a neighboring target in $\mathcal{N}_{j}$, it is said to be \emph{covered}. Now, an agent $a\in\mathcal{A}$ residing in target $i$ can prevent simultaneous target sharing at $j\in\mathcal{N}_{i}$ by simply modifying the neighbor set $\mathcal{N}_{i}$ used in its RHCP solved at time $t$ to exclude all such covered targets. Let us use $\mathcal{N}_{i}(t)$ to indicate a \emph{time-varying} neighbor set of $i$. Then, if target $j$ becomes covered at $t$, we set  
$
\mathcal{N}_{i}(t)=\mathcal{N}_{i}(t^-)\backslash\{j\}.
$
Most importantly, note that as soon as an agent $a$ is en route to $j^{\ast}$, $j^{\ast}$ becomes covered - preventing any other agent from visiting $j^{\ast}$ prior to agent $a$'s subsequent departure from $j^{\ast}$.

Based on this discussion, we define the following two additional \emph{neighbor-induced local events} at $j\in\mathcal{N}_{i}$ affecting an agent $a$ residing at target $i$:

\textbf{3. Covering Event }$C_{j}\mathbf{,}$ $j\in\mathcal{N}_{i}$\textbf{: } 
This event causes $\mathcal{N}_{i}(t)$ to be modified to $\mathcal{N}_{i}(t) \backslash \{j\}$.

\textbf{4. Uncovering Event }$\bar{C}_{j}\mathbf{,}$ $j\in\mathcal{N}_{i}$\textbf{: }
This event causes $\mathcal{N}_{i}(t)$ to be modified to $\mathcal{N}_{i}(t)\cup\{j\}$.

If one of these two events takes place while an agent remains active at target $i$ (i.e., prior to the occurrence of event $[h\rightarrow u_i^*]$), then the RHCP is re-solved to account for the updated $\mathcal{N}_{i}(t)$. This may affect the optimal solution's values $U_{i}^{\ast}$ compared to the previous solution. Note, however, that the new solution will still give rise to a subsequent event $[h\rightarrow u_{i}^{\ast}]$.

\paragraph{\textbf{Two Forms of RHCPs}} The exact form of the RHCP (\eqref{Eq:RHCGenSolStep1}-\eqref{Eq:RHCGenSolStep2}) that needs to be solved at a certain event time depends on the event that triggered the end of the previous action horizon. In particular, corresponding to the two controllable event types, there are two possible RHCP forms:

\textbf{1. RHCP1:} This problem is solved by an agent when an event $[h\rightarrow\rho_{ki}]$ occurs at time $t$ at target $i$ for any $k\in\mathcal{N}_{i}(t)$, i.e., at the arrival of the agent at target $i$. The solution $U_{i}^{\ast}(t)$ includes $u_{i}^{\ast}(t)\geq0$, representing the active time to be spent at $i$. This problem may also be solved while the agent is active at $i$ if a $C_{j}$ or $\bar{C}_{j}$ event occurs at any neighbor $j\in\mathcal{N}_{i}(t)$.

\textbf{2. RHCP2: } This problem is solved by an agent residing at target $i$ when an event $[h\rightarrow u_{i}^{\ast}]$ occurs at time $t$. The solution $U_{i}^{\ast}(t)$ is now constrained to include $u_{i}^{\ast}(t)=0$ by default, implying that the agent must immediately depart from $i$.

\begin{figure}[!t]
\centering
\includegraphics[width=0.85\columnwidth]{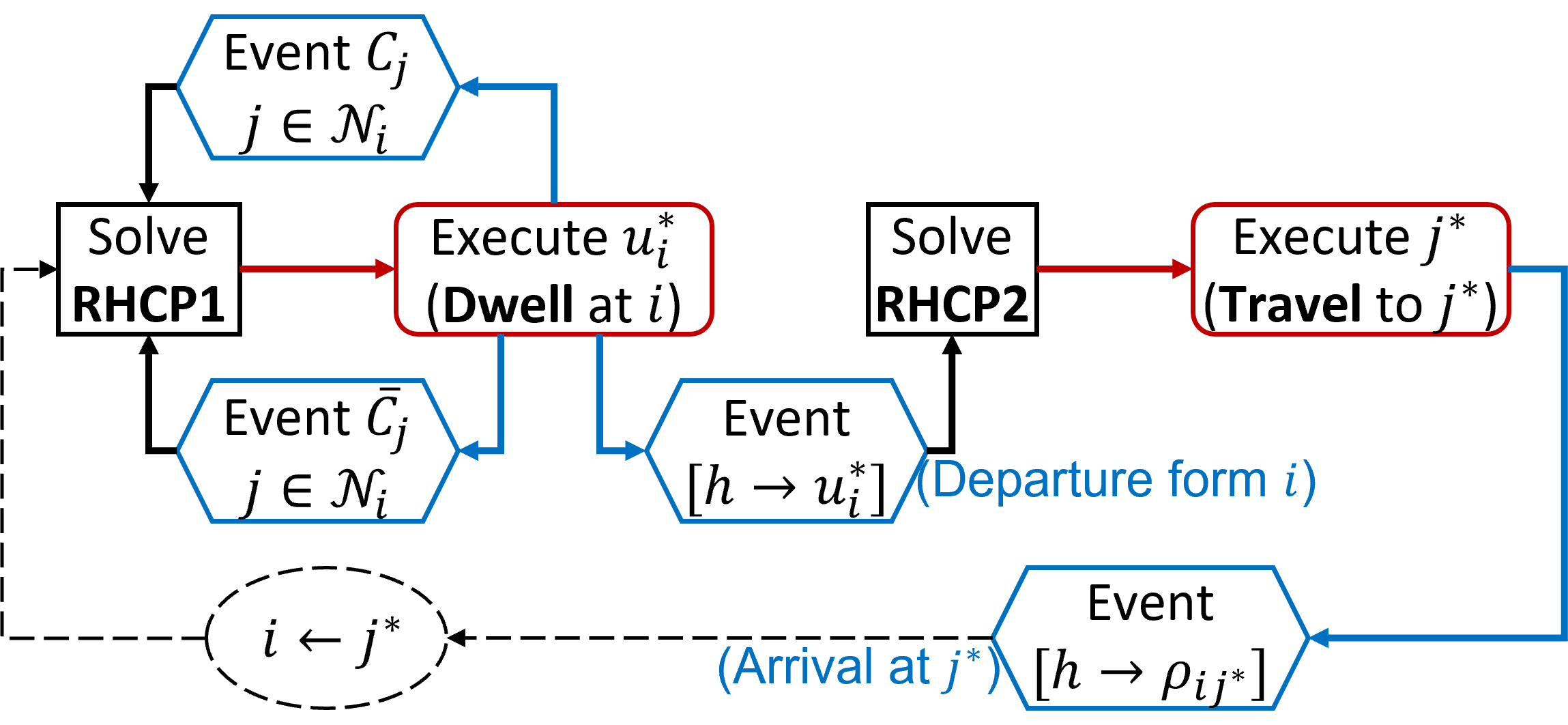} \caption{The proposed event-driven receding horizon control (RHC) strategy (focusing on an agent at a target $i \in \mathcal{V}$).}%
\label{Fig:EventDiagram}%
\end{figure}

For an agent at a target $i \in \mathcal{V}$, the interconnection between the aforementioned types of events and RHCPs involved in the proposed event-driven receding horizon control strategy is illustrated in Fig. \ref{Fig:EventDiagram}.

\section{Solving the Event-Driven Receding Horizon Control Problems}
\label{Sec:RHCPSolution}

\subsection{Solution of \textbf{RHCP2}}
We begin with \textbf{RHCP2} as it is the simplest RHCP given that in this case $u_{i}^{\ast}(t)=0$ by default. Therefore, $U_{ij}$ in \eqref{Eq:RHCGenSolStep1} is limited to $U_{ij}=u_{j}$ and the planning horizon length $w(U_{ij})$ in \eqref{Eq:VariableHorizon} becomes $w(U_{ij})=\rho_{ij}+u_{j}$. Based on the control constraints: $w(U_{ij}) = \rho_{ij}+u_{j} \leq H$ and $u_j \geq 0$, any target $j\in\mathcal{N}_{i}(t)$ such that $\rho_{ij}>H$ will not result in a feasible dwell-time value $u_j$. Hence, such targets are directly omitted from \eqref{Eq:RHCGenSolStep1}.

\paragraph{\textbf{Constraints}}
Based on the control constraints mentioned above, note that in this \textbf{RHCP2}, $u_j$ is constrained by   
$
    0\leq u_j \leq H-\rho_{ij}.
$

\begin{figure}[!h]
\centering
\includegraphics[width=0.85\columnwidth]{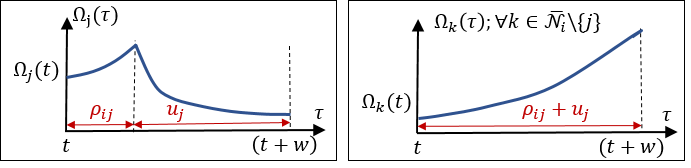}\caption{State
trajectories during $[t,t+w)$ for \textbf{RHCP2}.}%
\label{Fig:RHCP2Graphs}%
\end{figure}

\paragraph{\textbf{Objective}}
According to \eqref{Eq:RHCNewChoices}, the objective function of \textbf{RHCP2} is $\bar{J}_i(t,t+w)$. To obtain an exact expression for $\bar{J}_i$, it is decomposed using \eqref{Eq:LocalObjectiveFunction} as  
\begin{equation}
\bar{J}_{i}=- 
\frac{J_{j}^A+\sum_{k\in\bar{\mathcal{N}}_{i}(t)\backslash\{j\}}J_{k}^A}
{(J_{j}^A+J_j^I)+\sum_{k\in\bar{\mathcal{N}}_{i}(t)\backslash\{j\}}(J_{k}^A + J_k^I)}
=
- 
\frac{J_{j}^A}
{J_{j}^A+J_j^I+\sum_{k}J_k^I}
,\label{Eq:ObjDerivationOP3Step1}%
\end{equation}
where the last equality follows from the fact that $J_k^A = J_k^A(t,t+w) = 0$ as any target $k\in\bar{\mathcal{N}}_i\backslash\{j\}$ will not be visited during the planning horizon $[t,t+w]$ (see Fig. \ref{Fig:RHCP2Graphs}).
Similarly, using Fig. \ref{Fig:RHCP2Graphs}, we can write $J_j^A = J_j(t+\rho_{ij},t+\rho_{ij}+u_j)$, $J_j^I = J_j(t,t+\rho_{ij})$ and $J_k^I = J_k(t,t+\rho_{ij}+u_j)$. Each of these terms can be evaluated using Lemmas \ref{Lm:ErrorCovarianceActive} and \ref{Lm:ErrorCovarianceInactive}. 
These results together with \eqref{Eq:ObjDerivationOP3Step1} give the objective function $\bar{J}_{i}(t,t+w)$ required for \textbf{RHCP2} in the form $\bar{J}_i(u_j)$ (with a slight abuse of notation) as 
\begin{equation}\label{Eq:RHCP2ObjectiveFunction}
    \bar{J}_i(u_j) = -\frac{A(u_j)}{A(u_j) + B(u_j)},
\end{equation}
where we define
\begin{eqnarray}\label{Eq:RHCP2ObjectiveActivePart}
A(u_j) &\triangleq&  J_j^A = c_1 + c_2\log(1 + c_3 e^{-\lambda_ju_j}) + c_4 u_j,\\ \label{Eq:RHCP2ObjectiveInactivePart}
B(u_j) &\triangleq& J_j^I + \sum_k J_k^I = c_5 + c_{6} u_j + \sum_{k} c_{7k} e^{2A_ku_j},
\end{eqnarray}
with $\lambda_j = 2\sqrt{A_j^2+Q_jG_j}$ and 
\begin{equation}\label{Eq:RHCP2Coefficients}
    \begin{aligned}
        c_1 &=
        -c_2\log(1+c_3),\ \ 
        \ \ 
        c_2 = \frac{1}{G_j},\ \ 
        c_3 = -\frac{G_j\Omega_j^{\prime}+Q_jv_{j2}}{G_j\Omega_j^{\prime}+Q_jv_{j1}},\\
        \Omega_j^{\prime} &= \Omega_j(t+\rho_{ij}),\ \ 
        v_{j1}, v_{j2} = \frac{1}{Q_j}(-A_j \pm \sqrt{A_j^2+Q_jG_j}),\\
        c_4 &= \frac{1}{v_{j1}},\ \
        c_5 = \frac{1}{2A_j}\Big(\Omega_j + \frac{Q_j}{2A_j}\Big)\times(e^{2A_j\rho_{ij}}-1)-\frac{Q_j\rho_{ij}}{2A_j}\\
        &-\sum_k \frac{1}{2A_k}\Big(\Omega_k + \frac{Q_k}{2A_k} + Q_k\rho_{ij}\Big),\ \ 
        \Omega_j = \Omega_j(t),\\
        \Omega_k &= \Omega_k(t),\ \
        c_{6} = -\sum_k\frac{Q_k}{2A_k},\ \  
        c_{7k} = \frac{1}{2A_k}\Big(\Omega_k+\frac{Q_k}{2A_k}\Big)e^{2A_k\rho_{ij}}.
    \end{aligned}
\end{equation}

\paragraph{\textbf{Unimodality of }$\bar{J}_i(u_j)$}
We first prove the following lemma to eventually establish the unimodality of $\bar{J}_i(u_j)$. 

\begin{lemma}\label{Lm:RHCP2ObjectiveLimits}
The \textbf{RHCP2} objective function $\bar{J}_i(u_j)$ \eqref{Eq:RHCP2ObjectiveFunction} satisfies the following properties: 
\begin{eqnarray}\label{Eq:RHCP2ObjectiveLimits1}
    \lim_{u_j \rightarrow 0} \bar{J}_i(u_j) &=& 0 \ \ \mbox{ and }\\\label{Eq:RHCP2ObjectiveLimits2}
    \lim_{u_j \rightarrow \infty} \bar{J}_i(u_j) &=&
    \begin{cases} 
    L_i\ &\mbox{if } A_k<0, \forall k \in \bar{\mathcal{N}}_i\backslash\{j\},\\ 
    0 &\mbox{otherwise, } 
    \end{cases} 
\end{eqnarray}
where $L_i = -1/(1+\frac{c_6}{c_4})$ (with $c_4,\,c_6$ are as defined in \eqref{Eq:RHCP2Coefficients}), and all the limits are approached from below.
\end{lemma}

\emph{Proof: }
To establish these two results, we exploit the $A(u_j),\,B(u_j)$ notation introduced in \eqref{Eq:RHCP2ObjectiveFunction}. The result in \eqref{Eq:RHCP2ObjectiveLimits1} is proved using the relationships: $\lim_{u_j \rightarrow 0} A(u_j) = 0$ (note that, from \eqref{Eq:RHCP2Coefficients}: $c_1 + c_2\log(1+c_3)=0$) and $\lim_{u_j \rightarrow 0} B(u_j) = c_5 + \sum_{k} c_{7k} \geq 0$. Similarly, \eqref{Eq:RHCP2ObjectiveLimits2} is proved using the limit (given by L'Hospital's rule):
$$
\lim_{u_j\rightarrow\infty} \frac{B(u_j)}{A(u_j)} = \begin{cases}
\frac{c_6}{c_4} &\mbox{if } A_k<0, \forall k \in \bar{\mathcal{N}}_i\backslash\{j\},\\
\infty &\mbox{otherwise.}
\end{cases}
$$
The directions in which each of these limits are approached can be established using the inequalities: $A(u_j)\geq 0,\ B(u_j) > 0$ (since $A(u_j)$ and $B(u_j)$ defined respectively in \eqref{Eq:RHCP2ObjectiveActivePart} and  \eqref{Eq:RHCP2ObjectiveInactivePart} are contributions of the targets and the travel time $\rho_{ij}>0$) and $\bar{J}_i(u_j)\leq 0$ (based on the definition in \eqref{Eq:RHCP2ObjectiveFunction}), for all $u_j \geq 0$. \hfill $\blacksquare$

Second, in the following lemma, we establish two possible steady-state target error covariance values. 
\begin{lemma}\label{Lm:SteadyStateErrorCovariance}
If a target $i\in\mathcal{V}$ is sensed by an agent for an infinite duration of time, its error covariance $\Omega_i(t)$ is such that 
\begin{equation}\label{Eq:SteadyStateErrorCovarianceActive}
    \lim_{t\rightarrow\infty}\,\Omega_i(t) = \Omega_{i,ss} \triangleq \frac{Q_i}{-A_i + \sqrt{A_i^2+Q_iG_i}}. 
\end{equation}
On the other hand, if a target $i\in\mathcal{V}$ is not sensed by an agent for an infinite duration of time, 
\begin{equation}\label{Eq:SteadyStateErrorCovarianceInactive}
    \lim_{t\rightarrow\infty}\,\Omega_i(t) =
    \begin{cases}
    \bar{\Omega}_{i,ss} \triangleq -\frac{Q_i}{2A_i} &\mbox{ if } A_i < 0, \\
    \infty              &\mbox{ if } A_i > 0. 
    \end{cases}
\end{equation}
\end{lemma}
\emph{Proof: }
The relationships in \eqref{Eq:SteadyStateErrorCovarianceActive} and \eqref{Eq:SteadyStateErrorCovarianceInactive} can be obtained by simply evaluating the limit $t \rightarrow \infty$ of the $\Omega_i(t)$ expressions proved in Lemma \ref{Lm:ErrorCovarianceActive} and Lemma \ref{Lm:ErrorCovarianceInactive}, respectively.  
\hfill $\blacksquare$

It is worth noting that $\Omega_{i,ss}$ and $\bar{\Omega}_{i,ss}$ respectively defined in \eqref{Eq:SteadyStateErrorCovarianceActive} and \eqref{Eq:SteadyStateErrorCovarianceInactive} are two fixed characteristic values of target $i\in\mathcal{V}$. Note also that if $A_i<0$, $\Omega_{i,ss} < \bar{\Omega}_{i,ss}$. 

We next make the following assumption regarding the initial target error covariance values: $\{\Omega_i(0):i\in\mathcal{V}\}$.
\begin{assumption}\label{As:InitialErrorCovariance}
The initial error covariance value $\Omega_i(0)$ of a target $i\in\mathcal{V}$ is such that: 
\begin{equation}
\begin{cases}
\Omega_i(0) \in (\Omega_{i,ss},\ \bar{\Omega}_{i,ss})
&\mbox{ if } A_i<0,\\
\Omega_i(0) \in (\Omega_{i,ss},\ \infty) 
&\mbox{ if } A_i > 0. 
\end{cases}
\end{equation}
\end{assumption}

The mildness of the above assumption can be justified using the steady state target error covariance values established in Lemma \ref{Lm:SteadyStateErrorCovariance}. In particular, note that this assumption is violated by a target $i\in\mathcal{V}$ only if 1) $\Omega_i(0)<\Omega_{i,ss}$, or 2) $\Omega_i(0) > \bar{\Omega}_{i,ss}$ with $A_i<0$. In the first case, based on \eqref{Eq:SteadyStateErrorCovarianceActive}, there should exist a finite time $\bar{t}_{i,0}>0$ where $\Omega_i(\bar{t}_{i,0})>\Omega_{i,ss}$ occurs - if the target $i$ was not sensed by an agent during a finite interval $[t_{i,0},\ \bar{t}_{i,0}] \subseteq[0,\ \bar{t}_{i,0}]$. In the second case, based on \eqref{Eq:SteadyStateErrorCovarianceInactive} (with $A_i<0$), there should exist a finite time $\bar{t}_{i,0}>0$ where $\Omega_i(\bar{t}_{i,0})<\bar{\Omega}_{i,ss}$ occurs - if the target $i$ was sensed by an agent during a finite interval $[t_{i,0},\ \bar{t}_{i,0}]\subseteq[0,\bar{t}_{i,0}]$. This implies that even if Assumption \ref{As:InitialErrorCovariance} is violated at some target $i\in\mathcal{V}$ (at the initial time $t=0$), we can enforce it at an alternative initial time $t=\bar{t}_{i,0}$ simply by temporarily regulating agent visits to the target $i$. Finally, we point out that $0 < \bar{t}_{i,0} \ll T$ due to the exponentially fast error covariance dynamics proved in \eqref{Eq:ErrorCovarianceActive} and \eqref{Eq:ErrorCovarianceInactive}.

The following lemma establishes positive invariant sets for target error covariance values.
\begin{lemma}\label{Lm:InvarianceSets}
Under Assumption \ref{As:InitialErrorCovariance}, the error covariance value $\Omega_i(t)$ of a target $i\in\mathcal{V}$ at any time $t \geq 0$ satisfies 
\begin{equation}
\begin{cases}
\Omega_i(t) \in (\Omega_{i,ss},\ \bar{\Omega}_{i,ss})
&\mbox{ if } A_i<0,\\
\Omega_i(t) \in (\Omega_{i,ss},\ \infty)
&\mbox{ if } A_i > 0. 
\end{cases}  
\end{equation}
\end{lemma}
\emph{Proof: }
The proof follows directly from the results established in Lemma \ref{Lm:SteadyStateErrorCovariance}. This is because, under Assumption \ref{As:InitialErrorCovariance}, the proved limiting values in \eqref{Eq:SteadyStateErrorCovarianceActive} and \eqref{Eq:SteadyStateErrorCovarianceInactive} can respectively be considered as minimum and maximum achievable $\Omega_i(t)$ values. Note that this result holds irrespective of how target $i$ is sensed by the agents, i.e., irrespective of the form of $\eta_i(t)$ signal in \eqref{Eq:ErrorCovarianceDynamics}. 
\hfill $\blacksquare$

\begin{remark}
Irrespective of Assumption 1, using the Lyapunov stability analysis of switched systems \cite{Lin2014}, it can be shown that, under arbitrarily switched $\eta_i(t)$ signals (i.e., agent visits, see \eqref{Eq:ErrorCovarianceDynamics}), the intervals $[\Omega_{i,ss},\ \bar{\Omega}_{i,ss}]$ and $[\Omega_{i,ss},\ \infty]$ are globally attractive positively invariant sets for the error covariance dynamics \eqref{Eq:ErrorCovarianceDynamics} under $A_i<0$ and $A_i>0$, respectively. 
Moreover, the said attractiveness can be proved to be a \emph{finite-time} attractiveness to the corresponding \emph{open} intervals if we omit the switching signals (agent visits) of the form: 
1) $\eta_i(t)=0, \forall t \geq 0$ if $A_i<0$ and $\Omega_i(0)>\bar{\Omega}_{i,ss}$, and  
2) $\eta_i(t)=1, \forall t \geq 0$ if $A_i>0$ and $\Omega_i(0)<\Omega_{i,ss}$, respectively. Note that this omission is in line with the previously proposed approach of temporarily regulating agent visits (below Assumption \ref{As:InitialErrorCovariance}).
In all, by temporarily regulating agent visits to a target $i\in\mathcal{V}$ where Assumption \ref{As:InitialErrorCovariance} is violated, we still can ensure the statement in Lemma \ref{Lm:InvarianceSets} for that target $i$, but for any time $t\geq \bar{t}_{i,0}$ where $0 < \bar{t}_{i,0} \ll T$. Finally, note that this same argument is valid for any other theoretical result established under Assumption \ref{As:InitialErrorCovariance} in the sequel.
\end{remark}

To establish the unimodality of $\bar{J}_i(u_j)$, we need one final lemma. 
\begin{lemma}\label{Lm:NeighborhoodStateAndParamters}
Under Assumption \ref{As:InitialErrorCovariance}, for any target $i\in\mathcal{V}$ and time $t \geq 0$,  
$2\Omega_i(t)A_i + Q_i > 0$. 
\end{lemma}
\emph{Proof: }
First, note that in general $\Omega_i(t) > 0$ and $Q_i > 0$. Therefore, $A_i>0 \implies 2\Omega_i(t)A_i + Q_i > 0$. On the other hand, if $A_i<0$, according to Lemma \ref{Lm:InvarianceSets}, $\Omega_i(t) < \bar{\Omega}_{i,ss}$, i.e., $\Omega_i(t) < -\frac{Q_i}{2A_i}$ as $\bar{\Omega}_{i,ss} = -\frac{Q_i}{2A_i}$ (from  \eqref{Eq:SteadyStateErrorCovarianceInactive}). Therefore, $A_i<0 \implies \Omega_i(t) < -\frac{Q_i}{2A_i} \iff 2\Omega_i(t)A_i + Q_i > 0$. This completes the proof. 
\hfill $\blacksquare$

\begin{theorem}\label{Th:RHCP2Unimodality}
Under {Assumption \ref{As:InitialErrorCovariance}}, the objective function of \textbf{RHCP2} in \eqref{Eq:RHCP2ObjectiveFunction} is unimodal.
\end{theorem}

\emph{Proof: }
Again we exploit the $A(u_j),\,B(u_j)$ notation introduced in \eqref{Eq:RHCP2ObjectiveFunction}. As argued in the proof of Lemma \ref{Lm:RHCP2ObjectiveLimits}, $A(u_j) \geq 0$, $B(u_j) > 0$ and $\bar{J}_i(u_j)\leq 0$ for all $u_j \geq 0$. Now, according to the limits of $\bar{J}_i(u_j)$ established in Lemma \ref{Lm:RHCP2ObjectiveLimits}, it is clear that $\bar{J}_i(u_j)$ has at least one or more local minimizers.

Through differentiating \eqref{Eq:RHCP2ObjectiveFunction}, we can obtain an equation for the stationary points of $\bar{J}_i(u_j)$ as:
$$
\frac{d\bar{J}_i(u_j)}{du_j} = 0 \iff A(u_j)\frac{dB(u_j)}{du_j} - B(u_j)\frac{dA(u_j)}{du_j}=0.
$$
For notational convenience, let us re-state the above equation as $AB'-BA' = 0$. Using the same notation, the second derivative of $\bar{J}_i(u_j)$ can be written as
$$
\frac{d^2\bar{J}_i(u_j)}{du_j^2} = \frac{AB''-BA''}{(A+B)^2} - \frac{(AB'-BA')(A'+B')}{(A+B)^3}.
$$
Therefore, the nature of a stationary point of $\bar{J}_i(u_j)$ is determined by the sign of the term $AB''-BA''$. Since we already know $A, B \geq 0$, let us focus on the $A''$ and $B''$ terms.

Using the $A(u_j)$ expression in \eqref{Eq:RHCP2ObjectiveActivePart}, we can write
\begin{equation}\label{Eq:RHCP2UnimodalityProofStep1}
    A'' = \frac{d^2A(u_j)}{du_j^2} = \frac{c_3c_2\lambda_j^2e^{\lambda_ju_j}}{(c_3+e^{\lambda_ju_j})^2}.
\end{equation}
From \eqref{Eq:RHCP2Coefficients}, clearly, $c_2>0$ and
\begin{align*}
c_3<0 \iff \Omega_j(t+\rho_{ij})>&-\frac{Q_jv_{j2}}{G_j}
= \frac{1}{v_{j1}} = \Omega_{j,ss}.
\end{align*}
The last two steps respectively used the relationships $v_{j1}v_{j2} = -\frac{G_j}{Q_j}$ and \eqref{Eq:SteadyStateErrorCovarianceActive}. Since $\Omega_j(t+\rho_{ij})>\Omega_{j,ss}$ (from Lemma \ref{Lm:InvarianceSets}), $c_3<0$ and thus \eqref{Eq:RHCP2UnimodalityProofStep1} implies that $A'' < 0$ for all $u_j \geq 0$.

Using the $B(u_j)$ expression in \eqref{Eq:RHCP2ObjectiveInactivePart}, we can write 
\begin{equation}\label{Eq:RHCP2UnimodalityProofStep2}
    B'' = \frac{d^2B(u_j)}{du_j^2} = \sum_{k\in\bar{\mathcal{N}}_i\backslash\{j\}}
    (2\Omega_k(t)A_k + Q_k)e^{2A_k(\rho_{ij}+u_j)}.
\end{equation}

Notice that $(2\Omega_k(t)A_k + Q_k)>0, \forall t \geq 0, k\in \mathcal{V} \supset \bar{\mathcal{N}}_i\backslash\{j\}$ (based on Lemma \ref{Lm:NeighborhoodStateAndParamters}, under Assumption \ref{As:InitialErrorCovariance}) and $e^{2A_k(\rho_{ij}+u_j)}>0, \forall u_j\geq 0$. Using these inequalities in \eqref{Eq:RHCP2UnimodalityProofStep2}, it can be concluded that $B''>0$ for all $u_j \geq 0$.

So far, we have shown that $A\geq0,\ B>0,\ B''>0$ while $A''<0,\ \forall u_j\geq 0$. Therefore, $AB''-BA'' > 0$ for all $u_j\geq 0$. Hence, all the stationary points of $\bar{J}_i(u_j)$ should be local minimizers. Since $\bar{J}_i(u_j)$ and all its derivatives are continuous, it cannot have two (or more) local minimizers without having a local maximizer(s). Therefore, $\bar{J}_i(u_j)$ has only one stationary point which is the global minimizer and thus $\bar{J}_i(u_j)$ is unimodal.   
\hfill $\blacksquare$

\paragraph{\textbf{Solving RHCP2 for optimal control} $u_{j}^{\ast}$}

The solution $U_{ij}^* = [u_{j}^{\ast}]$ of \eqref{Eq:RHCGenSolStep1} is given by $u_{j}^{\ast}$ where 
\begin{equation}
u_{j}^{\ast} = \underset{0\leq u_{j}\leq H-\rho_{ij}}{\arg\min}\bar{J}_i(u_{j}).
\label{Eq:OP2_Formal}%
\end{equation}

Since the objective function is unimodal and the feasible space is convex, we use the \emph{projected gradient descent} \cite{Bertsekas2016} algorithm to efficiently obtain the \emph{globally} optimal control decision $u_j^*$.

\paragraph{\textbf{Solving for Optimal Next-Visit Target $j^{\ast}$}}

Using the obtained $u_j^*$ values in \eqref{Eq:OP2_Formal} for all $j\in\mathcal{N}_{i}(t)$, we now know the optimal trajectory costs $\bar{J}_i(u_{j}^*),\,\forall j\in\mathcal{N}_{i}(t)$. Based on \eqref{Eq:RHCGenSolStep2}, the optimal target to visit next is 
$
    j^{\ast} = {\arg\min}_{j\in\mathcal{N}_{i}(t)}\ \bar{J}_i(u_{j}^{\ast}).
$

Thus, upon solving \textbf{RHCP2}, agent $a$ departs from target $i$ at time $t$ and follows the path $(i,j^{\ast})\in\mathcal{E}$ to visit target $j^{\ast}$. In the spirit of RHC, recall that the optimal control will be updated upon the occurrence of the next event, which, in this case, will be the arrival of the agent at $j^{\ast}$, triggering the solution of an instance of \textbf{RHCP1} at $j^*$.

\subsection{Solution of \textbf{RHCP1}}

We next consider the \textbf{RHCP1}, which is the most general version among the two RHCP forms. In \textbf{RHCP1}, $U_{ij}$ in \eqref{Eq:RHCGenSolStep1} is directly $U_{ij}=[u_i,u_j]$ and the planning horizon $w$ is the same as in \eqref{Eq:VariableHorizon}, where $w(U_{ij})=u_i+\rho_{ij}+u_j$. 

\paragraph{\textbf{Constraints}}
Based on the control constraints in \eqref{Eq:RHCGenSolStep1}, note that in this \textbf{RHCP1}, $u_i$ and $u_j$ are constrained by 
$
    0\leq u_i,\ \ 0 \leq u_j\ \mbox{ and }\ u_i+u_j\leq H-\rho_{ij},
$
where the last constraint follows form $w(U_{ij})\leq H$.

\begin{figure}[!h]
\centering
\includegraphics[width=\columnwidth]{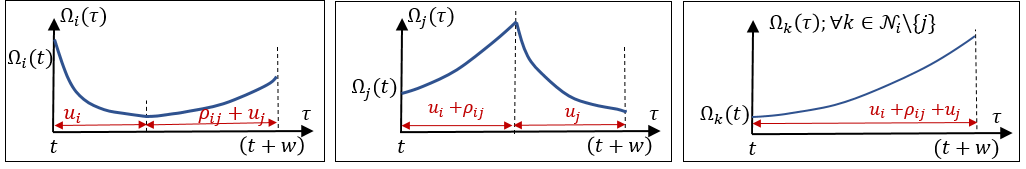}\caption{State
trajectories during $[t,t+w)$ for \textbf{RHCP1}.}%
\label{Fig:RHCP1Graphs}%
\end{figure}

\paragraph{\textbf{Objective}}
According to \eqref{Eq:RHCNewChoices}, the objective function of \textbf{RHCP1} is $\bar{J}_{i}(t,t+w)$. To obtain an exact expression for $\bar{J}_{i}$, it is decomposed using \eqref{Eq:LocalObjectiveFunction} as
\begin{align}
\bar{J}_{i}
&=- 
\frac{J_i^A + J_{j}^A+\sum_{k\in\mathcal{N}_{i}(t)\backslash\{j\}}J_{k}^A}
{(J_{i}^A+J_i^I) + (J_{j}^A+J_j^I)+\sum_{k\in\mathcal{N}_{i}(t)\backslash\{j\}}(J_{k}^A + J_k^I)}\nonumber\\
&=- 
\frac{J_i^A + J_{j}^A}
{(J_{i}^A+J_j^A) + (J_{i}^I+J_j^I + \sum_{k\in\mathcal{N}_{i}(t)\backslash\{j\}}J_k^I)}.\label{Eq:ObjDerivationOP1Step1}%
\end{align}
Similar to \eqref{Eq:ObjDerivationOP3Step1}, note that each term in \eqref{Eq:ObjDerivationOP1Step1} is also evaluated over the planning horizon $[t,t+w]$. Therefore $J_k^A = J_k^A(t,t+w) = 0$ as any target $k\in\mathcal{N}_i\backslash\{j\}$ will not be visited during the planning horizon (see Fig. \ref{Fig:RHCP1Graphs}). Moreover, using Fig. \ref{Fig:RHCP1Graphs} we can write  
$J_i^A = J_i(t,t+u_i)$, 
$J_j^A=J_j(t+u_i+\rho_{ij},t+u_i+\rho_{ij}+u_j)$, 
$J_i^I=J_i(t+u_i,t+u_i+\rho_{ij}+u_j)$, 
$J_j^I=J_j(t,t+u_i+\rho_{ij})$ and 
$J_k^I=J_k(t,t+u_i+\rho_{ij}+u_j)$. Each of these terms can be evaluated using Lemmas \ref{Lm:ErrorCovarianceActive} and \ref{Lm:ErrorCovarianceInactive}. 
These results together with \eqref{Eq:ObjDerivationOP1Step1} give the objective function $\bar{J}_{i}(t,t+w)$ required for \textbf{RHCP2} in the form $\bar{J}_{i}(u_i,u_j)$ (again with a slight abuse of notation) as \begin{equation}\label{Eq:RHCP1ObjectiveFunction}
    \bar{J}_i(u_i,u_j) = - \frac{A(u_i,u_j)}{A(u_i,u_j)+B(u_i,u_j)},
\end{equation}
where $A(u_i,u_j)\triangleq J_i^A + J_{j}^A$, $B(u_i,u_j)\triangleq J_{i}^I+J_j^I + \sum_{k\in\mathcal{N}_{i}\backslash\{j\}}J_k^I$. Specifically, $A(u_i,u_j)$ and $B(u_i,u_j)$ take the following forms:
\begin{align}
    A(u_i,u_j) =& 
    a_1 + a_2\log(1 + a_3e^{-\lambda_iu_i})\nonumber\\
    &+ a_4\log(1 + a_5e^{2A_ju_i} + a_6e^{-\lambda_ju_j}
    + a_7e^{2A_ju_i-\lambda_ju_j})\nonumber\\
    &+ a_8u_i + a_9u_j,
    \label{Eq:RHCP1ObjectiveActivePart}\\
    B(u_i,u_j) =& 
    b_1 + b_2u_i + b_3u_j 
    + b_4e^{2A_ju_i} + b_5e^{2A_iu_j}\nonumber\\
    &+ \sum_{k\in\mathcal{N}_i\backslash\{j\}}b_{6k}e^{2A_k(u_i+u_j)} + C(u_i,u_j),
    \label{Eq:RHCP1ObjectiveInactivePart}\\
    C(u_i,u_j) =& c_1\Big[
    \frac{1 + c_2e^{-\lambda_iu_i} + c_3e^{2A_iu_j}
    + c_4e^{-\lambda_iu_i+2A_iu_j}}
    {1 + c_5e^{-\lambda_iu_i}}
    \Big],\nonumber
\end{align}
where the coefficients $a_l, b_l, c_l, \forall l$ present in \eqref{Eq:RHCP1ObjectiveActivePart} and \eqref{Eq:RHCP1ObjectiveInactivePart} are given in appendix \ref{SubSec:CoeffsOfRHCP1}.

\paragraph{\textbf{Unimodality of } $\bar{J}_i(u_i,u_j)$}

\begin{figure}[!h]
     \centering
     \begin{subfigure}[b]{0.32\columnwidth}
         \centering
         \includegraphics[width=\textwidth]{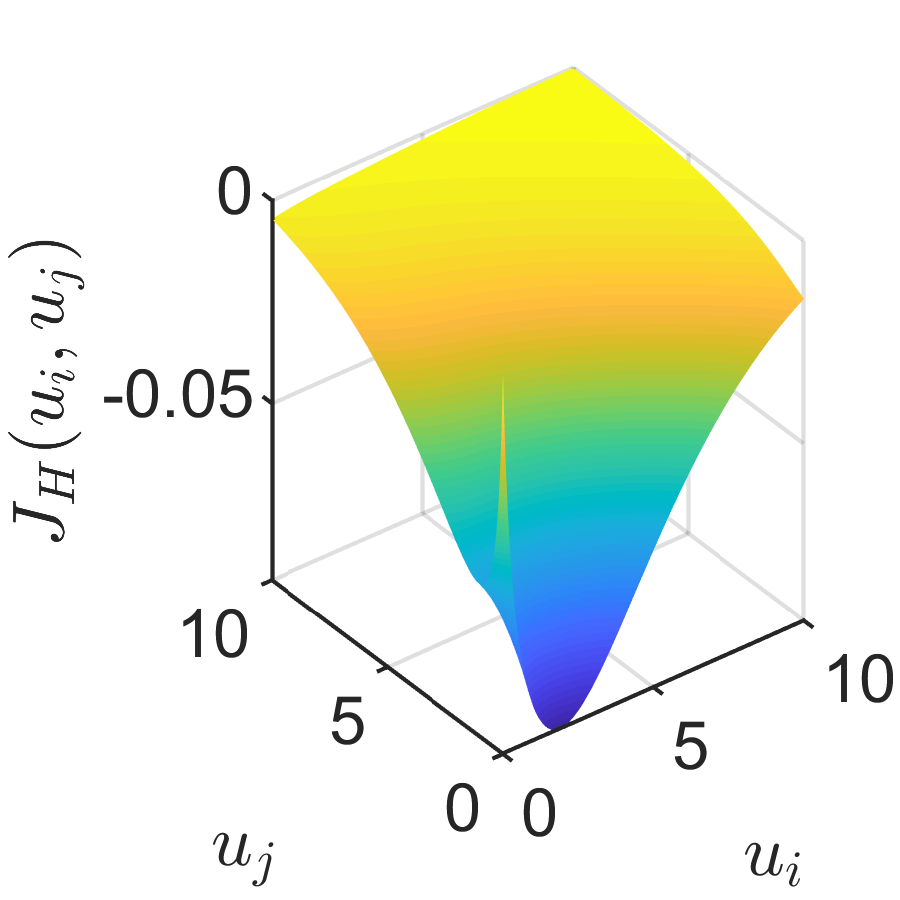}
         \caption{$u_j^* = 0$}
         \label{SubFig:1}
     \end{subfigure}
     \begin{subfigure}[b]{0.32\columnwidth}
         \centering
         \includegraphics[width=\textwidth]{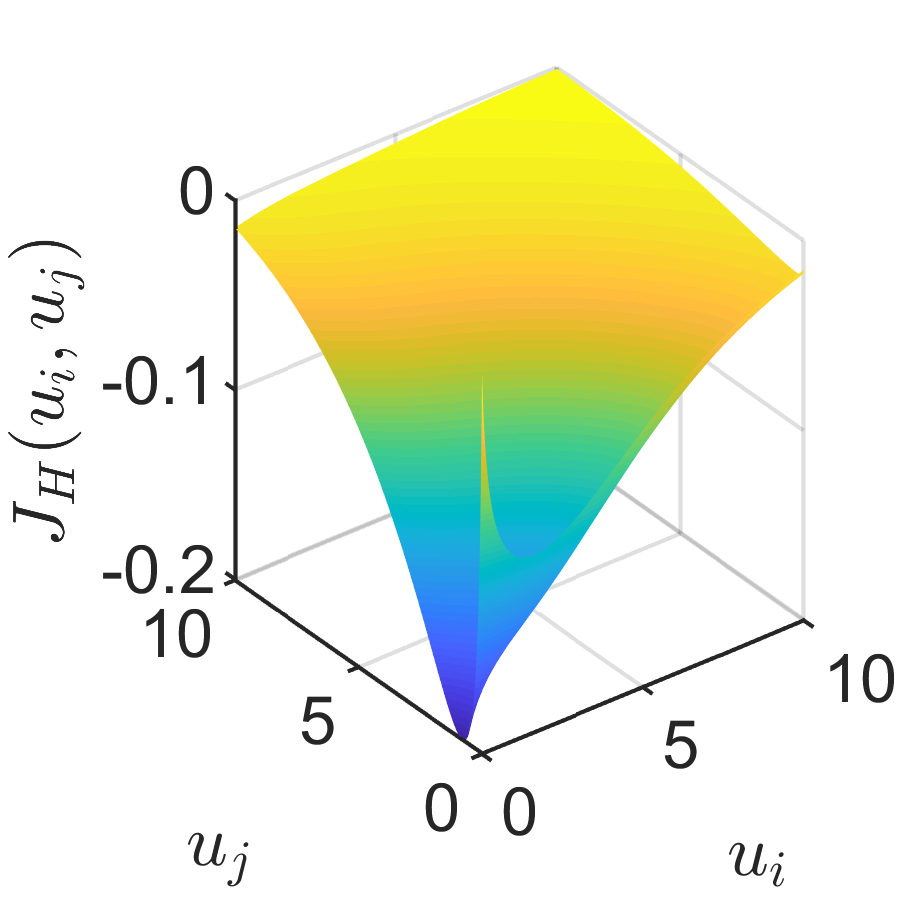}
         \caption{$u_i^* = 0$}
         \label{SubFig:2}
     \end{subfigure}
     \begin{subfigure}[b]{0.32\columnwidth}
         \centering
         \includegraphics[width=\textwidth]{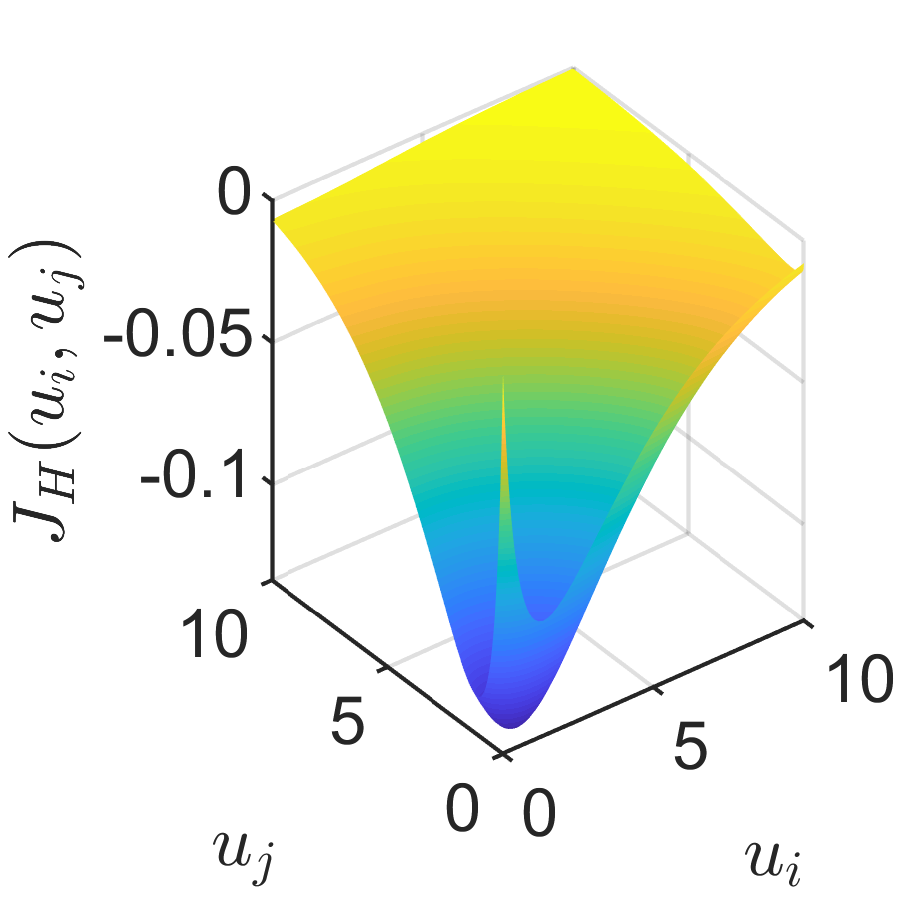}
         \caption{$u_i^*,\,u_j^* > 0$}
         \label{SubFig:3}
     \end{subfigure}
        \caption{Three example cases of \textbf{RHCP1} objective function $\bar{J}_i(u_i,u_j)$ plots (location of the minimizer: $(u_i^*,u_j^*)$).}
        \label{Fig:RHCP1ObjectiveFunction}
\end{figure}

Proving the unimodality of $\bar{J}_{i}(u_i,u_j)$ is a challenging task due to the complexity of the $A(u_i,u_j)$ and $B(u_i,u_j)$ expressions in \eqref{Eq:RHCP1ObjectiveFunction}. However, we establish that $\bar{J}_{i}(u_i,u_j)$ is unimodal along the lines $u_i=0$ and $u_j=0$. Further, we show that $\bar{J}_{i}(u_i,u_j)\rightarrow 0$ whenever $(u_i,u_j)\rightarrow (0,0)$, $u_i\rightarrow \infty$ or $u_j\rightarrow \infty$. Based on these theoretical observations and the experimental results (see Fig. \ref{Fig:RHCP1ObjectiveFunction}), we conjecture that $\bar{J}_{i}(u_i,u_j)$ is unimodal. However, to date, we have not provided a formal proof of this.

\begin{lemma}\label{Lm:RHCP1ObjectiveFunctionLimits}
The \textbf{RHCP1} objective function $\bar{J}_i(u_i,u_j)$ satisfies the following properties:
\begin{equation}\label{Eq:RHCP1ObjectiveFunctionLimits}
\begin{aligned}
    \lim_{(u_i,u_j)\rightarrow(0,0)}\bar{J}_i(u_i,u_j) &= 0,\\ 
    \lim_{u_i\rightarrow\infty}\bar{J}_i(u_i,0) &= 
    \begin{cases} L_i &\mbox{if } A_k<0,\,\forall k\in\mathcal{N}_i\backslash\{j\},\\
    0 &\mbox{otherwise,}
    \end{cases}\\
    \lim_{u_j\rightarrow\infty}\bar{J}_i(0,u_j) &= 
    \begin{cases} L_j &\mbox{if } A_k<0,\,\forall k\in\mathcal{N}_i\backslash\{j\},\\
    0 &\mbox{otherwise,}
    \end{cases}\\
    \lim_{(u_i,u_j)\rightarrow(\infty,\infty)}\bar{J}_i(u_i,u_j) &= 0.\\ 
\end{aligned}
\end{equation}
where $L_i = -1/(1+\frac{b_2}{a_8})$ and $L_j = -1/(1+\frac{b_3}{a_9})$ (with $b_2,\,a_8,\,b_3,\,a_9$ are as defined in Appendix \ref{SubSec:CoeffsOfRHCP1}).
\end{lemma}

\emph{Proof: }The result in \eqref{Eq:RHCP1ObjectiveFunctionLimits} can be obtained by following the same steps used in the proof of Lemma \ref{Lm:RHCP2ObjectiveLimits}.
\hfill $\blacksquare$

\begin{theorem}\label{Th:RHCP1Unimodality}
Under Assumption \ref{As:InitialErrorCovariance}, the functions $\bar{J}_{i}(u_i,0)$ and $\bar{J}_{i}(0,u_j)$ are unimodal.
\end{theorem}

\emph{Proof: }This proof basically follows the same steps as the proof of Theorem \ref{Th:RHCP2Unimodality}. As an example, let us consider proving the unimodality of $\bar{J}_i(u_i,0)$. First, $\bar{J}_i(u_i,0)$ is written as $\bar{J}_i(u_i,0)=-\bar{A}(u_i)/(\bar{A}(u_i)+\bar{B}(u_i))$ where $\bar{A}(u_i)=A(u_i,0)$ and $\bar{B}(u_i) = B(u_i,0)$. Therefore, similar to before, the nature of the stationary points of $\bar{J}_i(u_i,0)$ is dependent on the sign of $\bar{A}\bar{B}''-\bar{B}\bar{A}''$. Next, using $A(u_i,u_j)$ and $B(u_i,u_j)$ expressions in \eqref{Eq:RHCP1ObjectiveActivePart} and \eqref{Eq:RHCP1ObjectiveInactivePart}, we can write
\begin{align*}
\bar{B}'' = \frac{d^2B(u_i,0)}{du_i^2} =&  
4A_j^2b_4e^{2A_ju_i} + \sum_{k\in\mathcal{N}_i\backslash\{j\}}4A_k^2b_{6k}e^{2A_ku_i}\\
&+ \frac{c_1c_5(1+c_3)\lambda_i^2e^{-\lambda_iu_i}(-1+c_5e^{-\lambda_iu_i})}{(1+c_5e^{-\lambda_iu_i})^3},\\
\bar{A}'' = \frac{d^2A(u_i,0)}{du_i^2} =& 
\frac{\lambda_i^2a_2a_3e^{-\lambda_iu_i}}{(1+a_3e^{-\lambda_iu_i})^2}.
\end{align*}
Finally, using the above two expressions, the coefficients shown in Appendix \ref{SubSec:CoeffsOfRHCP1} and the results established in Lemma  \ref{Lm:InvarianceSets} and Lemma \ref{Lm:NeighborhoodStateAndParamters} (under Assumption \ref{As:InitialErrorCovariance}), it can be proven that $\bar{B}''>0$ and $\bar{A}''<0$ for all $u_i\geq 0$. Since $\bar{A}\geq 0,\ \bar{B}>0,\,\forall u_i\geq0$, we now can conclude that $\bar{A}\bar{B}''-\bar{B}\bar{A}''>0$ for all $u_i\geq 0$.

This result, together with the limits established in Lemma \ref{Lm:RHCP1ObjectiveFunctionLimits} implies that there exists only one stationary point in $\bar{J}_i(u_i,0)$, which is the global minimizer. Further, since $\bar{J}_i(u_i,0)$ and all of its derivatives are continuous, it can also be concluded that $\bar{J}_i(u_i,0)$ is a unimodal function. Following the same steps, the unimodality of $\bar{J}_i(0,u_j)$ can also be established.
\hfill $\blacksquare$

\paragraph{\textbf{Solving RHCP1 for Optimal Controls} $u_i^*,\ u_{j}^*$}

The solution $U_{ij}^* = [u_i^*,\, u_{j}^*]$ of \eqref{Eq:RHCGenSolStep1} is given by $(u_i^{\ast},\,u_{j}^{\ast})$ where
\begin{equation}
\begin{aligned}
(u_i^{\ast},\,u_{j}^{\ast}) = &\ \underset{u_i,u_j}{\arg\min} \ \bar{J}_i(u_i,u_{j}),\\
&\ 0\leq u_{i},\ 0\leq u_j,\\
&\ u_i + u_j \leq H-\rho_{ij}
\end{aligned}
\label{Eq:OP1_Formal}%
\end{equation}

In \eqref{Eq:OP1_Formal}, the feasible space is convex and we already conjectured that the objective function is unimodal. Therefore, again, we use the projected gradient descent algorithm \cite{Bertsekas2016} to efficiently obtain the optimal control decisions $(u_i^*,u_j^*)$ in \eqref{Eq:OP1_Formal}.

\paragraph{\textbf{Solving for Optimal Next-Visit Target $j^{\ast}$}}

Using the obtained $U_{ij}^* = [u_i^*, u_j^*]$ values in \eqref{Eq:OP1_Formal} for all $j\in\mathcal{N}_{i}(t)$, we now have at our disposal the optimal trajectory costs $\bar{J}_i(u_i^*, u_{j}^*)$ for all $j\in\mathcal{N}_{i}(t)$. Based on \eqref{Eq:RHCGenSolStep2}, the optimal neighbor to plan as the next-visit target is given by 
$
j^{\ast} = {\arg\min}_{j\in\mathcal{N}_{i}(t)}\ \bar{J}_i(u_i^*,u_{j}^*).    
$

Upon solving \textbf{RHCP1}, the agent remains stationary (active) on target $i$ for a duration of $u_i^*$ or until any other event occurs. If the agent completes the determined active time $u_i^*$ (i.e., if the corresponding event $[h\rightarrow u_i^*]$ occurs), the agent will have to subsequently solve an instance of \textbf{RHCP2} to determine the next-visit target and depart from target $i$. However, if a different event occurred before the anticipated event $[h\rightarrow u_i^*]$, the agent will have to re-solve \textbf{RHCP1} to re-compute the remaining active time at target $i$.

\begin{remark}
In the proposed RHC solution, there are only three tunable parameters: 1) the upper bound $H$ to the planning horizon,  
2) the gradient descent step size, and 
3) the gradient descent initial condition. 
For $H$, as we have mentioned before, selecting a sufficiently large value (e.g., $H=T-t$) ensures that it does not affect the RHC solutions. For the gradient descent step size, there are established standard choices \cite{Bertsekas2016} as well as specialized ones \cite{Anescu2018}. Finally, based on the established unimodality properties (that guarantees global convergence of gradient descent processes), it is clear that the RHC solutions will not be affected by the choice of the gradient descent initial condition. 
\end{remark}

\section{Improving Computational Efficiency Using Machine Learning}
\label{Sec:RHCLSolution}

Recall that when solving a RHCP, an agent $a\in\mathcal{A}$ (residing on a target $i\in\mathcal{V}$ at some time $t\in[0,T]$) has to solve the optimization problems \eqref{Eq:RHCGenSolStep1}-\eqref{Eq:RHCGenSolStep2}. The problem in \eqref{Eq:RHCGenSolStep1} involves solving $\vert \mathcal{N}_i \vert$ different optimization problems (one for each neighbor $j\in\mathcal{N}_i$) to get the optimal \emph{continuous} (real-valued) controls: $\{U_{ij}^*:j\in\mathcal{N}_i\}$. The problem in \eqref{Eq:RHCGenSolStep2} is only a simple numerical comparison that determines the optimal \emph{discrete} control: the next-visit $j^*\in\mathcal{N}_i$. Upon solving this RHCP, the agent will only use $U_{ij^*}^*$ and $j^*$ to make its immediate decisions. Hence, the continuous controls: $\{U_{ij}^*:j\in\mathcal{N}_i\backslash\{j^*\}\}$ found when solving \eqref{Eq:RHCGenSolStep1} are wastefully discarded. 
This motivates the use of derived information $\{U_{ij}^*:j\in\mathcal{N}_i\}$ (part of which will surely be discarded) to introduce a learning component as explained next.

Note that if $j^*$ can be determined ahead of solving \eqref{Eq:RHCGenSolStep1}, we can prevent this waste of computational resources by limiting the evaluation of \eqref{Eq:RHCGenSolStep1} only for the pre-determined neighbor $j^*$ to directly get $U_{ij^*}^*$. Roughly speaking, this approach should save a fraction $\frac{\vert \mathcal{N}_i\vert-1}{\vert \mathcal{N}_i\vert}$ of the processing (CPU) time required to solve the RHCP (i.e., \eqref{Eq:RHCGenSolStep1}-\eqref{Eq:RHCGenSolStep2}).

\paragraph{\textbf{Ideal Classification Function}} 
The aim here is to approximate an ideal classification function $F_i:\R^{\vert \bar{\mathcal{N}}_i \vert}_{\geq0}\rightarrow \mathcal{N}_i$ of the form  
\begin{equation}\label{Eq:ClassifierFunction}
\begin{aligned}
    j^* 
    =& F_i(X_i(t)) \\
    \triangleq& \underset{j\in\mathcal{N}_i}{\arg\min}\ J_H\big(X_i(t),\, \underbrace{\underset{U_{ij}\in\mathbb{U}}{\arg\min}\ J_H(X_i(t),U_{ij};\,w(U_{ij}))}_{U_{ij}^*};\,w(U_{ij}^*)\big)
\end{aligned}
\end{equation}
where $X_i(t) = \{\Omega_j(t):j\in\bar{\mathcal{N}}_i\}$ is the local state at target $i$ at time $t$ and $F_i(X_i(t))$ (explicitly expressed in the second line) is the result of combining equations \eqref{Eq:RHCGenSolStep1} and \eqref{Eq:RHCGenSolStep2}.

In the machine learning literature, this kind of an ideal classification function $F_i(X_i)$ is commonly known as an \emph{underlying function} (or a \emph{target function}) and $X_i$ is considered as a \emph{feature vector} \cite{Zhang2000}.

We emphasize that $F_i$ is strictly dependent on: (i) the current target $i\in\mathcal{V}$, (ii) the agent $a\in \mathcal{A}$ and (iii) the RHCP type. Therefore, $F_i$ in actuality should be written as $F_i^{a,l}$ (where $l\in\{1,2\}$ represents the RHCP type) even though we omit doing so for notational simplicity.

\paragraph{\textbf{Classifier Function}} 
Due to the complexity of this ideal classification function $F_i(X_i)$ in \eqref{Eq:ClassifierFunction}, we cannot analytically simplify it to obtain a closed form solution for a generic input (feature) $X_i$. 
Therefore, we propose to use machine learning techniques to model $F_i(X_i)$ by an estimate of it - which we denote as $f_i(X_i; \mathcal{D}_i)$. Here, $\mathcal{D}_i$ represents a collected \emph{data set} of size $L$ and the notation $f_i(X_i; \mathcal{D}_i)$ implies that this \emph{classifier function} has been constructed (trained) based on $\mathcal{D}_i$. 
Note that similar to $F_i$, both $f_i$ and $\mathcal{D}_i$ depend not only on the target $i$ but also on the agent $a$ and the RHCP type $l$.

Since our aim is to develop a distributed on-line persistent monitoring solution, the agent $a$ itself has to collect this data set $\mathcal{D}_i$ based on its very first $L$ instants where a RHCP of type $l$ was fully solved at target $i$. Specifically, $\mathcal{D}_i$ can be thought of as a set of input-output pairs: $\mathcal{D}_i = \{(X_i(\tau),j^*(\tau)):\tau\in \Gamma_i^{a,l}\}$ where $\Gamma_i^{a,l}$ is the set of the first $L$ event times where  agent $a$ fully solved a RHCP of type $l$ while residing at target $i$.

\paragraph{\textbf{Application of Neural Networks}} 
In order to construct the classifier function $f_i(X_i;\mathcal{D}_i)$, among many commonly used classification techniques such as linear classifiers, support vector machines, kernel estimation techniques, etc., we chose an \emph{Artificial Neural Networks} (ANN) based approach. This choice was made because of the key advantages that an ANN-based classification approach holds \cite{Zhang2000}: (i) generality, (ii) data-driven nature, (iii) non-linear modeling capability and (iv) the ability to provide posterior probabilities.

Let us denote a shallow feed forward ANN model as $y = h_i(x;\Theta)$ where $x$ is the $\vert \bar{\mathcal{N}}_i \vert$-dimensional input feature vector and $y$ (or $h_i(x;\Theta)$) is the $\vert \mathcal{N}_i\vert$-dimensional output vector under the ANN weight parameters $\Theta$. For simplicity, we propose to use only one hidden layer with ten neurons with hyperbolic tangent sigmoid (\emph{tansig}) activation functions. At the output layer, we propose to use the \emph{softmax} activation function so that each component of the output (denoted as $h_{i,k}(x;\Theta),\, k\in\mathcal{N}_i$) will be in the interval $(0,1)$. 

Based on this ANN model, the classifier function is 
\begin{equation}\label{Eq:Classifier}
\hat{j}^* = f_i(X_i; \mathcal{D}_i) = \underset{k\in\mathcal{N}_i}{\arg\max} \ h_{i,k}(X_i;\Theta^*),    
\end{equation}
where $\Theta^*$ represents the optimal set of ANN weights obtained by training the ANN model $y = h_i(x;\Theta)$ using the data set $\mathcal{D}_i$. Specifically, these optimal weights $\Theta^*$ are determined through back-propagation (and gradient descent) \cite{Zhang2000} such that the (standard) cross-entropy based cost function $H(\Theta)$ evaluated over the data set $\mathcal{D}_i = \{(X_i(\tau),j^*(\tau)):\tau\in \Gamma_i^{a,l}\}$ given by 
\begin{equation}
\begin{aligned}
    H(\Theta) = \frac{1}{L}\sum_{\tau\in\Gamma_i^{a,l}}\sum_{k\in\mathcal{N}_i} \left[
    \textbf{1}\{j^*(\tau)=k\}\log \left(h_{i,k}(X_i(\tau);\Theta)\right)\right. \\
    \left.+ \textbf{1}\{j^*(\tau)\neq k\}\log\left(1-h_{i,k}(X_i(\tau);\Theta)\right)\right] + \frac{\lambda}{2L}\Vert \Theta \Vert^2,
\end{aligned}
\end{equation}
is minimized ($\lambda$ represents the regularization constant).

\paragraph{\textbf{RHC with Learning (RHC-L)}} 
Needless to say, the optimal weights $\Theta^*$ (and hence the classifier function $f_i(X_i; \mathcal{D}_i)$ in \eqref{Eq:Classifier}) are determined only when the agent $a$ has accumulated a data set $\mathcal{D}_i$ of length $L$. In other words, the agent $a$ has to be familiar enough with solving the RHCP type $l$ at target $i$ in order to learn $f_i(X_i; \mathcal{D}_i)$. 

Upon learning $f_i(X_i; \mathcal{D}_i)$, the RHCP given in \eqref{Eq:RHCGenSolStep1} and \eqref{Eq:RHCGenSolStep2} can be solved very efficiently by simply evaluating:
\begin{eqnarray}\label{Eq:RHCLGenStep1}
\hat{j}^* &=& f_i(X_i(t);\mathcal{D}_i),\\ \label{Eq:RHCLGenStep2}
U_{i\hat{j}^*}^* &=& \underset{U_{i\hat{j}^*}\in\mathbb{U}}{\arg\min}\ J_H(X_i(t),U_{i\hat{j}^*};H),
\end{eqnarray}
to directly obtain the optimal controls $U_i^*(t)=\{U_{i\hat{j}^*}^*,\hat{j}^*\}$ (i.e., without having to solve \eqref{Eq:RHCGenSolStep1} associated with targets $j\in \mathcal{N}_i\backslash\{\hat{j}^*\}$). For convenience, we call this approach the RHC-L method.

Notice that \eqref{Eq:RHCLGenStep2} (when compared to \eqref{Eq:RHCGenSolStep1}) only involves a single continuous optimization problem - which may even be a redundant one to solve if the underlying RHCP is of type $2$ (i.e., a \textbf{RHCP2}) where knowing the next-visit target (i.e., $j^*$ now approximated by $\hat{j}^*$) is sufficient to take the immediate action. Therefore, the proposed RHC-L method can be expected to have significantly lower processing times for evaluating the RHCPs faced by the agents compared to the RHC method - upon the completion of the learning phase.

The only drawback in this RHC-L approach when compared to the original RHC method is the performance degradation that can be expected due to learning-related errors, i.e., due to the mismatch between $F_i(X_i)(=j^*)$ in \eqref{Eq:ClassifierFunction} and its estimate $f_i(X_i;\mathcal{D}_i)(=\hat{j}^*)$ learned in  \eqref{Eq:Classifier}.

\paragraph{\textbf{RHC with Active Learning (RHC-AL)}}
We next propose a technique to suppress the aforementioned performance degradation that stems from the learning-related errors. For this purpose, we exploit the fact that ANN outputs are actually estimates of the \emph{posterior probabilities} \cite{Zhang2000}. This simply means 
\begin{equation}\label{Eq:PosteriorProbability}
    h_{i,k}(X_i;\Theta^*) \simeq P( j^*=k \vert X_i),
\end{equation}
where, $h_{i,k}(X_i;\Theta^*)$ is the output of the ANN corresponding to the neighbor $k\in\mathcal{N}_i$ and $P(j^*=k \vert X_i)$ is the probability of the ideal classification function $j^*=F_i(X_i)$ in \eqref{Eq:ClassifierFunction} resulting $F_i(X_i)=k\in\mathcal{N}_i$, given the feature vector $X_i$. Note that $F_i(\cdot)$ here is an unknown function that we try to estimate and hence $j^*(=F_i(X_i))$ is a random variable.

Based on \eqref{Eq:PosteriorProbability} and \eqref{Eq:Classifier}, the \emph{mismatch error} between $F_i(X_i)$ and $f_i(X_i;\mathcal{D}_i)$ given the feature vector $X_i$ can be estimated as $e_i(X_i)$ where 
\begin{equation*}
    e_i(X_i) \triangleq P(F_i(X_i) \neq f_i(X_i;\mathcal{D}_i)\vert X_i) = 1 - \max_{k\in\mathcal{N}_i}\ h_{i,k}(X_i;\Theta^*). 
\end{equation*}
Clearly, prior to solving the RHC-L problems \eqref{Eq:RHCLGenStep1} and \eqref{Eq:RHCLGenStep2}, the agent can evaluate this mismatch error metric $e_i(X_i)$ and if it falls above a certain threshold, it can resort to follow the original RHC approach and solve \eqref{Eq:RHCGenSolStep1} and \eqref{Eq:RHCGenSolStep2}, instead. Moreover, in such a case, the obtained RHC solutions can be incorporated into the data set $\mathcal{D}_i$ and re-train the ANN (to update the weight parameters $\Theta^*$ in $h_i(X_i;\Theta^*)$ \eqref{Eq:Classifier}).

We call this ``active learning'' approach the RHC-AL method. It is important to highlight that this RHC-AL approach helps agents to make correct decisions in the face of unfamiliar scenarios. Therefore, we can expect the RHC-AL method to perform well compared to the RHC-L method - only at the expense of trading off the advantage that the RHC-L method had in terms of the processing times compared to the RHC method.

\begin{remark}\label{Rm:RHCParameters}
The proposed on-line learning process can alternatively be carried out off-line (if the system allows it) as each agent (for each target $i\in\mathcal{V}$ and each RHCP type) can synthetically generate data sets $\mathcal{D}_i$ exploiting the relationship \eqref{Eq:ClassifierFunction} with a set of randomly generated features $X_i$. Moreover, if the agents are homogeneous, the proposed distributed learning process can be made centralized by allowing agents to share their data sets (pertaining to the same targets and RHCP types). However, the effectiveness of a such ``shared data based learning'' scenario is debatable as each agent's optimal trajectory decisions might be unique even though the agents are homogeneous. 
\end{remark}

\begin{remark}\label{Rm:RHCParameters}
In addition to the three tunable parameters summarized in Rm. \ref{Rm:RHCParameters}, there are several more tunable parameters associated with the proposed RHC-L and RHC-AL solutions in this section. In particular, the new tunable parameters are related to the used: 
1) ANN architecture,
2) ANN learning process, and
3) the data set (size). 
As stated before, regarding these new tunable parameters, our choices have been made mainly to promote simplicity. Thus, clearly, a hyperparameter tuning process can be used to obtain further improved results. 
\end{remark}

\section{Simulation Results}
\label{Sec:SimulationResults}

This section contains the details of three different simulation studies. In the first, we explore how the RHC-based agent control method performs compared to four other agent control techniques, in terms of the performance metric $J_T$ in \eqref{Eq:GlobalObjective} evaluated over a relatively short period: $T=50\,s$. Second, we study how well the persistent target state estimates provided by different agent control methods can facilitate local target state tracking control tasks. Finally, we explore the long-term performance of agent controllers by selecting $T=750\,s$. In particular, we compare the agent control methods: RHC, RHC-L and RHC-AL in terms of the performance metric $J_T$ and the average processing (CPU) time taken to solve each RHCP.

\paragraph*{\textbf{Persistent Monitoring Problem Configurations}}

\begin{figure}[!h]
     \centering
     \begin{subfigure}[b]{0.4\columnwidth}
         \centering
         \includegraphics[width=\textwidth]{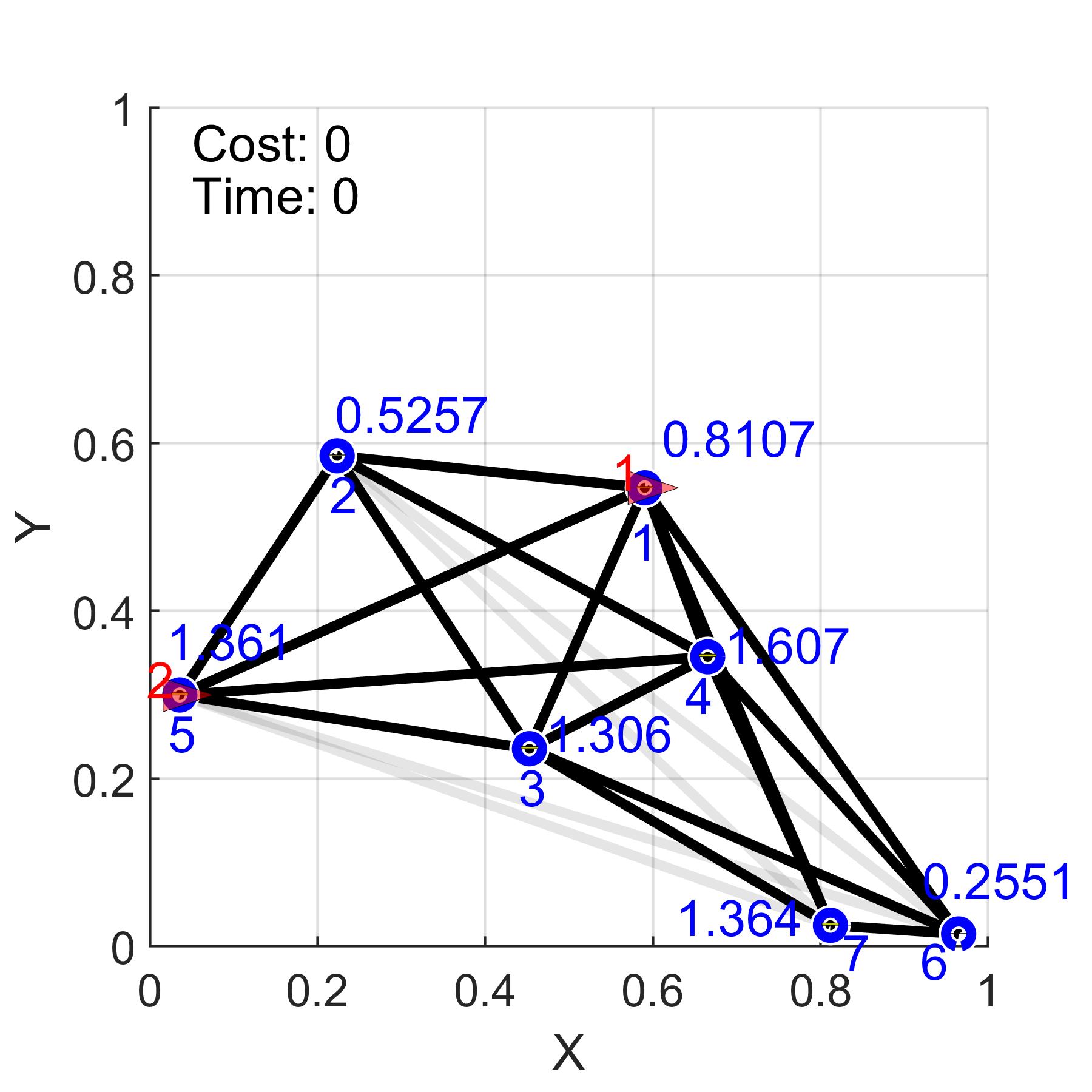}
         \caption{PC 1}
         \label{SubFig:1}
     \end{subfigure}
     \begin{subfigure}[b]{0.4\columnwidth}
         \centering
         \includegraphics[width=\textwidth]{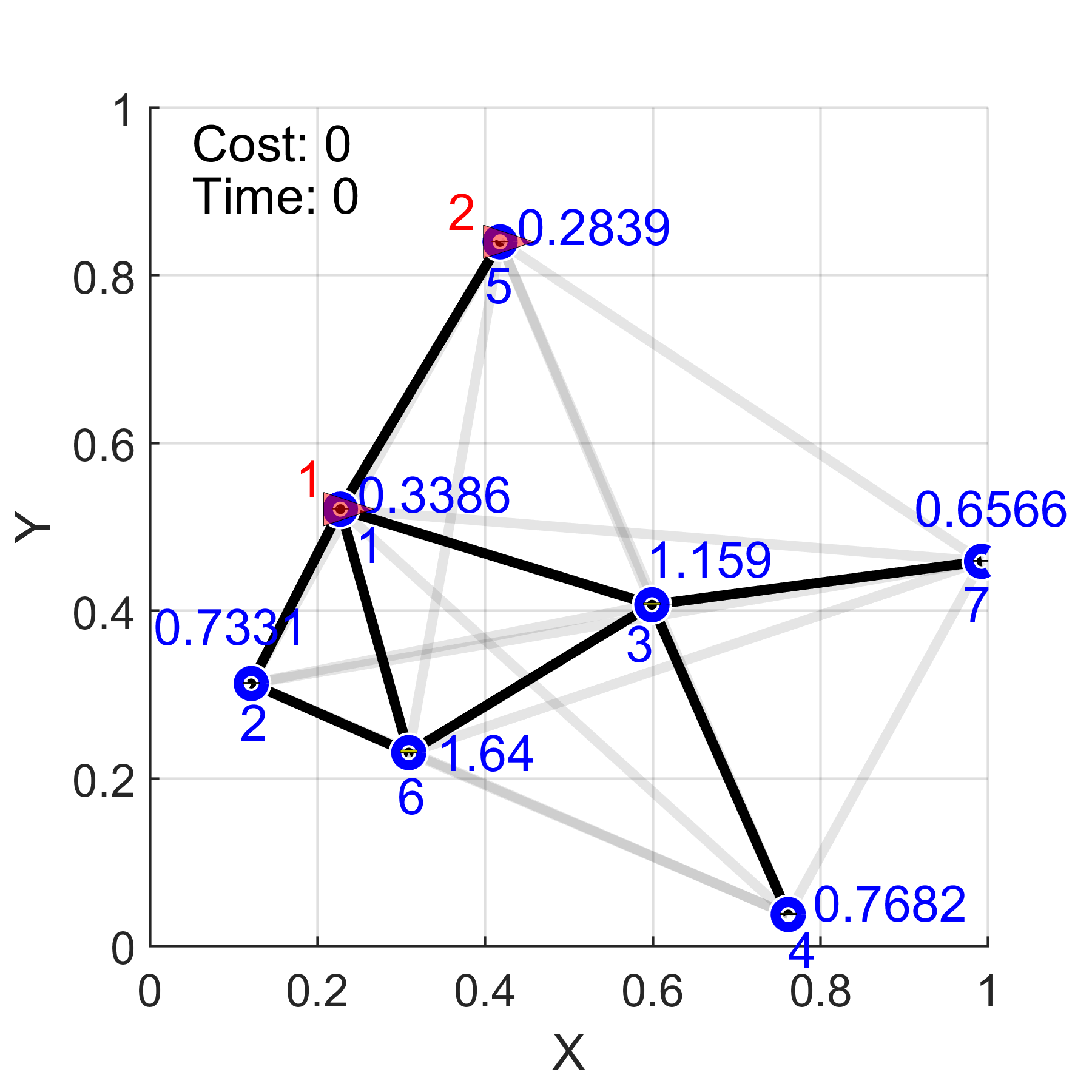}
         \caption{PC 2}
         \label{SubFig:2}
     \end{subfigure}
     \begin{subfigure}[b]{0.4\columnwidth}
         \centering
         \includegraphics[width=\textwidth]{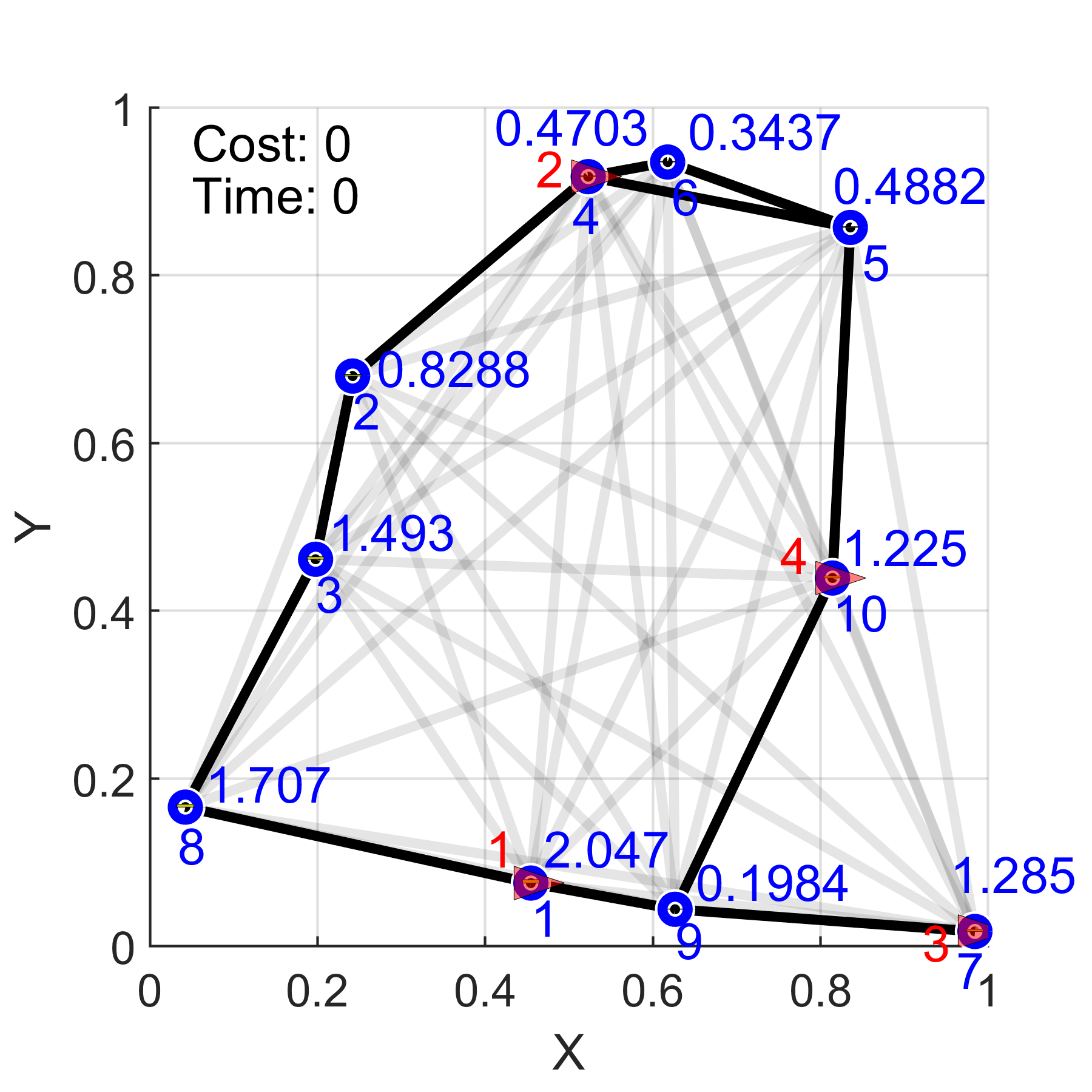}
         \caption{PC 3}
         \label{SubFig:3}
     \end{subfigure}
     \begin{subfigure}[b]{0.4\columnwidth}
         \centering
         \includegraphics[width=\textwidth]{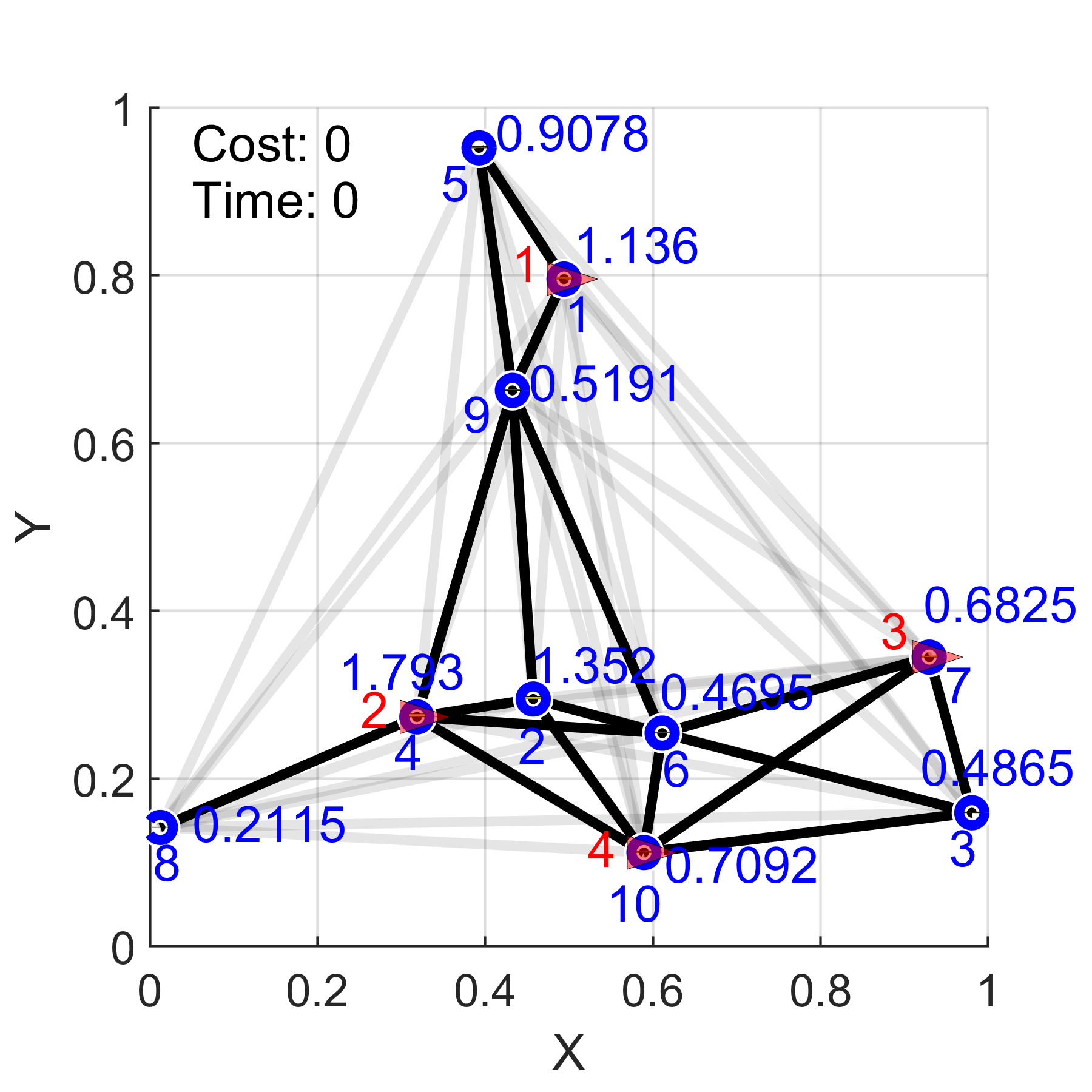}
         \caption{PC 4}
         \label{SubFig:4}
     \end{subfigure}
        \caption{
        Four randomly generated persistent monitoring problem configurations (PCs) and their initial conditions.
        }
        \label{Fig:InitialConditions}
\end{figure}

In this section, we consider the four randomly generated persistent monitoring problem configurations (PCs) shown in Fig. \ref{Fig:InitialConditions}. In there, blue circles represent the targets and dark black lines indicate the trajectory segments that are available for the agents to travel between targets. Agents and target error covariance values at $t=0$ are represented by red triangles and yellow vertical bars/blue texts, respectively (see also Fig. \ref{Fig:FinalConditions} for PCs at $t=T$). The PCs 1,2 have seven targets and two agents each and the PCs 3,4 have ten targets and four agents each. In each PC, the target parameters were selected using the uniform distribution $U[\cdot,\cdot]$ as follows: $Y_i \sim U[0,1]$, $A_i \sim U[0.01,0.41]$, $B_i \sim U[0.01,0.41]$, $Q_i\sim U[0.1,2.1]$, $R_i\sim U[2,10]$ and we set $H_i=1$ for all $i \in \mathcal{V}$. If the distance between any two targets is less than a certain threshold $\sigma$, a linear shaped trajectory segment was deployed between those targets. For PC 1, $\sigma = 0.7$ (dense) was used and for the reset, $\sigma=0.45$ (sparse) was used. Each agent is assumed to travel with a unit speed on these trajectory segments and the fixed planning horizon length $H=10$ is used.

\subsection{Simulation Study 1: The effect of agent controls on target state estimation over a relatively short period.}

In this section, we compare the performance metric $J_{T}$ (defined in \eqref{Eq:GlobalObjective} with $T=50$) observed for the four PCs shown in Fig. \ref{Fig:InitialConditions} when using five different agent control methods: 
(\romannum{1}) the centralized off-line periodic control (MTSP) method proposed in \cite{Pinto2021}
(\romannum{2}) a basic distributed on-line control (BDC) method (which is an ad-hoc control method), 
(\romannum{3}) the proposed RHC method,
(\romannum{4}) a periodic version of the BDC (BDC-P) method and
(\romannum{5}) a periodic version of the RHC (RHC-P) method. 
In essence, this simulation study is aimed to observe the effectiveness of the overall \emph{target state estimation} process (as it is directly reflected by the metric $J_T$) rendered by the aforementioned different agent controllers. According to \eqref{Eq:ErrorCovarianceDynamics}, the choice of target controls $\upsilon_i(t)$ in \eqref{Eq:TargetStateDynamics} do not affect $J_T$. Hence in this study, we set: $\upsilon_i(t)=0,\ \forall i\in\mathcal{V},\ \forall t\in[0,T]$.

\paragraph{\textbf{The Basic Distributed Control (BDC) Method}}
The BDC method uses the same event-driven control architecture as the RHC method. However, instead of $u_i^*$ and $j^*$ given in \eqref{Eq:RHCGenSolStep1} and \eqref{Eq:RHCGenSolStep2}, it uses: 
\begin{equation}\label{Eq:BDC}
\begin{aligned}
    u_i^* =& \underset{\tau \geq 0}{\arg\min} \ \textbf{1}\{\Omega_i(t+\tau)\leq (1+\epsilon)\Omega_{i,ss}\},\\
    j^* =& \underset{j\in \mathcal{N}_i(t)}{\arg\max}\ \Omega_j(t)
\end{aligned}
\end{equation}
with $\epsilon = 0.075$ and $\Omega_{i,ss}$ defined in \eqref{Eq:SteadyStateErrorCovarianceActive}. In a nut shell, the BDC method forces an agent to dwell at each visited target $i$ until its error covariance $\Omega_i(t)$ drops to an $\epsilon$ fraction closer to the corresponding $\Omega_{i,ss}$ value. Upon completing this requirement, the next-visit target is determined as the neighbor $j \in \mathcal{N}_i(t)$ with the maximum $\Omega_j(t)$ value.

\paragraph{\textbf{The Centralized Off-line Control (MTSP) Method \cite{Pinto2021}}}
Unlike the distributed on-line agent control methods: RHC and BDC, the MTSP method proposed in \cite{Pinto2021} fully computes the agent trajectories in a centralized off-line stage, focusing on minimizing an infinite horizon objective function:
\begin{equation}\label{Eq:InfiniteHorizonObjective}
    \underset{i\in\mathcal{V}}{\max}\ \underset{t\rightarrow\infty}{\lim\sup}\ \mbox{tr}(\Omega_i(t))
\end{equation}
via selecting appropriate periodic agent trajectories. Nevertheless, this objective function is in the same spirit of $J_T$ \eqref{Eq:GlobalObjective} as it also aims to maintain the target error covariances as low as possible.

The MTSP method first uses the spectral clustering algorithm \cite{Luxburg2007} to decompose the target topology $\mathcal{G}$ into sub-graphs among the agents. Then, on each sub-graph, starting from the \emph{traveling salesman problem} (TSP) solution, a greedy target visitation cycle is constructed. Essentially, this set of target visitation cycles is a candidate solution for the famous multi-TSP \cite{Bektas2006} (hence the acronym: MTSP). Finally, the dwell-time spent at each target (on the constructed target visitation cycle) is found using a golden ratio search algorithm exploiting many interesting mathematical properties.

\begin{figure}[!h]
     \centering
     \begin{subfigure}[b]{0.4\columnwidth}
         \centering
         \includegraphics[width=\textwidth]{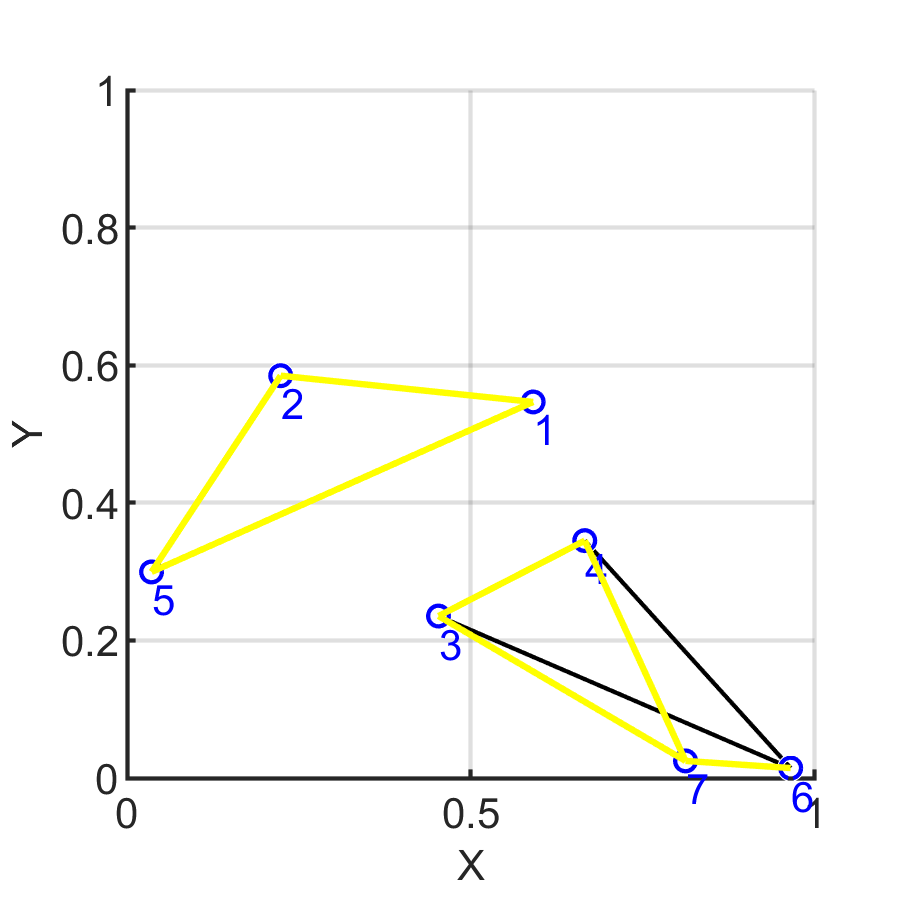}
         \caption{PC 1}
         \label{SubFig:1}
     \end{subfigure}
     \begin{subfigure}[b]{0.4\columnwidth}
         \centering
         \includegraphics[width=\textwidth]{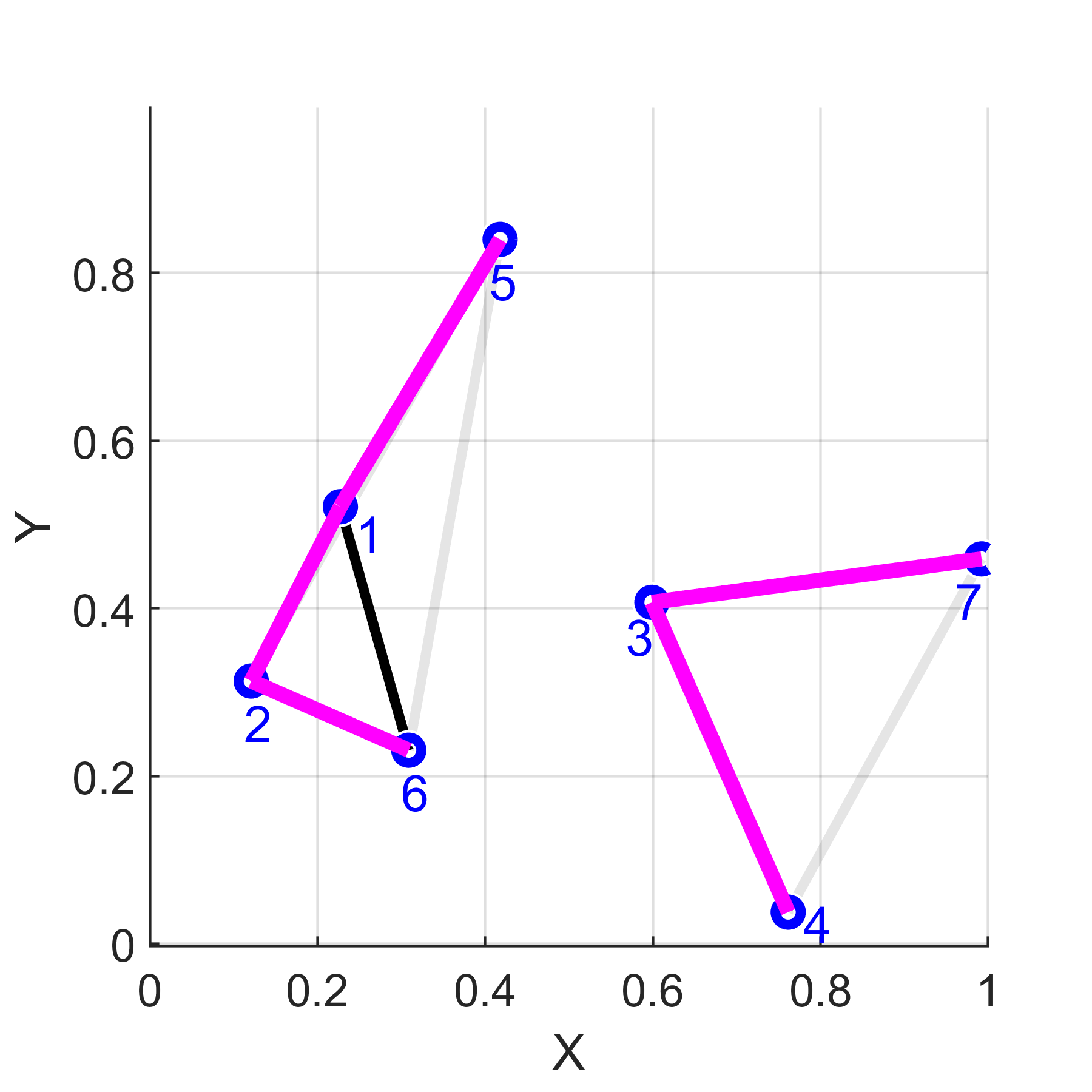}
         \caption{PC 2}
         \label{SubFig:2}
     \end{subfigure}
     \begin{subfigure}[b]{0.4\columnwidth}
         \centering
         \includegraphics[width=\textwidth]{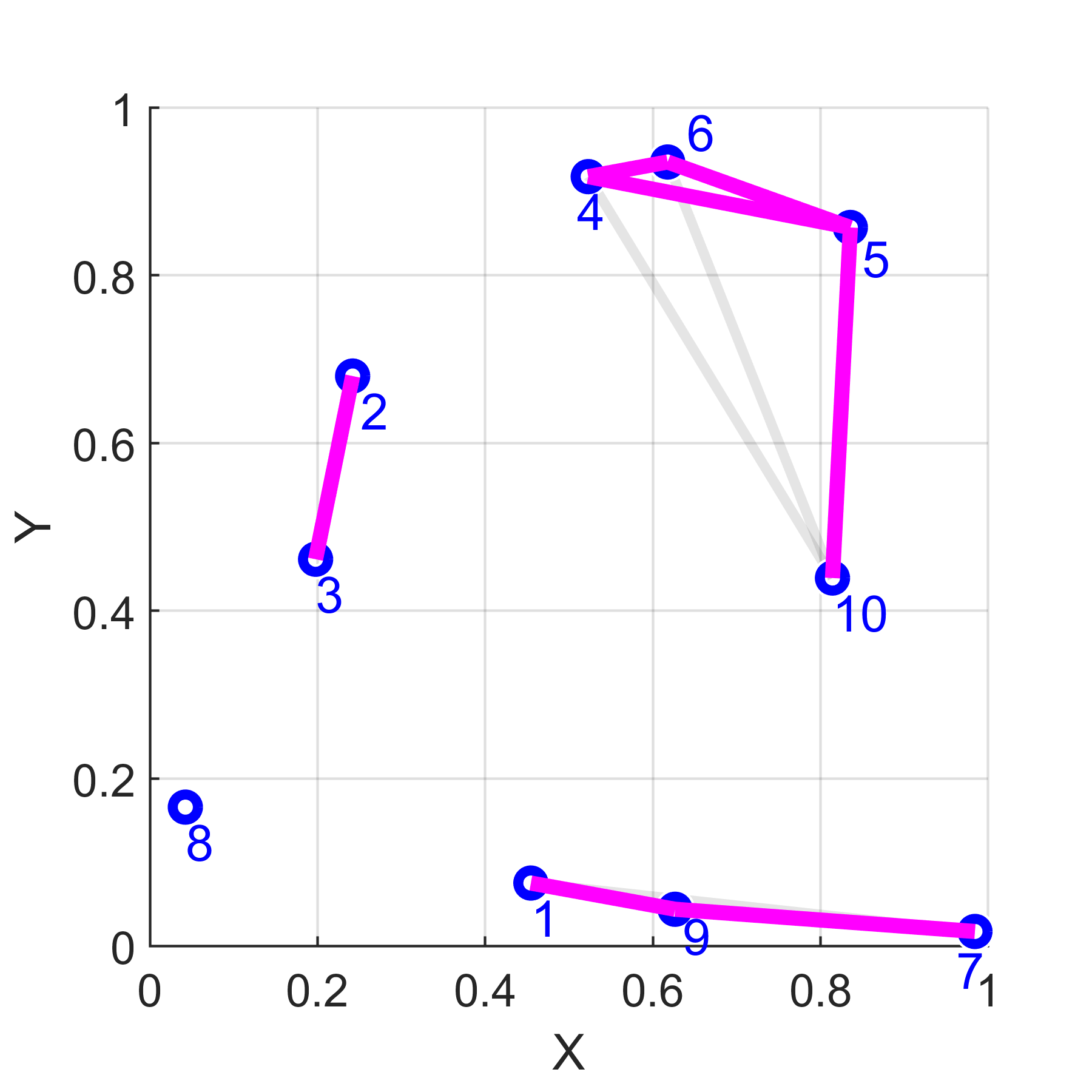}
         \caption{PC 3}
         \label{SubFig:3}
     \end{subfigure}
     \begin{subfigure}[b]{0.4\columnwidth}
         \centering
         \includegraphics[width=\textwidth]{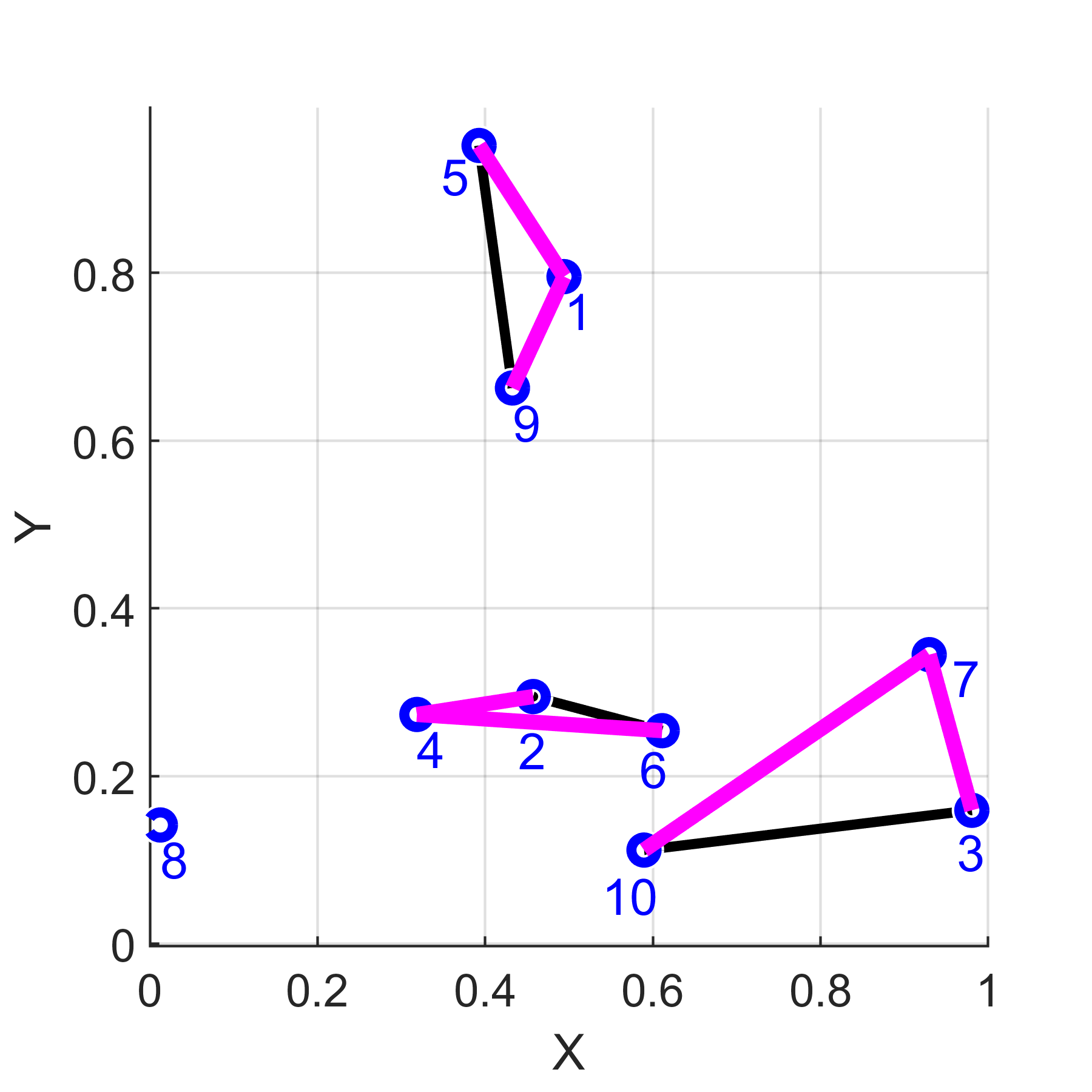}
         \caption{PC 4}
         \label{SubFig:4}
     \end{subfigure}
        \caption{
        Target clusters and periodic agent trajectories (magenta contours) used by the methods: MTSP, BDC-P and RHC-P.}
        \label{Fig:Cycles}
\end{figure}

\paragraph{\textbf{Hybrid Methods: BDC-P and RHC-P}}
In some applications, having a periodicity in visiting targets can be a crucial constraint (e.g., bus routes). Even in such cases, the proposed RHC method (or the BDC method) can still be used to make the dwell-time decisions at each visited target instead of using a fixed set of predetermined dwell-times like in the MTSP method. However, the optimal next-visit target, i.e., $j^*$ in \eqref{Eq:RHCGenSolStep2} (or in \eqref{Eq:BDC}) would now be given by the off-line computed target visitation cycles (similar to the MTSP method). We use the label RHC-P (or BDC-P) to represent such a hybrid periodic agent control method. Note that in this RHC-P method, when solving for the dwell-times, (i.e., \eqref{Eq:RHCGenSolStep1}), the RHCP objective \eqref{Eq:LocalObjectiveFunction} should only consider neighboring targets in the agent's target visitation cycle. Pertaining to the PCs shown in Fig. \ref{Fig:InitialConditions}, target clusters and corresponding periodic target visitation cycles used by the periodic agent control methods (MTSP, BDC-P and RHC-P) are shown in Fig. \ref{Fig:Cycles}.

\begin{figure}[!h]
     \centering
     \begin{subfigure}[b]{0.4\columnwidth}
         \centering
         \includegraphics[width=\textwidth]{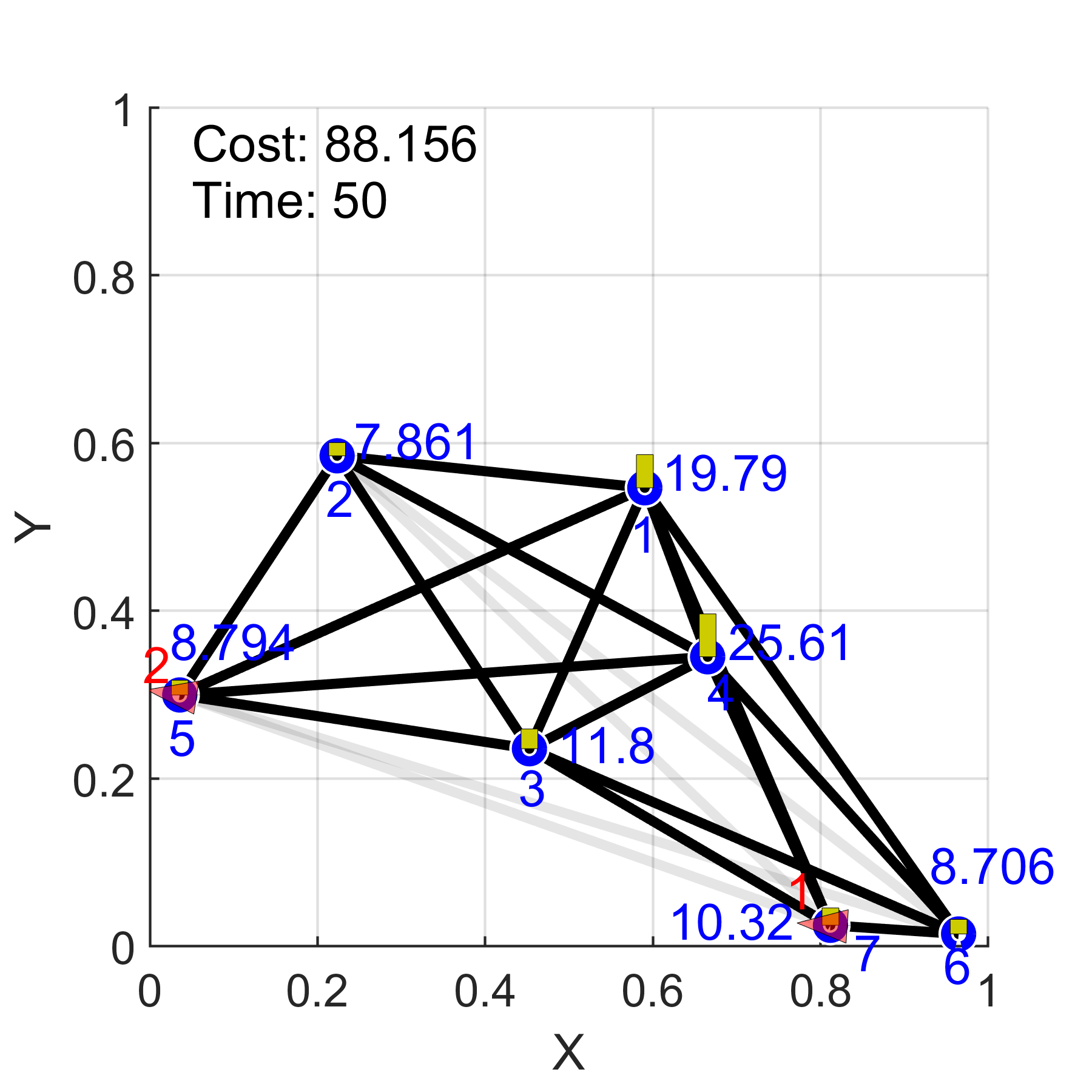}
         \caption{PC 1}
         \label{SubFig:1}
     \end{subfigure}
     \begin{subfigure}[b]{0.4\columnwidth}
         \centering
         \includegraphics[width=\textwidth]{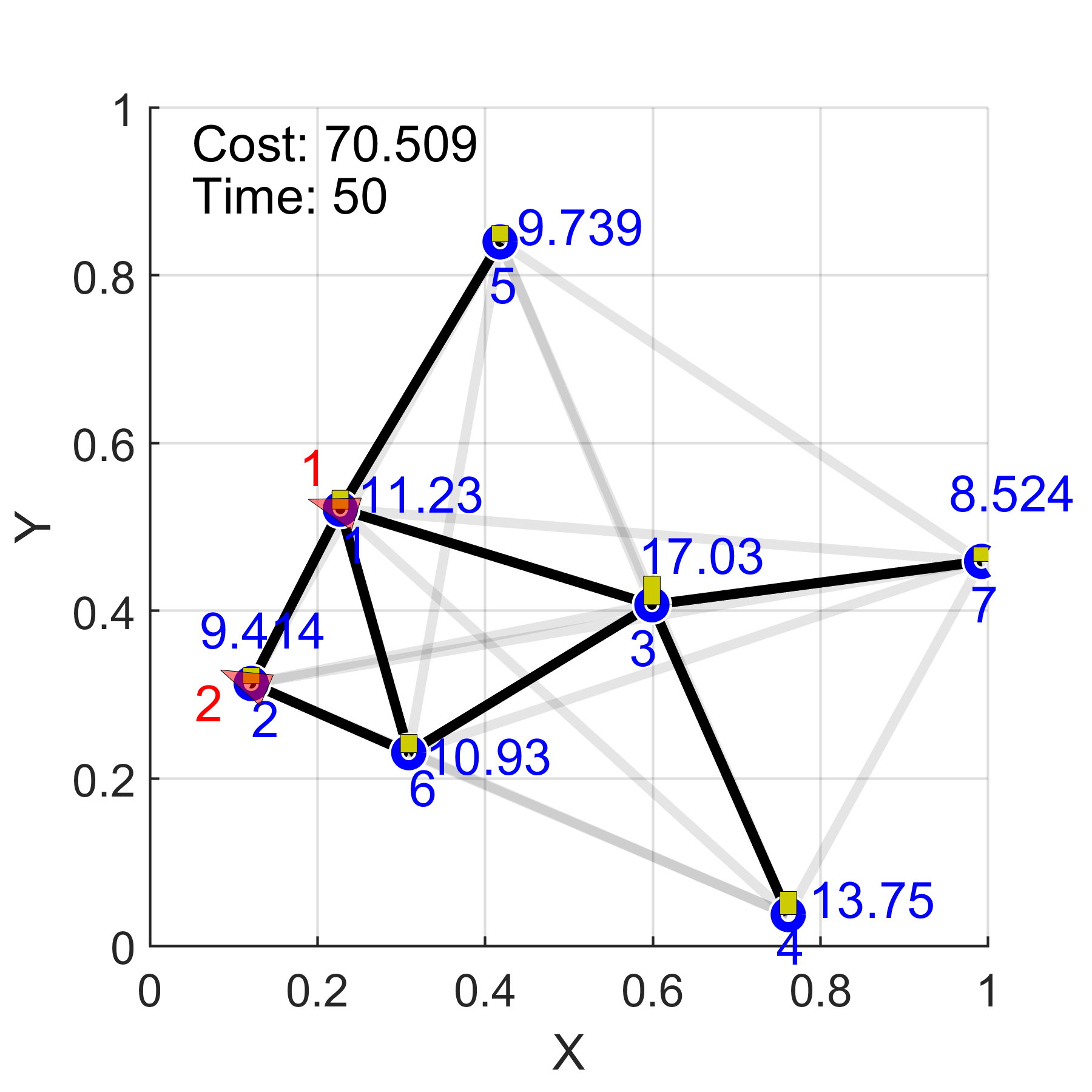}
         \caption{PC 2}
         \label{SubFig:2}
     \end{subfigure}
     \begin{subfigure}[b]{0.4\columnwidth}
         \centering
         \includegraphics[width=\textwidth]{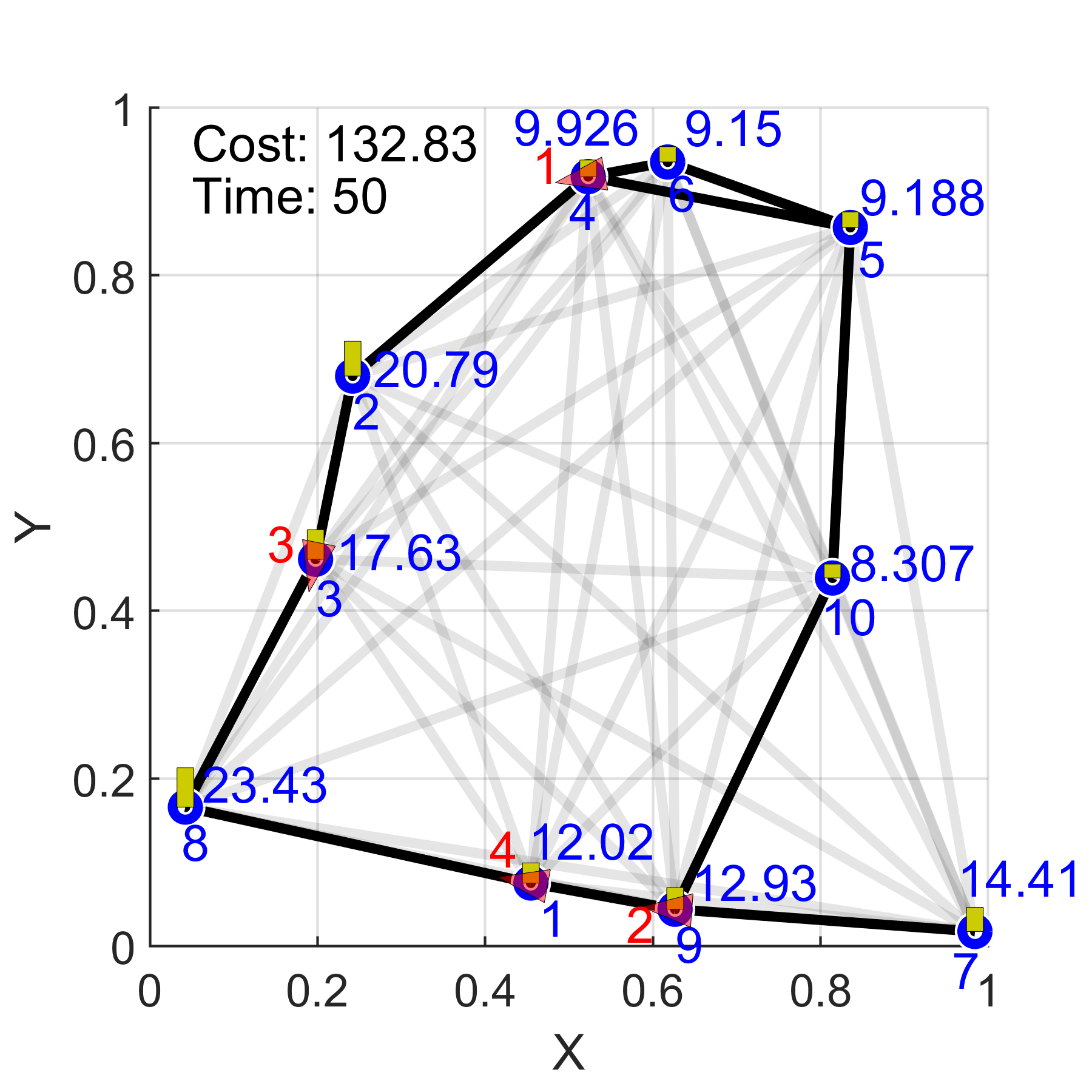}
         \caption{PC 3}
         \label{SubFig:3}
     \end{subfigure}
     \begin{subfigure}[b]{0.4\columnwidth}
         \centering
         \includegraphics[width=\textwidth]{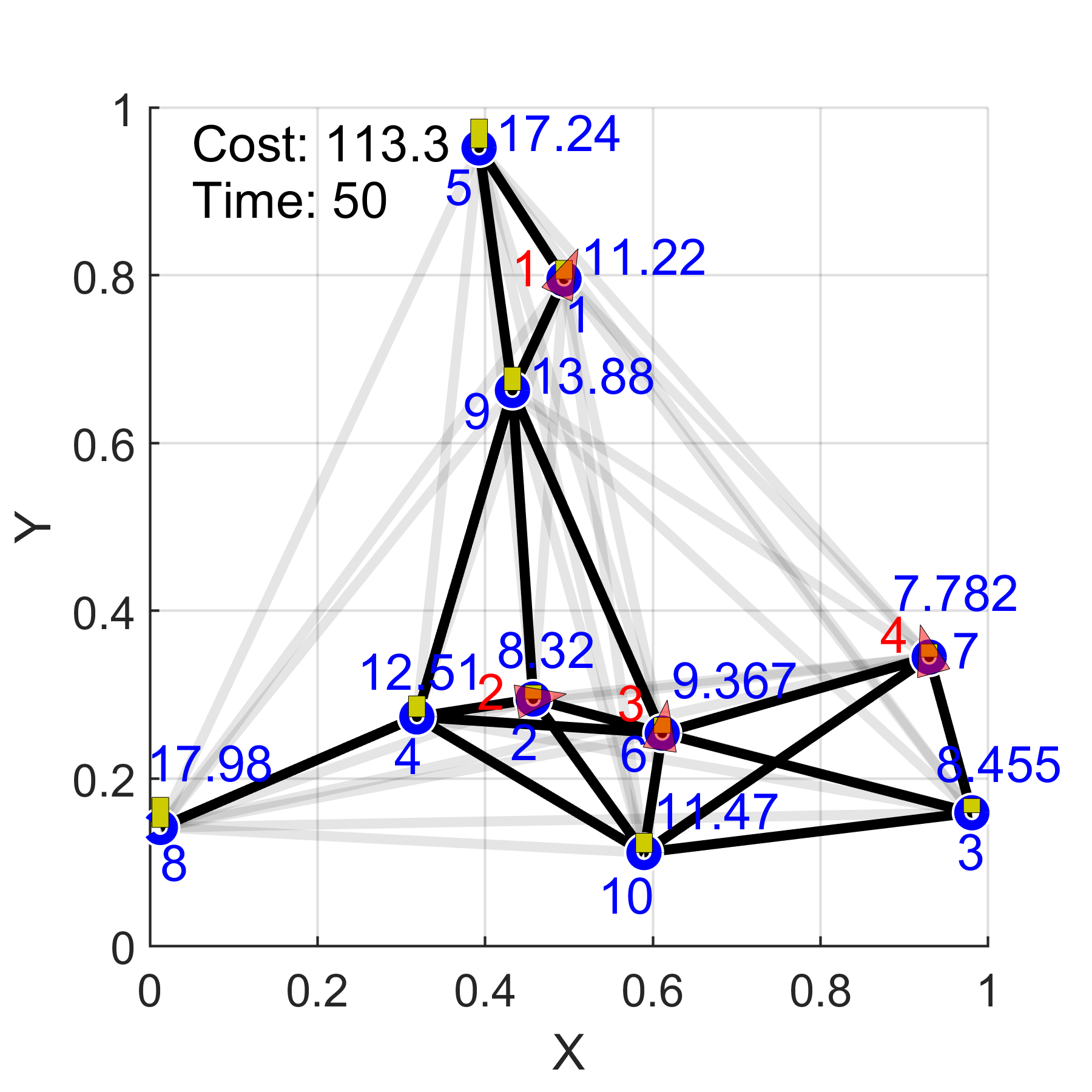}
         \caption{PC 4}
         \label{SubFig:4}
     \end{subfigure}
        \caption{
        Final state of the PCs after using the RHC method.}
        \label{Fig:FinalConditions}
\end{figure}

\paragraph*{\textbf{Results and Discussion}}
Obtained results from the comparison are summarized in Tab. \ref{Tab:TargetStateEstimatorPerformances}. According to these results, on average, the RHC method has outperformed all the other agent control methods. It can be seen that the RHC-P method has the second-best performance level, and it has even performed slightly better than the RHC method in two cases. This observation is justifiable because the RHC-P method has a significant centralized and off-line component compared to the RHC method, which is completely distributed and on-line. Corresponding final states of the PCs given by the RHC method are shown in Fig. \ref{Fig:FinalConditions}.

\begin{table}[!h]
\centering
\caption{Performance comparison of target state estimation (i.e., $J_T$ in \eqref{Eq:GlobalObjective}) under five agent control methods in four PCs.}
\label{Tab:TargetStateEstimatorPerformances}
\resizebox{\columnwidth}{!}{%
\begin{tabular}{|c|c|r|r|r|r|r|}
\hline
\multicolumn{2}{|c|}{\multirow{3}{*}{\begin{tabular}[c]{@{}c@{}}Target State Estimator \\ Performance ($J_T$)\end{tabular}}} & \multicolumn{5}{c|}{Agent Control Mechanism} \\ \cline{3-7} 
\multicolumn{2}{|c|}{} & \multicolumn{1}{c|}{Off-line} & \multicolumn{2}{c|}{Off-line/On-line} & \multicolumn{2}{c|}{On-line} \\ \cline{3-7} 
\multicolumn{2}{|c|}{} & \multicolumn{1}{c|}{MTSP} & \multicolumn{1}{c|}{BDC-P} & \multicolumn{1}{c|}{RHC-P} & \multicolumn{1}{c|}{BDC} & \multicolumn{1}{c|}{RHC} \\ \hline
\multirow{4}{*}{PC No.} & 1 & 99.88 & 119.23 & \textbf{84.41} & 88.68 & 88.16 \\ \cline{2-7} 
 & 2 & 90.08 & 155.25 & 77.80 & 101.75 & \textbf{70.51} \\ \cline{2-7} 
 & 3 & 133.50 & 268.85 & \textbf{128.90} & 162.48 & 132.83 \\ \cline{2-7} 
 & 4 & 187.88 & 231.28 & 123.70 & 174.32 & \textbf{113.30} \\ \hline
\multicolumn{2}{|c|}{\textbf{Average:}} & \multicolumn{1}{r|}{127.83} & \multicolumn{1}{r|}{193.65} & \multicolumn{1}{r|}{103.70} & \multicolumn{1}{r|}{131.81} & \multicolumn{1}{r|}{\textbf{101.20}} \\ \hline
\end{tabular}%
}
\end{table}

\paragraph*{\textbf{The Performance in Terms of $\hat{J}_T$ in \eqref{Eq:GlobalObjectiveHat}}} 

\begin{figure}[!t]
    \centering
    \includegraphics[width=0.7\columnwidth]{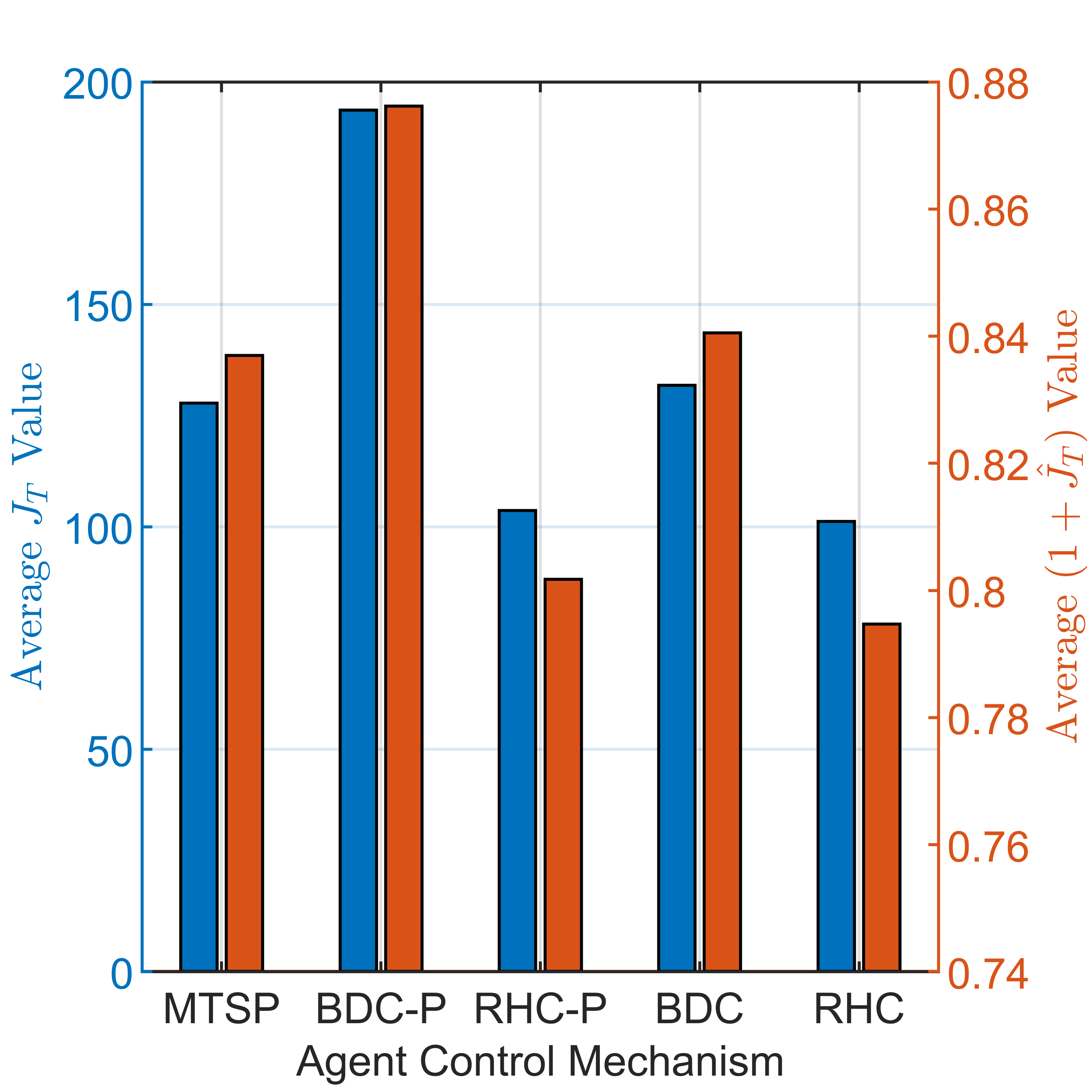}
    \caption{Comparison of global objective functions $J_T$ in \eqref{Eq:GlobalObjective} and $\hat{J}_T$ in \eqref{Eq:GlobalObjectiveHat} - based on average performance values reported in Tab. \ref{Tab:TargetStateEstimatorPerformances} and Tab. \ref{Tab:TargetStateEstimatorPerformancesHat}.}
    \label{Fig:ObjectiveComparison}
\end{figure}

Recall that we earlier claimed both the original and the alternative global objective functions (i.e., $J_T$ in \eqref{Eq:GlobalObjective} and $\hat{J}_T$ in \eqref{Eq:GlobalObjectiveHat}, respectively) behave similarly in experiments under same agent controls (e.g., see Fig. \ref{Fig:ObjectiveFunctionComparison}(a)). To further validate this claim, we computed the $\hat{J}_T$ values corresponding to the same experiments that generated the results reported in Tab. \ref{Tab:TargetStateEstimatorPerformances}. These new results are provided in Tab. \ref{Tab:TargetStateEstimatorPerformancesHat}. In there, to improve the readability, note that we have actually provided $(1+\hat{J}_T)$ values (that represent the \emph{inefficiency} of the overall agent sensing effort) rather than $\hat{J}_T$ values. Figure \ref{Fig:ObjectiveComparison} compares the average performance values reported in Tab. \ref{Tab:TargetStateEstimatorPerformances} and Tab. \ref{Tab:TargetStateEstimatorPerformancesHat}. 
While these results validate our stated claim, they also show that the proposed RHC method is capable of allocating agent sensing resources more efficiently (by $\sim 4\%$) compared to other agent control methods.

\begin{table}[!h]
\centering
\caption{Performance comparison of target state estimation in terms of the alternative global objective $\hat{J}_T$ in \eqref{Eq:GlobalObjectiveHat} under five agent control methods in four PCs.}
\label{Tab:TargetStateEstimatorPerformancesHat}
\resizebox{\columnwidth}{!}{%
\begin{tabular}{|c|c|r|r|r|r|r|}
\hline
\multicolumn{2}{|c|}{\multirow{3}{*}{\begin{tabular}[c]{@{}c@{}}Target State Estimator \\ Performance ($1+\hat{J}_T$)\end{tabular}}} & \multicolumn{5}{c|}{Agent Control Mechanism} \\ \cline{3-7} 
\multicolumn{2}{|c|}{} & \multicolumn{1}{c|}{Off-line} & \multicolumn{2}{c|}{Off-line/On-line} & \multicolumn{2}{c|}{On-line} \\ \cline{3-7} 
\multicolumn{2}{|c|}{} & \multicolumn{1}{c|}{MTSP} & \multicolumn{1}{c|}{BDC-P} & \multicolumn{1}{c|}{RHC-P} & \multicolumn{1}{c|}{BDC} & \multicolumn{1}{c|}{RHC} \\ \hline
\multirow{4}{*}{PC No.} & 1 & 0.864 & 0.879 & \textbf{0.837} & 0.842 & 0.843 \\ \cline{2-7} 
 & 2 & 0.863 & 0.863 & 0.837 & 0.874 & \textbf{0.819} \\ \cline{2-7} 
 & 3 & 0.760 & 0.882 & \textbf{0.750} & 0.804 & 0.759 \\ \cline{2-7} 
 & 4 & 0.861 & 0.882 & 0.783 & 0.842 & \textbf{0.757} \\ \hline
\multicolumn{2}{|c|}{\textbf{Average:}} & \multicolumn{1}{r|}{0.837} & \multicolumn{1}{r|}{0.876} & \multicolumn{1}{r|}{0.802} & \multicolumn{1}{r|}{0.841} & \multicolumn{1}{r|}{\textbf{0.795}} \\ \hline
\end{tabular}%
}
\end{table}

\paragraph*{\textbf{The Worst-Case Performance}} 
Inspired by the objective function \eqref{Eq:InfiniteHorizonObjective} used in \cite{Pinto2021} (and also to make the performance comparison with \cite{Pinto2021} fair), we define the \emph{worst-case performance} of an agent controller over the period $[0,T]$ as $J_W$ where
\begin{equation}\label{Eq:WorstCasePerformance}
   J_W = \max_{i\in\mathcal{V},\ t\in[0,T]} \mbox{tr}(\Omega_i(t)). 
\end{equation}
In our case, $J_W$ is simply the maximum recorded target error covariance value in the network over the period $[0, T]$.
For the same experiments that gave the results shown in Tab. \ref{Tab:TargetStateEstimatorPerformances}, we have evaluated the corresponding $J_W$ \eqref{Eq:WorstCasePerformance} value and the obtained results are summarized in Tab.  \ref{Tab:TargetStateEstimatorWorstCasePerformances}. According to these results, it is evident that the periodic agent control methods have an advantage compared to the fully distributed and on-line methods like RHC and BDC in terms of worst-case performance. Nevertheless, the fact that the RHC-P method has obtained the best average $J_W$ value (and the second-best average $J_T$ value as shown in Tab. \ref{Tab:TargetStateEstimatorPerformances}) implies that the proposed RHC method can successfully be adopted to address different problem settings (with different constraints, objectives, etc.). 

\begin{table}[!h]
\centering
\caption{The worst-case performance comparison (i.e., $J_W$ in \eqref{Eq:WorstCasePerformance}) under five agent control methods in four PCs.}
\label{Tab:TargetStateEstimatorWorstCasePerformances}
\resizebox{\columnwidth}{!}{%
\begin{tabular}{|c|c|r|r|r|r|r|}
\hline
\multicolumn{2}{|c|}{\multirow{3}{*}{\begin{tabular}[c]{@{}c@{}}The Worst-Case \\ Target State Estimator \\ Performance ($J_W$)\end{tabular}}} & \multicolumn{5}{c|}{Agent Control Mechanism} \\ \cline{3-7} 
\multicolumn{2}{|c|}{} & \multicolumn{1}{c|}{Off-line} & \multicolumn{2}{c|}{Off-line/On-line} & \multicolumn{2}{c|}{On-line} \\ \cline{3-7} 
\multicolumn{2}{|c|}{} & \multicolumn{1}{c|}{MTSP} & \multicolumn{1}{c|}{BDC-P} & \multicolumn{1}{c|}{RHC-P} & \multicolumn{1}{c|}{BDC} & \multicolumn{1}{c|}{RHC} \\ \hline
\multirow{4}{*}{PC No.} & 1 & \textbf{40.23} & 288.34 & 50.19 & 96.33 & 71.57 \\ \cline{2-7} 
 & 2 & 38.03 & 586.19 & 65.62 & 270.42 & \textbf{32.29} \\ \cline{2-7} 
 & 3 & 47.54 & 427.53 & \textbf{46.52} & 386.66 & 306.84 \\ \cline{2-7} 
 & 4 & 115.74 & 403.38 & \textbf{49.26} & 487.65 & 77.41 \\ \hline
\multicolumn{2}{|c|}{\textbf{Average:}} & 60.38 & 426.36 & \textbf{52.90} & 310.27 & 122.03 \\ \hline
\end{tabular}%
}
\end{table}

\begin{figure}[!h]
     \centering
     \begin{subfigure}[b]{0.32\columnwidth}
         \centering
         \includegraphics[width=\textwidth]{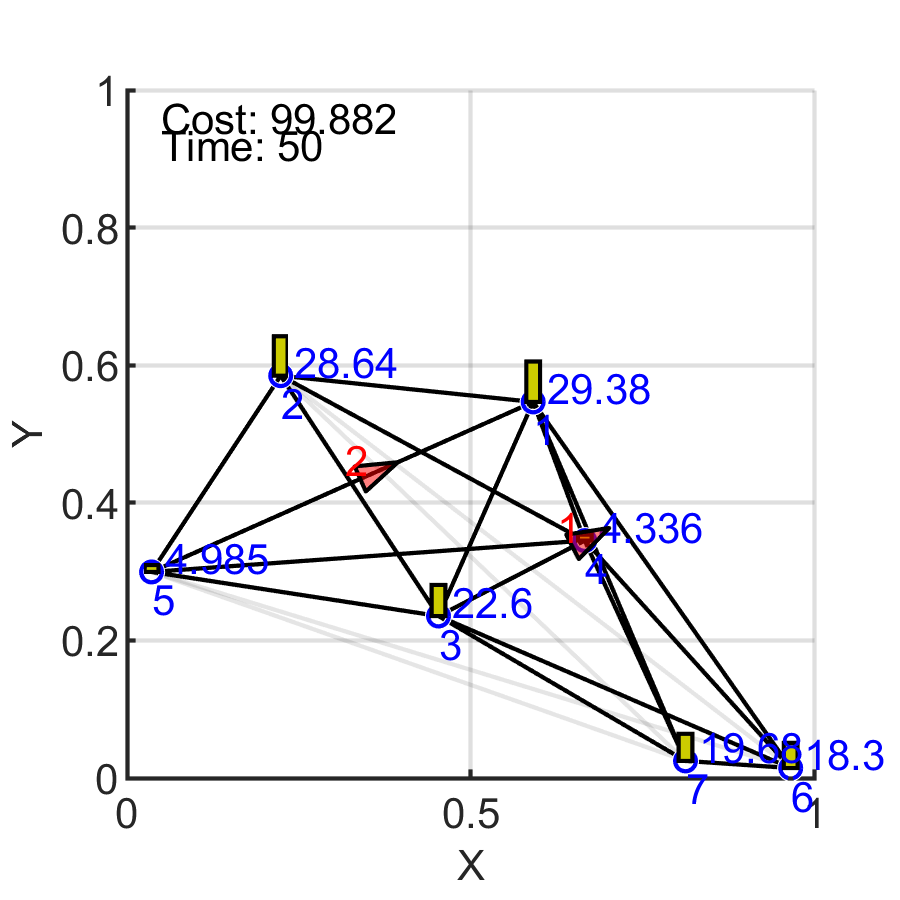}
         \caption{MTSP: \\ $J_T = 99.9$}
         \label{SubFig:1}
     \end{subfigure}
     \begin{subfigure}[b]{0.32\columnwidth}
         \centering
         \includegraphics[width=\textwidth]{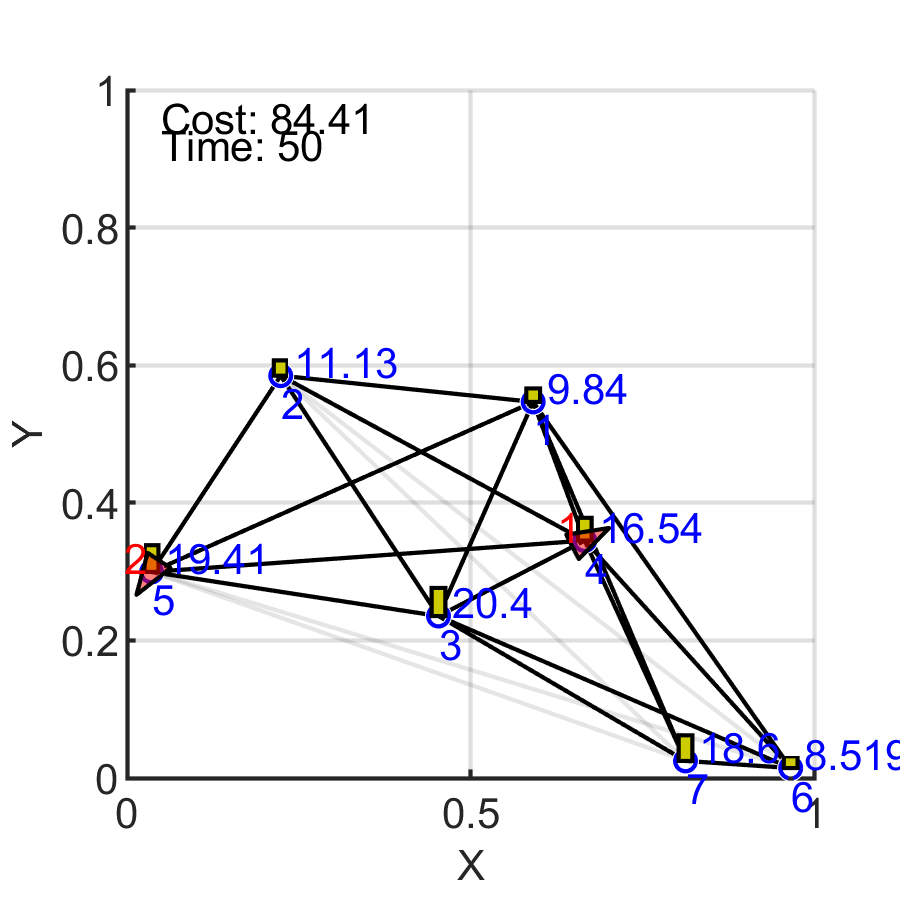}
         \caption{RHC-P: \\ $J_T = \textbf{84.4}$}
         \label{SubFig:2}
     \end{subfigure}
     \begin{subfigure}[b]{0.32\columnwidth}
         \centering
         \includegraphics[width=\textwidth]{results/STCase1/RHC.png}
         \caption{RHC: \\ $J_T = 88.2$}
         \label{SubFig:3}
     \end{subfigure}
        \caption{Performance of target state estimation in PC 1.}
        \label{Fig:ShortTerm1}
\end{figure}

\begin{figure}[!h]
     \centering
     \begin{subfigure}[b]{0.32\columnwidth}
         \centering
         \captionsetup{justification=centering}
         \includegraphics[width=\textwidth]{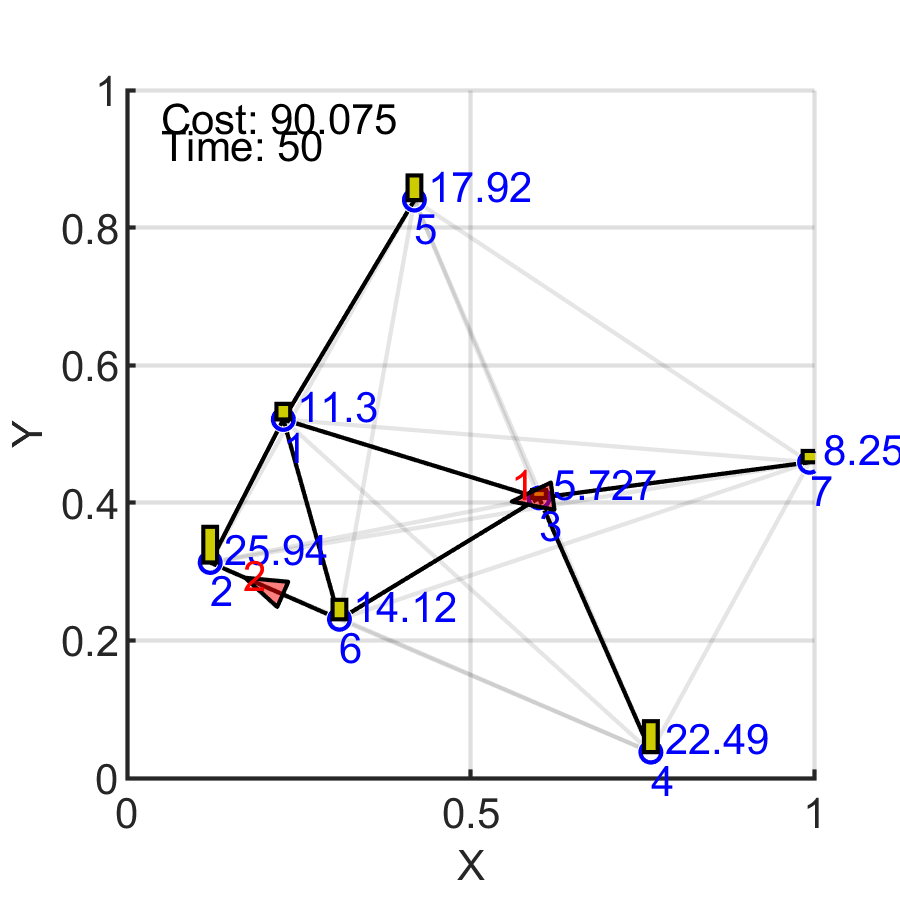}
         \caption{MTSP: \\ $J_T = 90.1$}
         \label{SubFig:1}
     \end{subfigure}
     \begin{subfigure}[b]{0.32\columnwidth}
         \centering
        \captionsetup{justification=centering}
         \includegraphics[width=\textwidth]{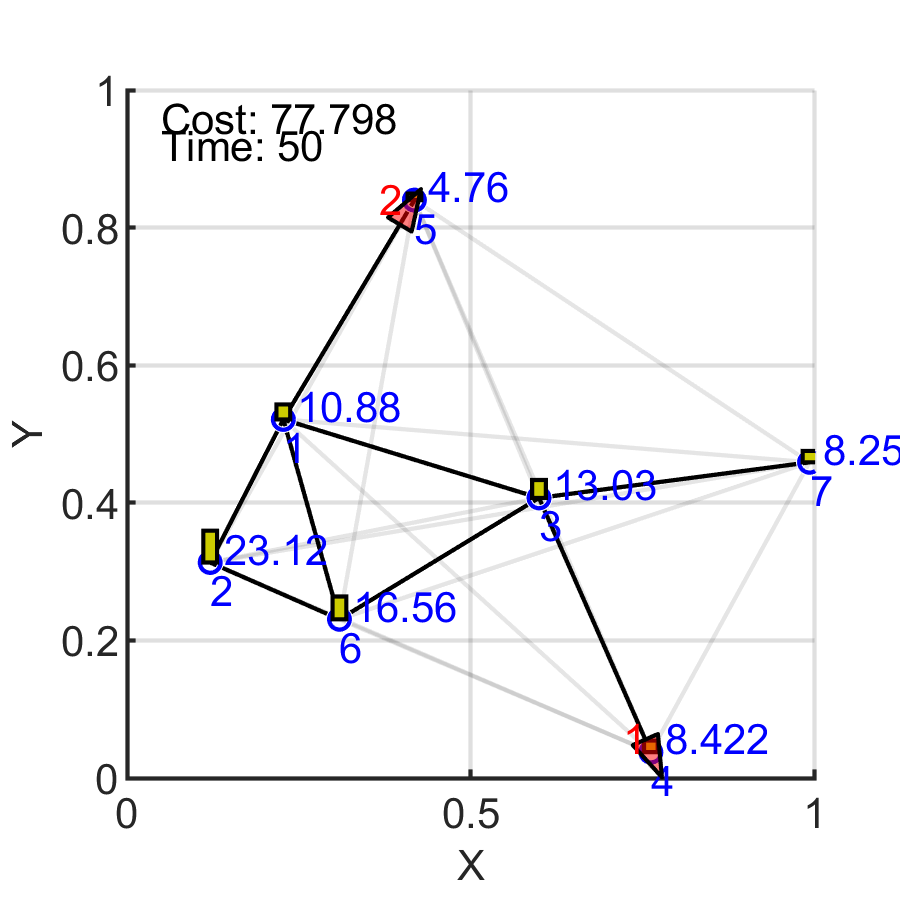}
         \caption{RHC-P: \\ $J_T = 77.8$}
         \label{SubFig:2}
     \end{subfigure}
     \begin{subfigure}[b]{0.32\columnwidth}
         \centering
         \captionsetup{justification=centering}
         \includegraphics[width=\textwidth]{results/STCase2/RHC.png}
         \caption{RHC: \\ $J_T = \textbf{70.5}$}
         \label{SubFig:3}
     \end{subfigure}
        \caption{Performance of target state estimation in PC 2.}
        \label{Fig:ShortTerm2}
\end{figure}

\begin{figure}[!h]
     \centering
     \begin{subfigure}[b]{0.32\columnwidth}
         \centering
         \captionsetup{justification=centering}
         \includegraphics[width=\textwidth]{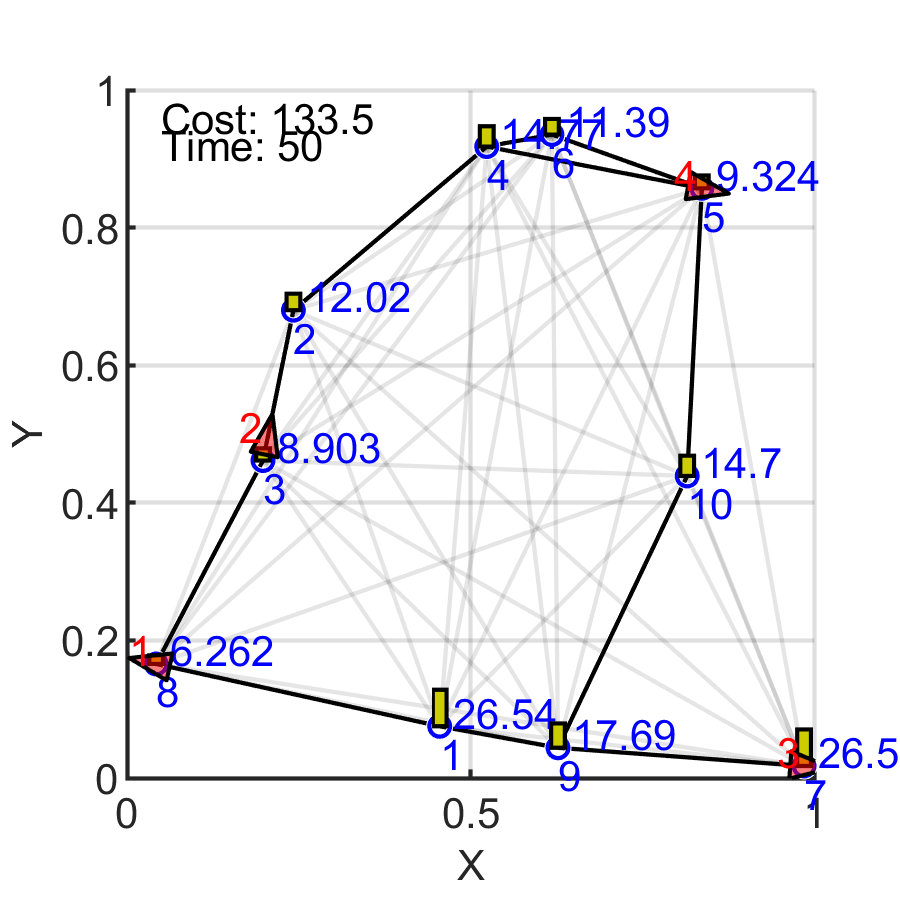}
         \caption{MTSP: \\ $J_T = 133.5$}
         \label{SubFig:1}
     \end{subfigure}
     \begin{subfigure}[b]{0.32\columnwidth}
         \centering
         \captionsetup{justification=centering}
         \includegraphics[width=\textwidth]{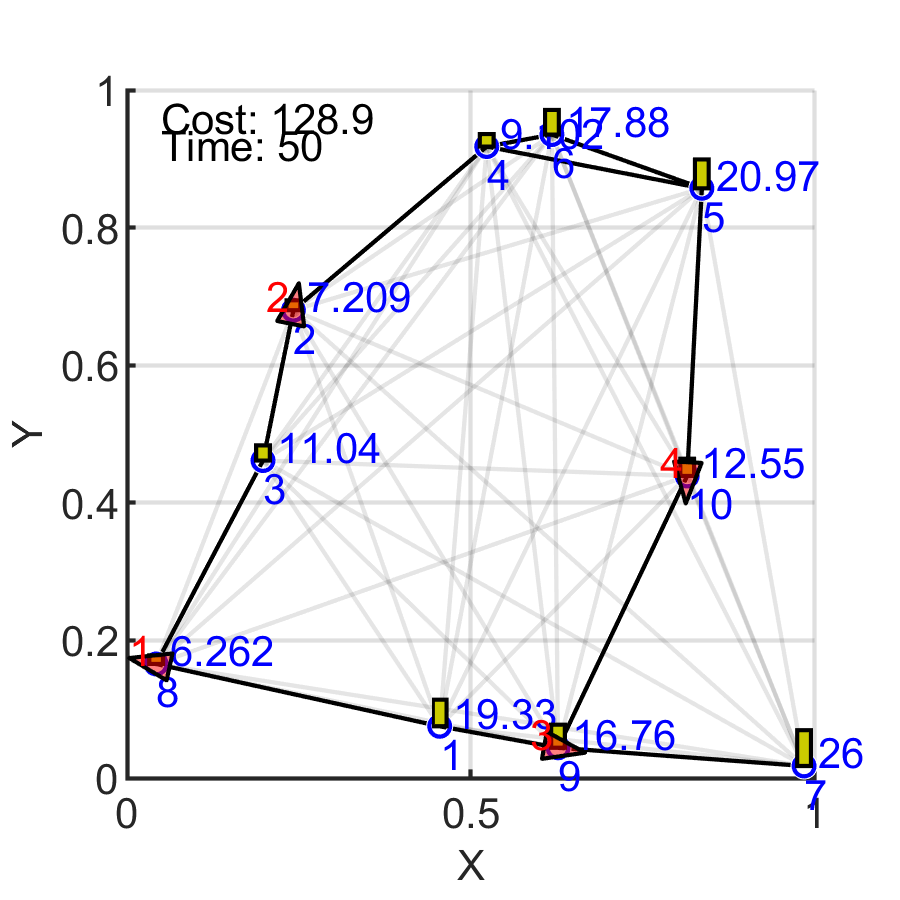}
         \caption{RHC-P: \\ $J_T = \textbf{128.9}$}
         \label{SubFig:2}
     \end{subfigure}
     \begin{subfigure}[b]{0.32\columnwidth}
         \centering
         \captionsetup{justification=centering}
         \includegraphics[width=\textwidth]{results/STCase3/RHC.png}
         \caption{RHC: \\ $J_T = 132.8$}
         \label{SubFig:3}
     \end{subfigure}
        \caption{Performance of target state estimation in PC 3.}
        \label{Fig:ShortTerm3}
\end{figure}

\begin{figure}[!h]
     \centering
     \begin{subfigure}[b]{0.32\columnwidth}
         \centering
         \captionsetup{justification=centering}
         \includegraphics[width=\textwidth]{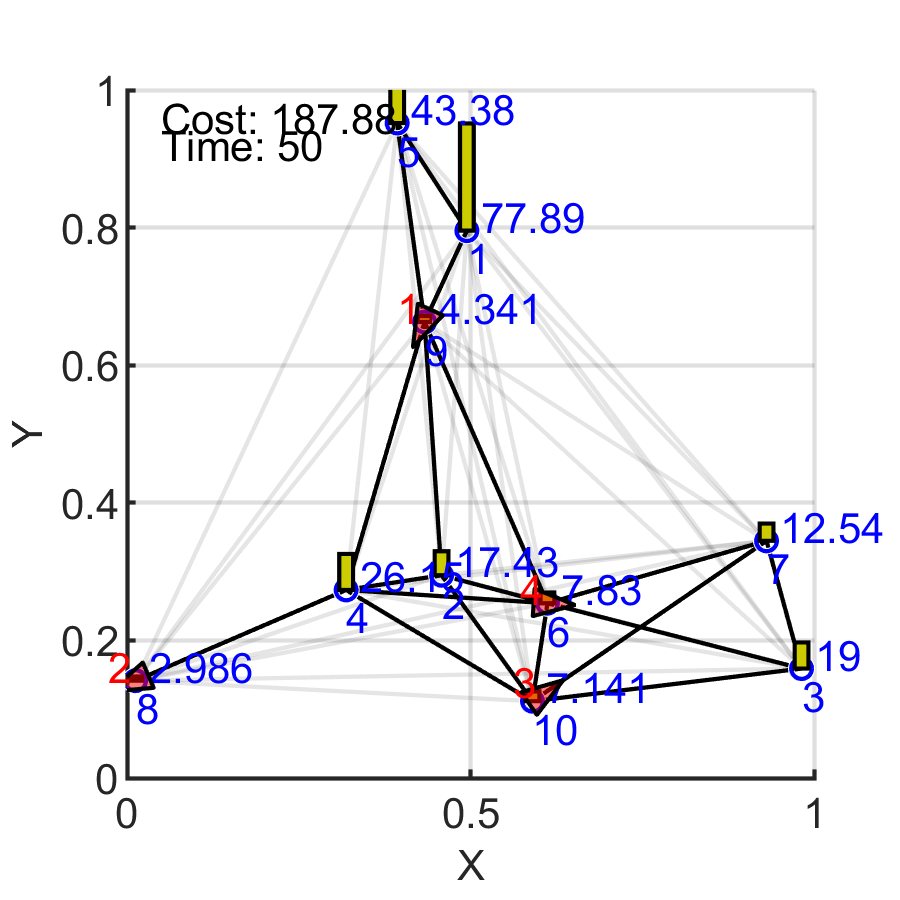}
         \caption{MTSP: \\ $J_T = 187.9$}
         \label{SubFig:1}
     \end{subfigure}
     \begin{subfigure}[b]{0.32\columnwidth}
         \centering
         \captionsetup{justification=centering}
         \includegraphics[width=\textwidth]{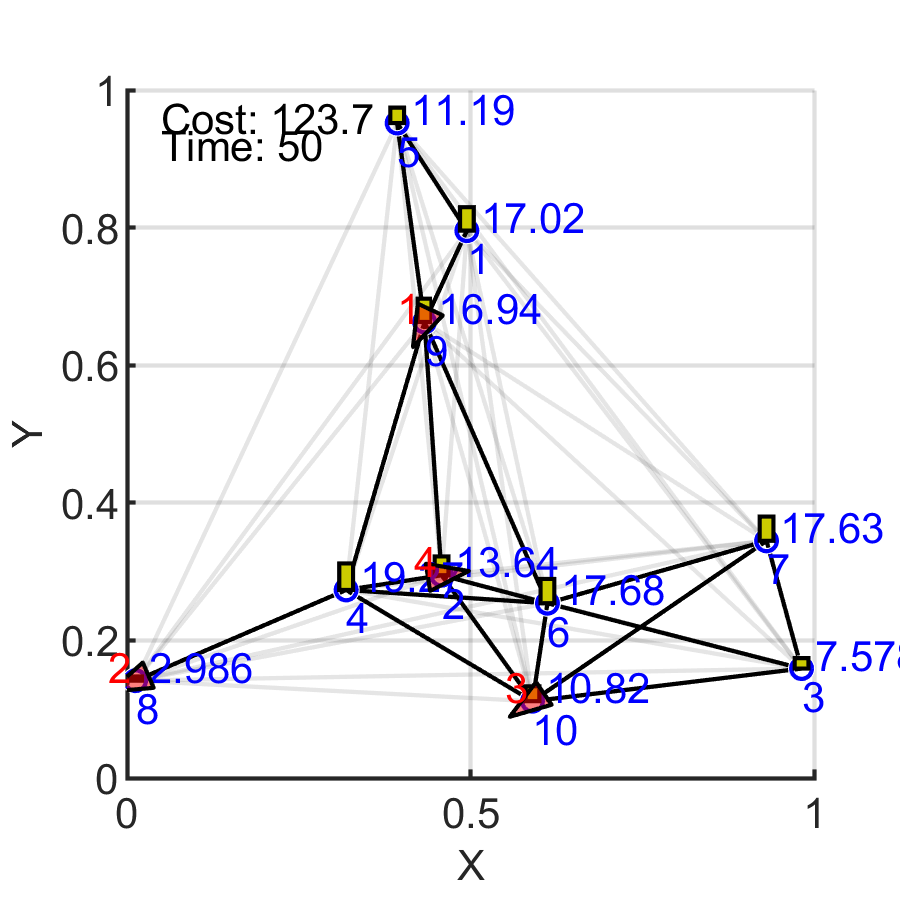}
         \caption{RHC-P: \\ $J_T = 123.7$}
         \label{SubFig:2}
     \end{subfigure}
     \begin{subfigure}[b]{0.32\columnwidth}
         \centering
         \captionsetup{justification=centering}
         \includegraphics[width=\textwidth]{results/STCase4/RHC.png}
         \caption{RHC: \\ $J_T = \textbf{113.3}$}
         \label{SubFig:3}
     \end{subfigure}
        \caption{Performance of target state estimation in PC 4.}
        \label{Fig:ShortTerm4}
\end{figure}

\subsection{Simulation Study 2: The effect of agent controls on local target state tracking control}

In this simulation study, we explore a byproduct of achieving reasonable target state estimates: the ability to control the target states effectively. Here, we assume each target has its own tracking control task that needs to be achieved through a simple local state feedback control mechanism. Clearly, for this purpose, each target has to rely on its own state estimate - of which the accuracy deteriorates when the target is not visited by an agent regularly. We define a new metric $J_C$ to represent the performance of the overall \emph{target state tracking control} process and compare the obtained $J_C$ values by different agent controllers under different PCs.

\paragraph{\textbf{Target Control Mechanism}}
In this simulation study, we assume that each target $i\in\mathcal{V}$ has to control its state $\phi_i(t)$ such that a signal
\begin{equation}\label{Eq:TargetStateOutput}
    y_i(t) = C_i \phi_i(t) + D_i,
\end{equation}
tracks a given reference signal $r_i(t)$ ($C_i,\,D_i$ are also given).

Let us define the tracking error as $e_i(t) = y_i(t)-r_i(t)$. In order to make $e_i(t)$ follow the asymptotically stable dynamics: $\dot{e}_i = -K_ie_i(t)$ (with $K_i>0$), the target $i$ needs to select its control input $\upsilon_i(t)$ in \eqref{Eq:TargetStateDynamics} as (also recall \eqref{Eq:OneDimensionConstraint})
\begin{equation*}
    \upsilon_i(t) = -\frac{1}{B_iC_i}\left(C_i(A_i + K_i)\phi_i(t) + K_iD_i - (\dot{r}_i(t)+K_ir_i(t))\right).
\end{equation*}
However, since target $i$ is unaware of its state $\phi_i$, naturally, the target state tracking controller can use the state estimate $\hat{\phi}_i$ in the above state feedback control law as
\begin{equation*}
    \upsilon_i(t) = -\frac{1}{B_iC_i}\left(C_i(A_i + K_i)\hat{\phi}_i(t) + K_iD_i - (\dot{r}_i(t)+K_ir_i(t))\right).
\end{equation*}
To measure the performance of this target state tracking control task, we propose to use the performance metric $J_C$ where
\begin{equation}\label{Eq:TargetStateTrackingControlPerformance}
    J_C \triangleq \frac{1}{T}\int_0^T \vert e_i(t) \vert dt.
\end{equation}

\begin{figure}[!h]
     \centering
     \begin{subfigure}[b]{0.4\columnwidth}
         \centering
         \includegraphics[width=\textwidth]{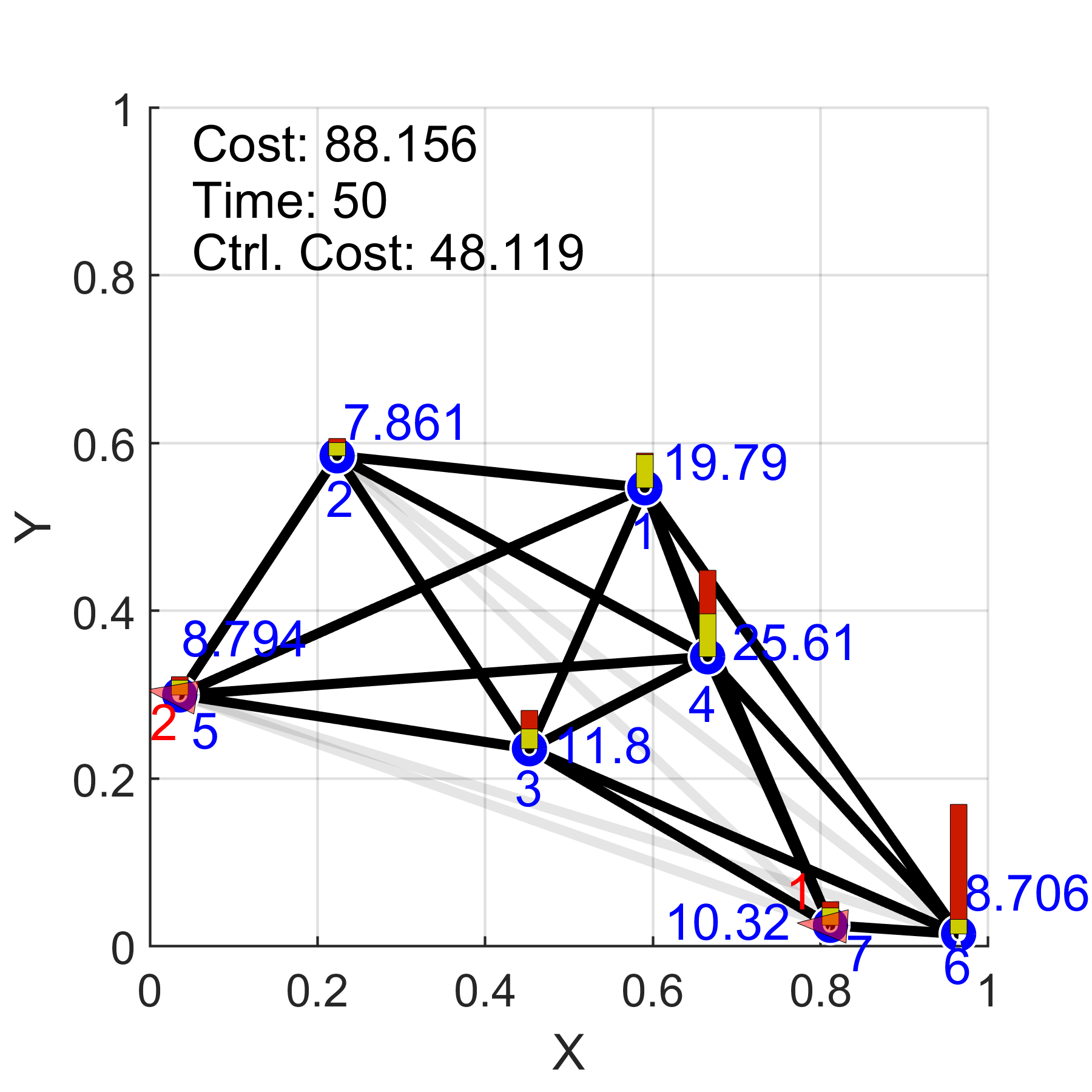}
         \caption{PC 1}
         \label{SubFig:1}
     \end{subfigure}
     \begin{subfigure}[b]{0.4\columnwidth}
         \centering
         \includegraphics[width=\textwidth]{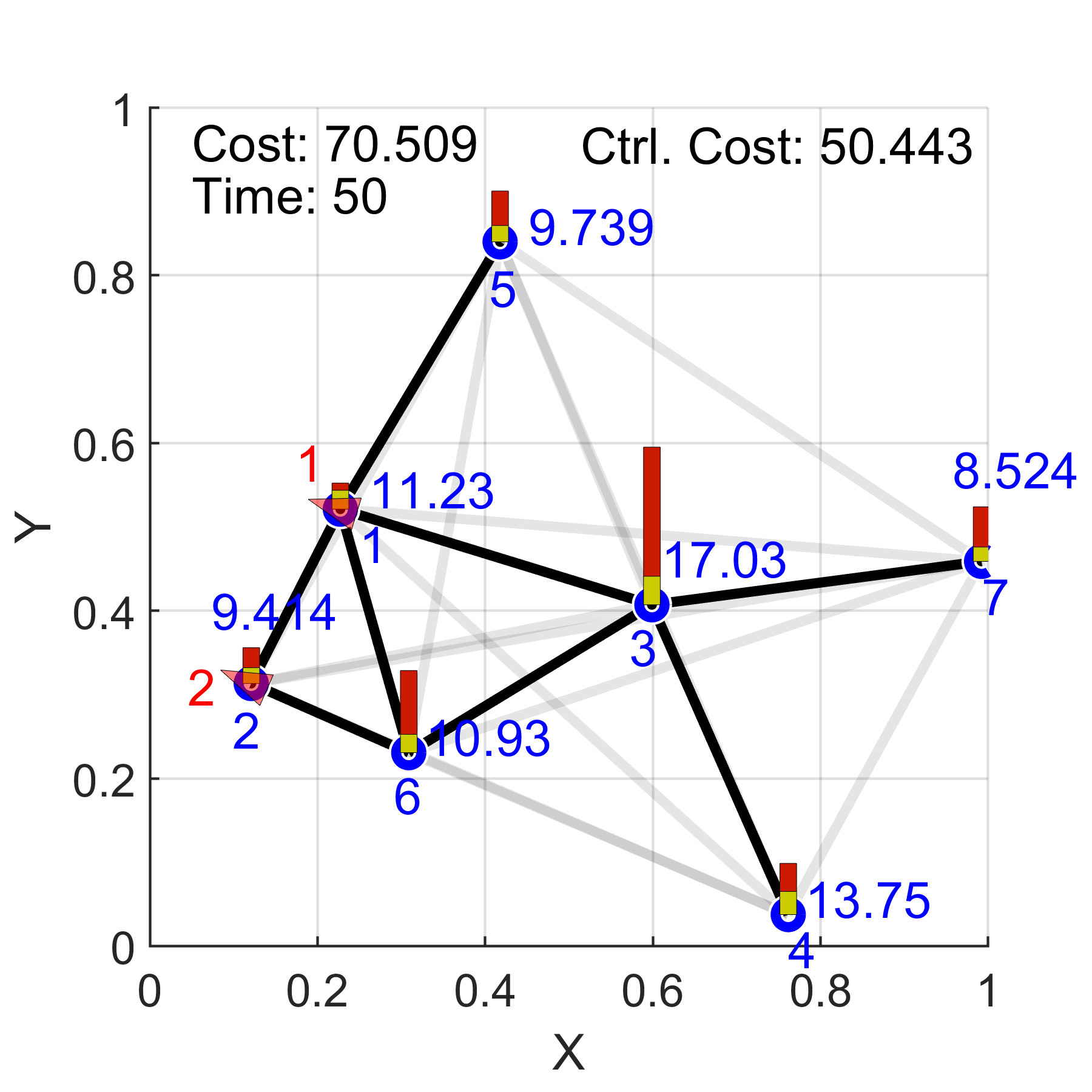}
         \caption{PC 2}
         \label{SubFig:2}
     \end{subfigure}
     \begin{subfigure}[b]{0.4\columnwidth}
         \centering
         \includegraphics[width=\textwidth]{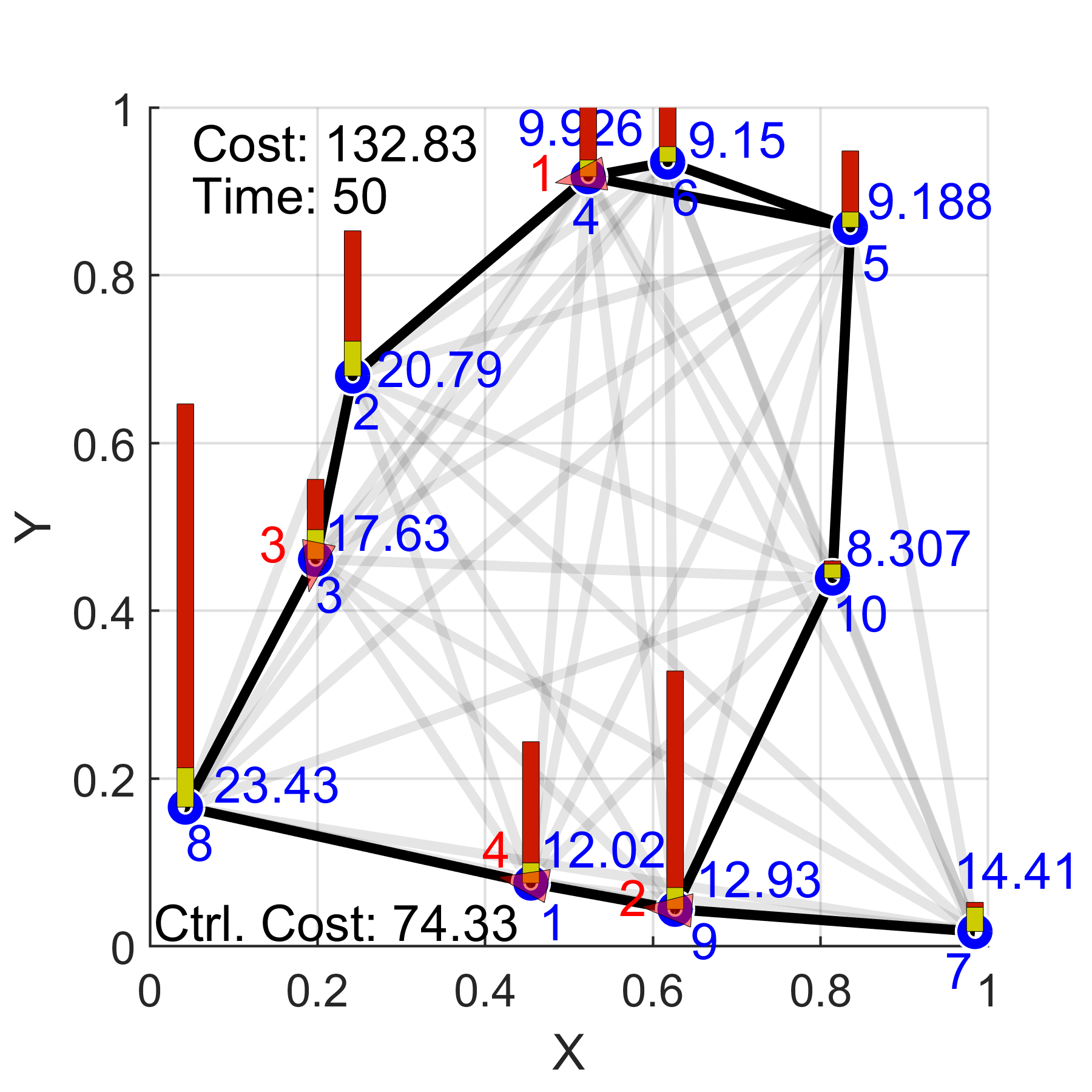}
         \caption{PC 3}
         \label{SubFig:3}
     \end{subfigure}
     \begin{subfigure}[b]{0.4\columnwidth}
         \centering
         \includegraphics[width=\textwidth]{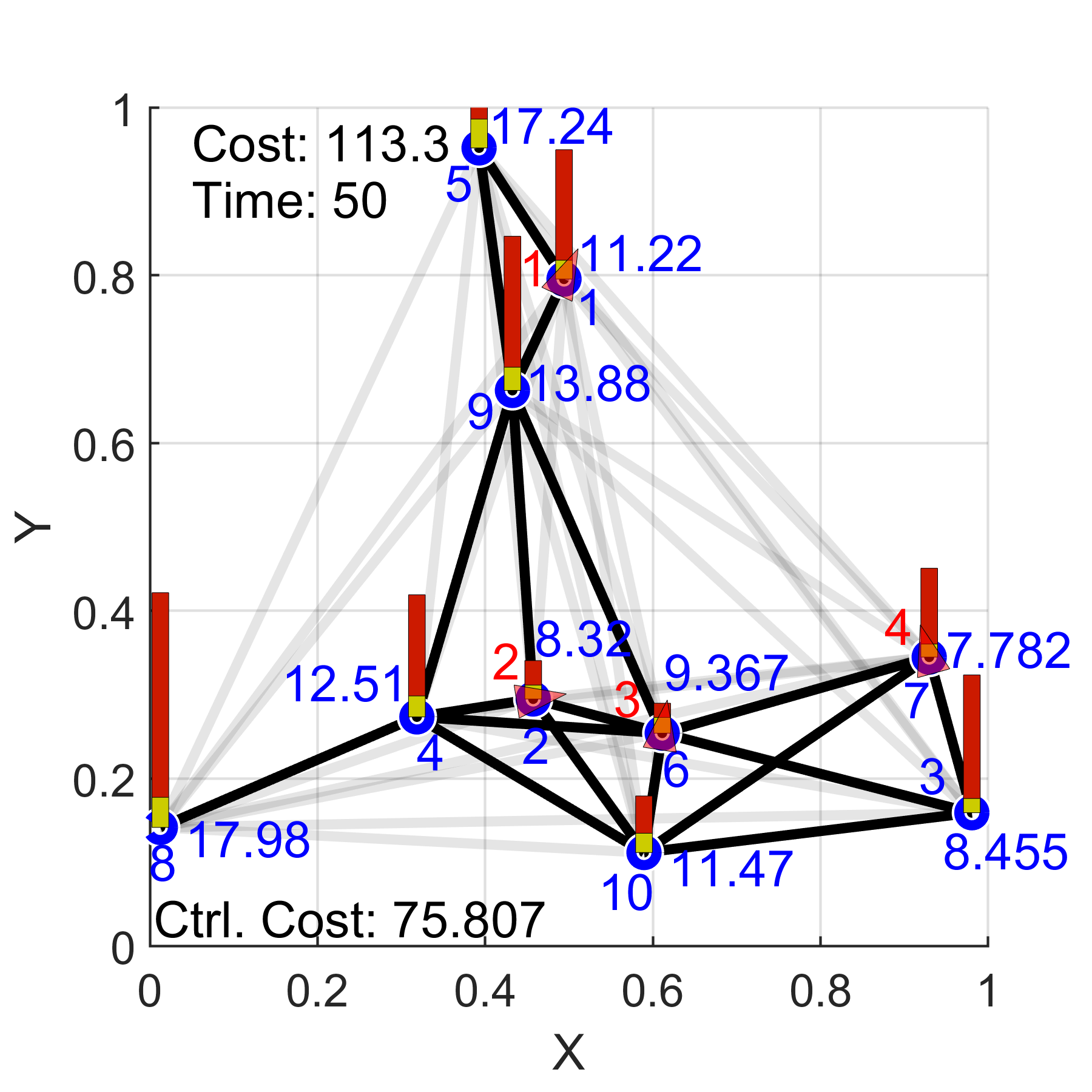}
         \caption{PC 4}
         \label{SubFig:4}
     \end{subfigure}
        \caption{
        Final state of the PCs after using the RHC method with target state tracking control.}
        \label{Fig:FinalConditionsWithTargetControl}
\end{figure}

\paragraph{\textbf{Results and Discussion}}
In this study, we set $C_i=1$, $D_i=0$, $K_i=2$ and select the reference signal that needs to be tracked as: $r_i(t) = 10\sin(2t+i),\,\forall i\in\mathcal{V}, t\in [0,T]$ with $T=50\,s$. The performance metric $J_C$ observed for different PCs with different agent controllers are summarized in Tab \ref{Tab:TargetStateControllerPerformances}. Similar to before, the obtained results show that the RHC method, on average, has outperformed all the other agent controllers. Corresponding final states of the PCs observed under the RHC method are shown in Fig. \ref{Fig:FinalConditionsWithTargetControl}. The red vertical bars (drawn on top of yellow vertical bars) represent the absolute tracking error $\vert e_i(t)\vert$ of each target $i$ at $t=T$. These results imply that having an agent control mechanism that provides superior target state estimation capabilities (i.e., lower $J_T$) indirectly enables the targets to have better control over their states (i.e., lower $J_C$).

\begin{table}[!h]
\centering
\caption{Performance comparison of target state tracking controllers (i.e., $J_C$ in \eqref{Eq:TargetStateTrackingControlPerformance}) under five agent control methods in four PCs.}
\label{Tab:TargetStateControllerPerformances}
\resizebox{\columnwidth}{!}{%
\begin{tabular}{|c|c|r|r|r|r|r|}
\hline
\multicolumn{2}{|c|}{\multirow{3}{*}{\begin{tabular}[c]{@{}c@{}}Target State Controller \\ Performance ($J_C\times T$)\end{tabular}}} & \multicolumn{5}{c|}{Agent Control Mechanism} \\ \cline{3-7} 
\multicolumn{2}{|c|}{} & \multicolumn{1}{c|}{Off-line} & \multicolumn{2}{c|}{Off-line/on-line} & \multicolumn{2}{c|}{On-line} \\ \cline{3-7} 
\multicolumn{2}{|c|}{} & \multicolumn{1}{c|}{MTSP} & \multicolumn{1}{c|}{BDC-P} & \multicolumn{1}{c|}{RHC-P} & \multicolumn{1}{c|}{BDC} & \multicolumn{1}{c|}{RHC} \\ \hline
\multirow{4}{*}{PC No.} & 1 & 57.97 & 51.99 & 50.99 & 48.40 & \textbf{48.12} \\ \cline{2-7} 
 & 2 & 51.66 & 56.80 & 52.35 & 55.54 & \textbf{50.14} \\ \cline{2-7} 
 & 3 & \textbf{73.81} & 81.40 & 74.08 & 80.89 & 74.33 \\ \cline{2-7} 
 & 4 & 85.76 & 86.34 & 77.35 & 86.61 & \textbf{75.81} \\ \hline
\multicolumn{2}{|c|}{\textbf{Average:}} & 67.30 & 69.13 & 63.69 & 67.86 & \textbf{62.10} \\ \hline
\end{tabular}%
}
\end{table}

\begin{figure}[!h]
     \centering
     \begin{subfigure}[b]{0.32\columnwidth}
         \centering
         \includegraphics[width=\textwidth]{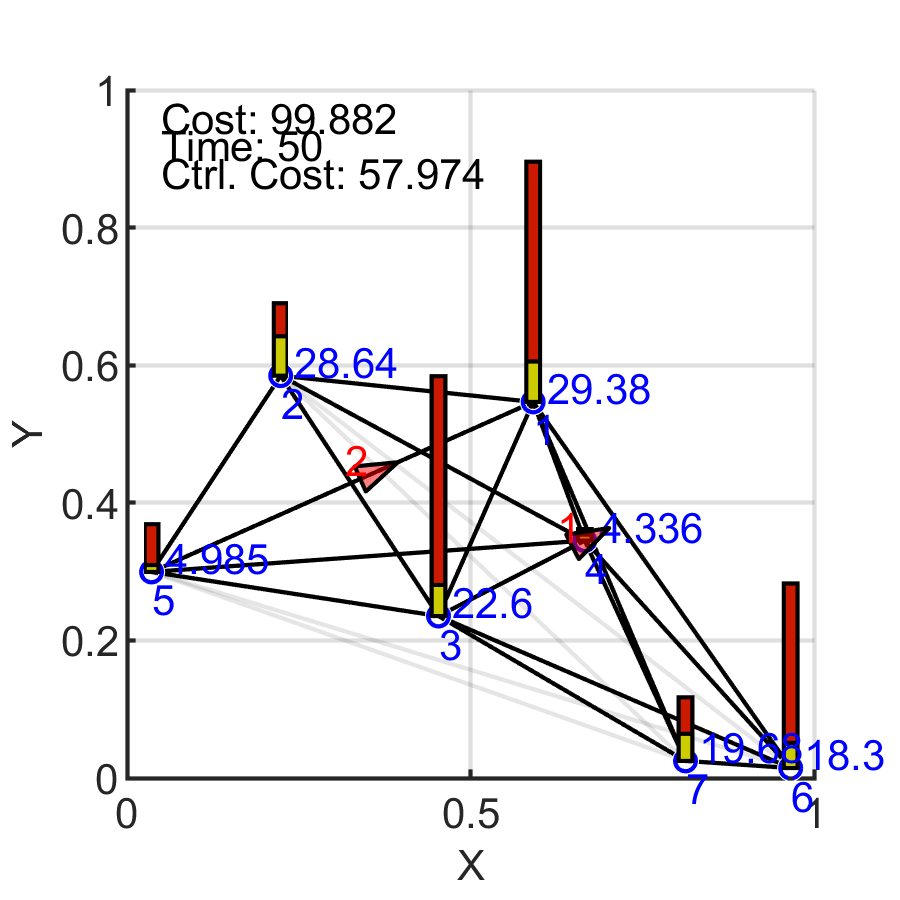}
         \caption{MTSP: $J_C = 58.0$}
         \label{SubFig:1}
     \end{subfigure}
     \begin{subfigure}[b]{0.32\columnwidth}
         \centering
         \includegraphics[width=\textwidth]{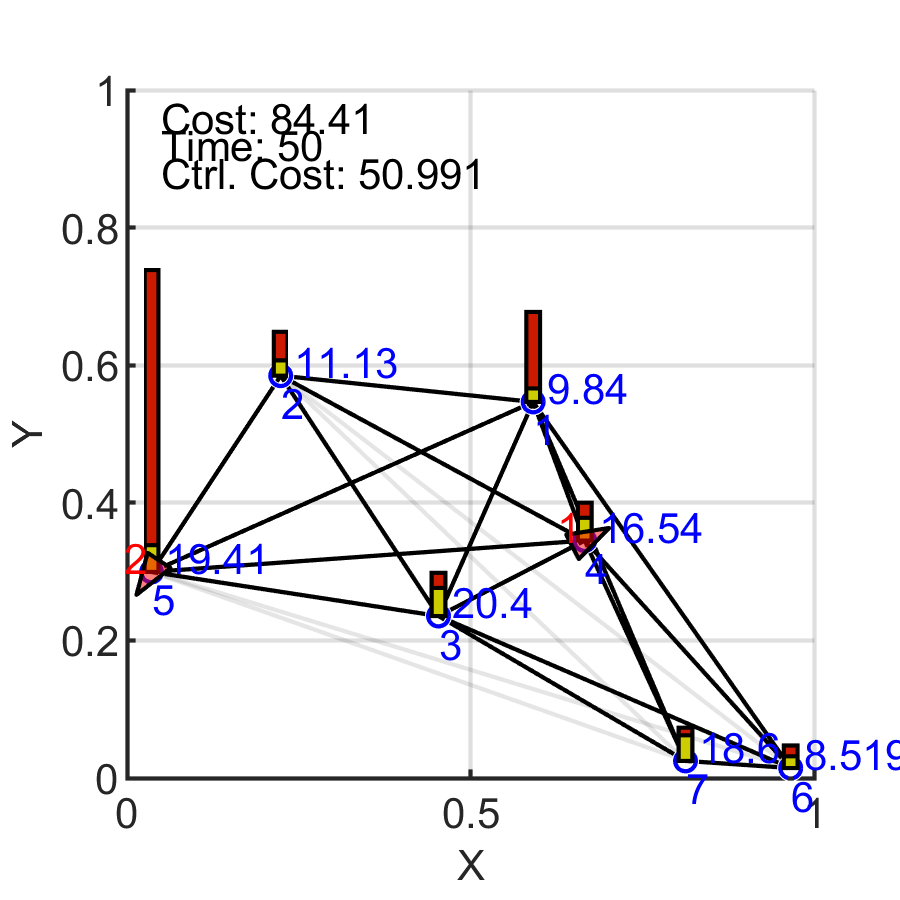}
         \caption{RHC-P: $J_C = 51.0$}
         \label{SubFig:2}
     \end{subfigure}
     \begin{subfigure}[b]{0.32\columnwidth}
         \centering
         \includegraphics[width=\textwidth]{results/TCCase1/RHC.png}
         \caption{RHC: $J_C = \textbf{48.1}$}
         \label{SubFig:3}
     \end{subfigure}
        \caption{Performance of target state controllers in PC 1.}
        \label{Fig:TargetControl1}
\end{figure}

\begin{figure}[!h]
     \centering
     \begin{subfigure}[b]{0.32\columnwidth}
         \centering
         \includegraphics[width=\textwidth]{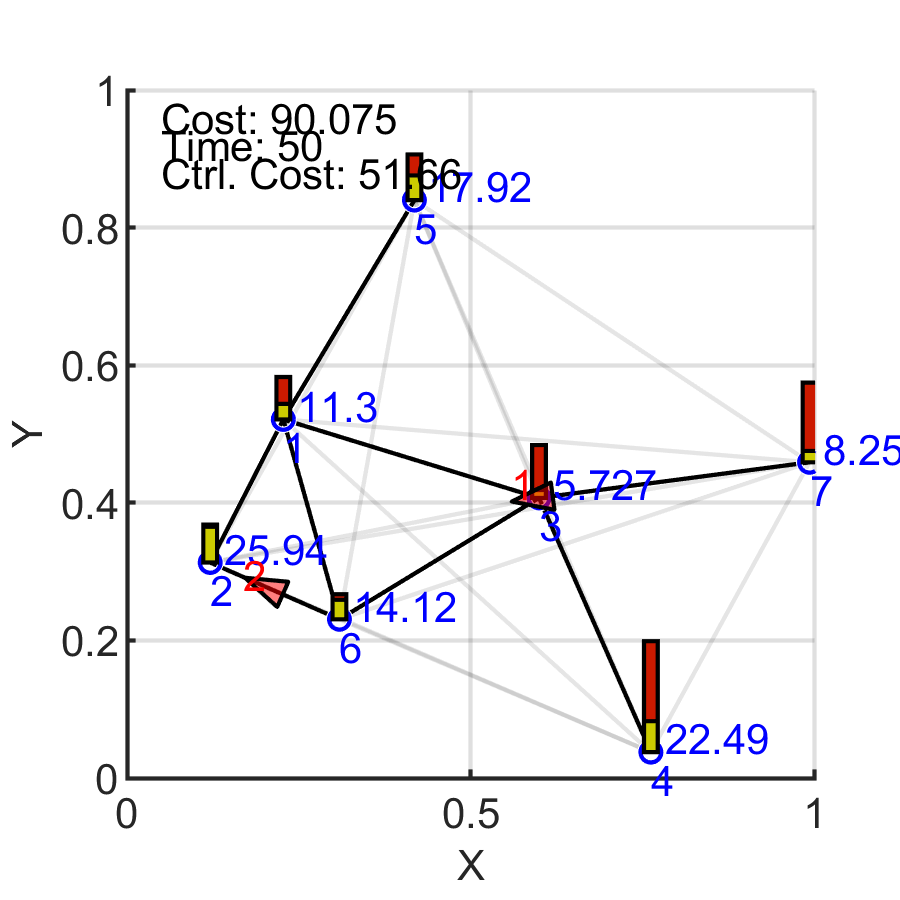}
         \caption{MTSP: $J_C = 51.7$}
         \label{SubFig:1}
     \end{subfigure}
     \begin{subfigure}[b]{0.32\columnwidth}
         \centering
         \includegraphics[width=\textwidth]{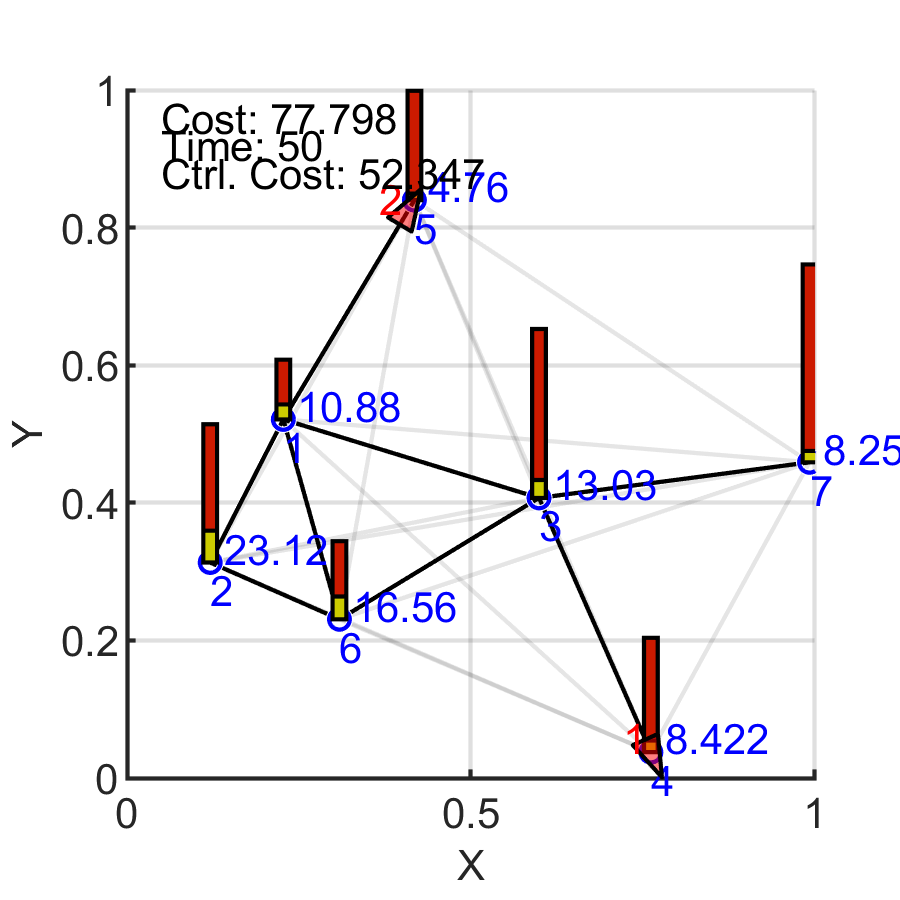}
         \caption{RHC-P: $J_C = 52.3$}
         \label{SubFig:2}
     \end{subfigure}
     \begin{subfigure}[b]{0.32\columnwidth}
         \centering
         \includegraphics[width=\textwidth]{results/TCCase2/RHC.png}
         \caption{RHC: $J_C = \textbf{50.1}$}
         \label{SubFig:3}
     \end{subfigure}
        \caption{Performance of target state controllers in PC 2.}
        \label{Fig:TargetControl2}
\end{figure}

\begin{figure}[!h]
     \centering
     \begin{subfigure}[b]{0.32\columnwidth}
         \centering
         \includegraphics[width=\textwidth]{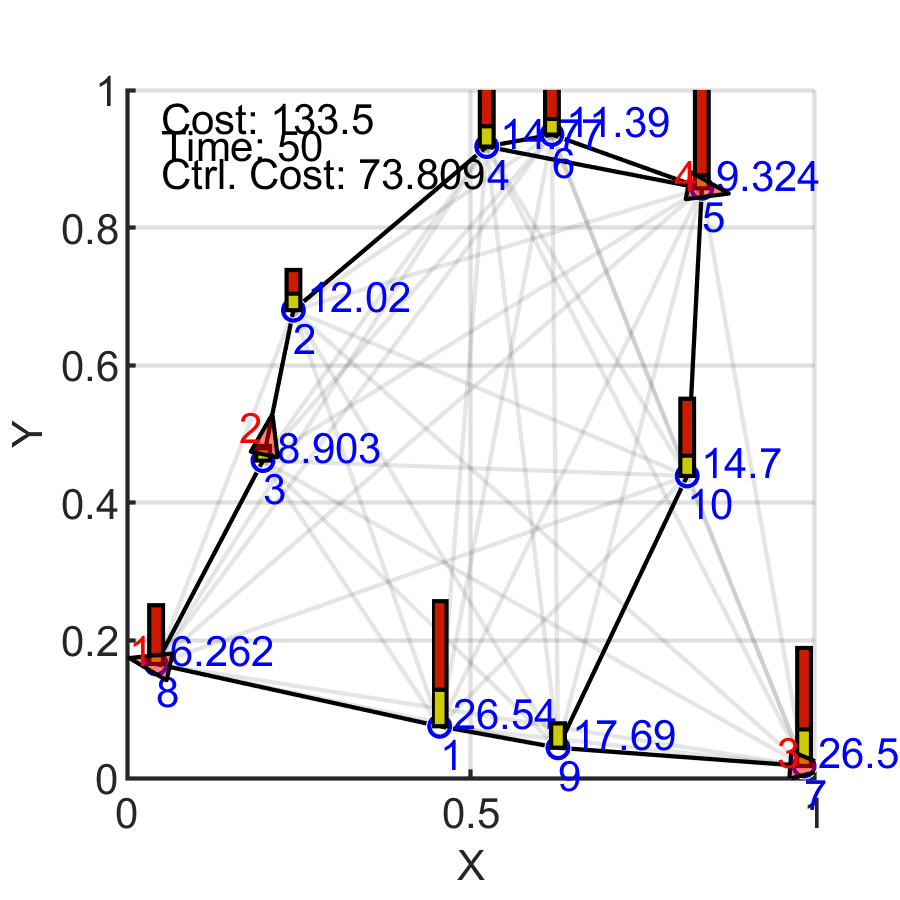}
         \caption{MTSP: $J_C = \textbf{73.8}$}
         \label{SubFig:1}
     \end{subfigure}
     \begin{subfigure}[b]{0.32\columnwidth}
         \centering
         \includegraphics[width=\textwidth]{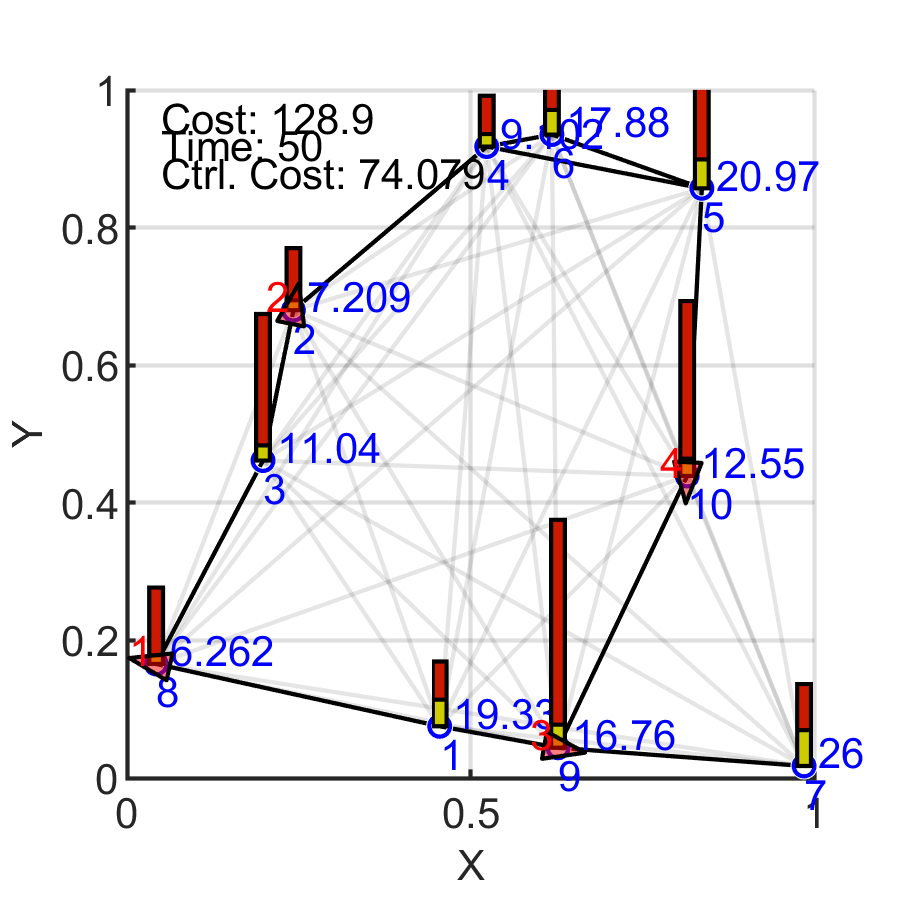}
         \caption{RHC-P: $J_C = 74.1$}
         \label{SubFig:2}
     \end{subfigure}
     \begin{subfigure}[b]{0.32\columnwidth}
         \centering
         \includegraphics[width=\textwidth]{results/TCCase3/RHC.png}
         \caption{RHC: $J_C = 74.3$}
         \label{SubFig:3}
     \end{subfigure}
        \caption{Performance of target state controllers in PC 3.}
        \label{Fig:TargetControl3}
\end{figure}

\begin{figure}[!h]
     \centering
     \begin{subfigure}[b]{0.32\columnwidth}
         \centering
         \includegraphics[width=\textwidth]{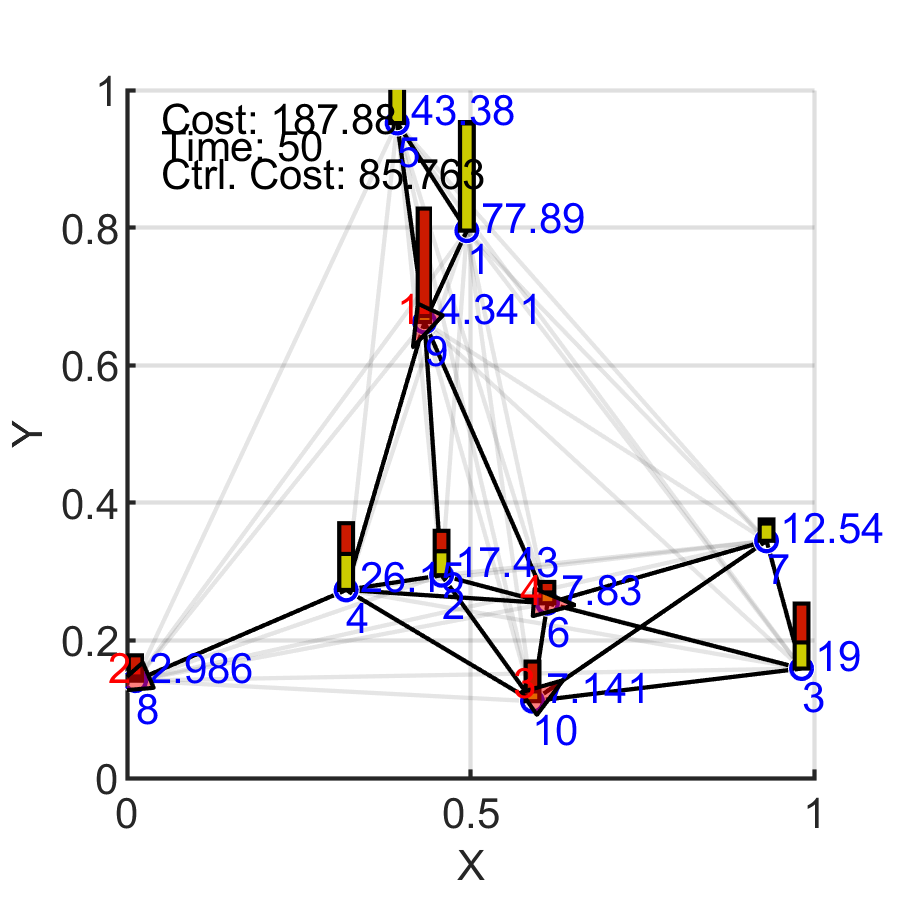}
         \caption{MTSP: $J_C = 85.8$}
         \label{SubFig:1}
     \end{subfigure}
     \begin{subfigure}[b]{0.32\columnwidth}
         \centering
         \includegraphics[width=\textwidth]{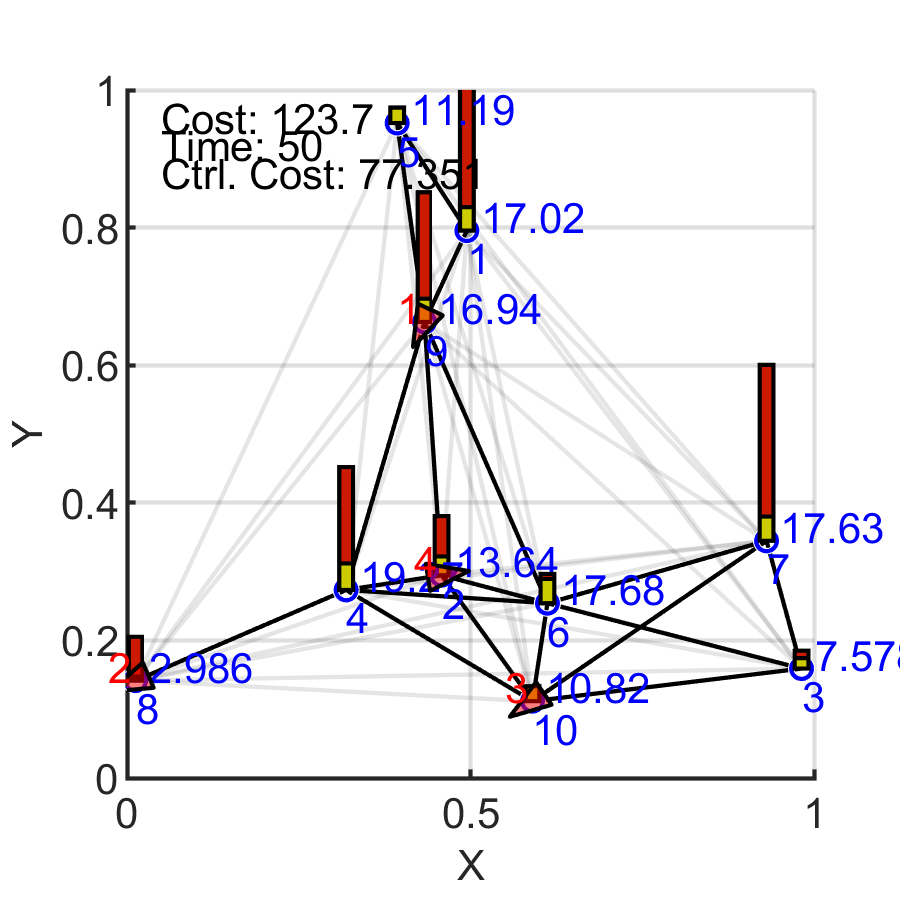}
         \caption{RHC-P: $J_C = 77.4$}
         \label{SubFig:2}
     \end{subfigure}
     \begin{subfigure}[b]{0.32\columnwidth}
         \centering
         \includegraphics[width=\textwidth]{results/TCCase4/RHC.png}
         \caption{RHC: $J_C = \textbf{75.8}$}
         \label{SubFig:3}
     \end{subfigure}
        \caption{Performance of target state controllers in PC 4.}
        \label{Fig:TargetControl4}
\end{figure}

\subsection{Simulation Study 3: Long term performance with learning}

In both previous simulation studies, we focused on a relatively short period ($T= 50\,s$) that essentially encompassed the transient phase of the PMN system (which, in general, is the most challenging part to control/regulate). However, in this final simulation study, we aim to explore the performance of agent controllers over a lengthy period ($T=750\,s$) that includes both the transient and the steady-state phases of the PMN system. Note that this kind of a problem setup is ideal for deploying the machine learning influenced RHC solutions: RHC-L and RHC-AL proposed in Section \ref{Sec:RHCLSolution}. Therefore, in this study, we specifically compare the three controllers: RHC, RHC-L and RHC-AL for PCs 1 and 2, in terms of the evolution of: (i) the performance metric $J_t$ \eqref{Eq:GlobalObjective} and (ii) the average processing time (commonly known as the ``CPU time'') taken to solve a RHCP, throughout the period $t\in[0, T]$. Note that these CPU times were recorded on an Intel Core i7-8700 CPU \@3.20 GHz Processor with a 32 GB RAM.

\begin{figure}[!t]
     \centering
     \begin{subfigure}[b]{0.49\columnwidth}
         \centering
         \includegraphics[width=\textwidth]{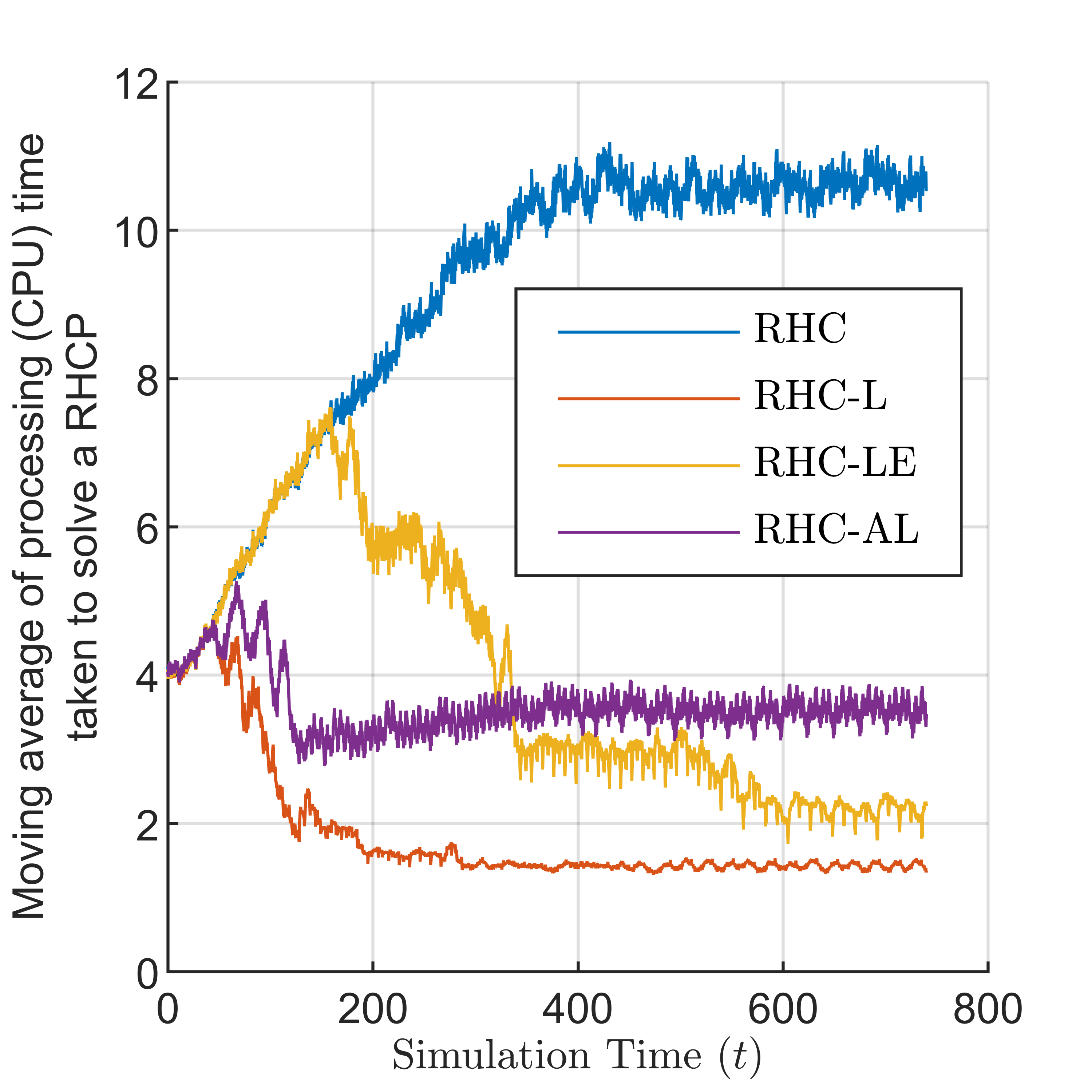}
         \caption{}
         \label{SubFig:1}
     \end{subfigure}
     \begin{subfigure}[b]{0.49\columnwidth}
         \centering
         \includegraphics[width=\textwidth]{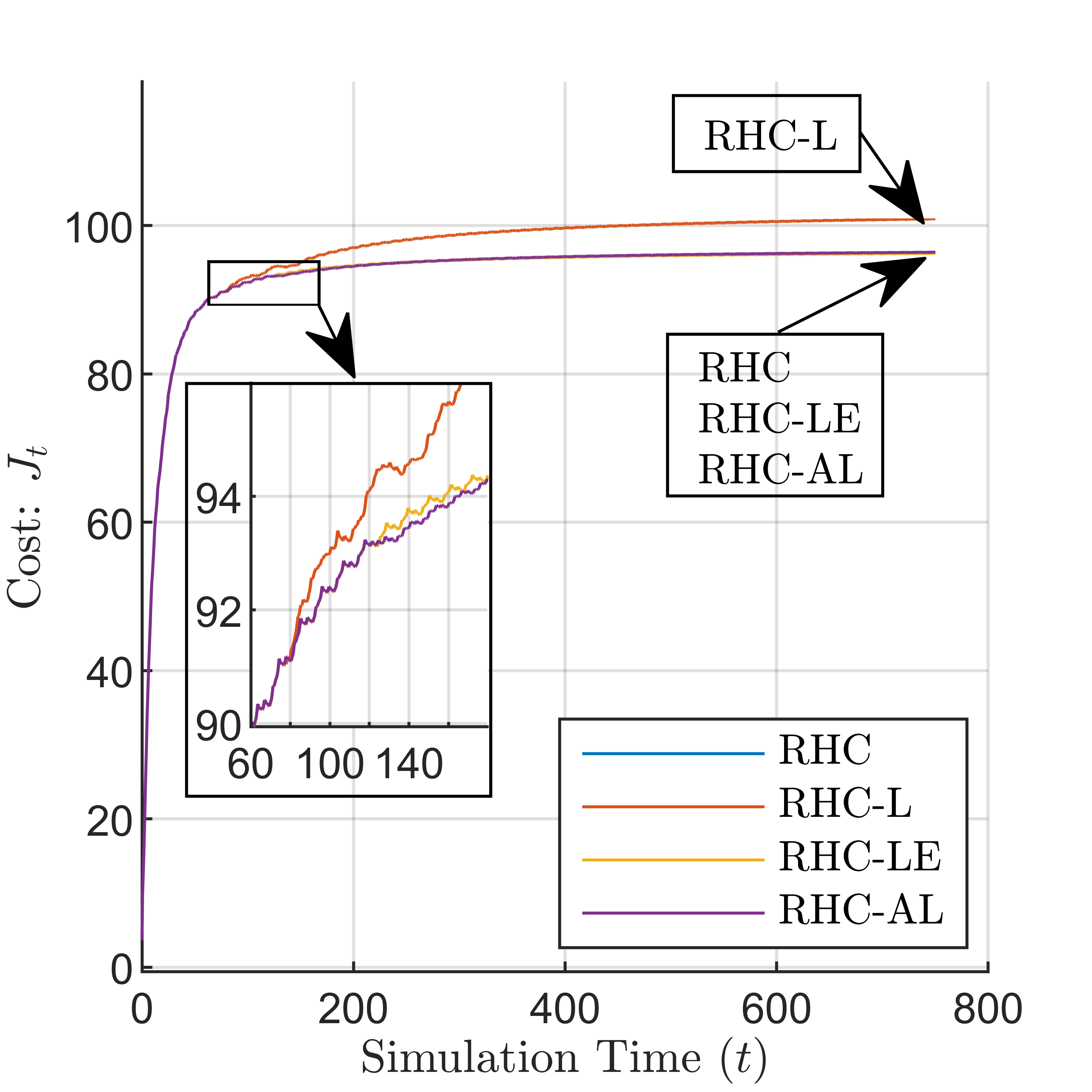}   
         \caption{}
         \label{SubFig:2}
     \end{subfigure}
        \caption{Evolution of the average processing time taken to solve a RHCP and the objective function value for PC 1.}
        \label{Fig:ExecutionTimes1}
\end{figure}

\begin{figure}[!t]
     \centering
     \begin{subfigure}[b]{0.49\columnwidth}
         \centering
         \includegraphics[width=\textwidth]{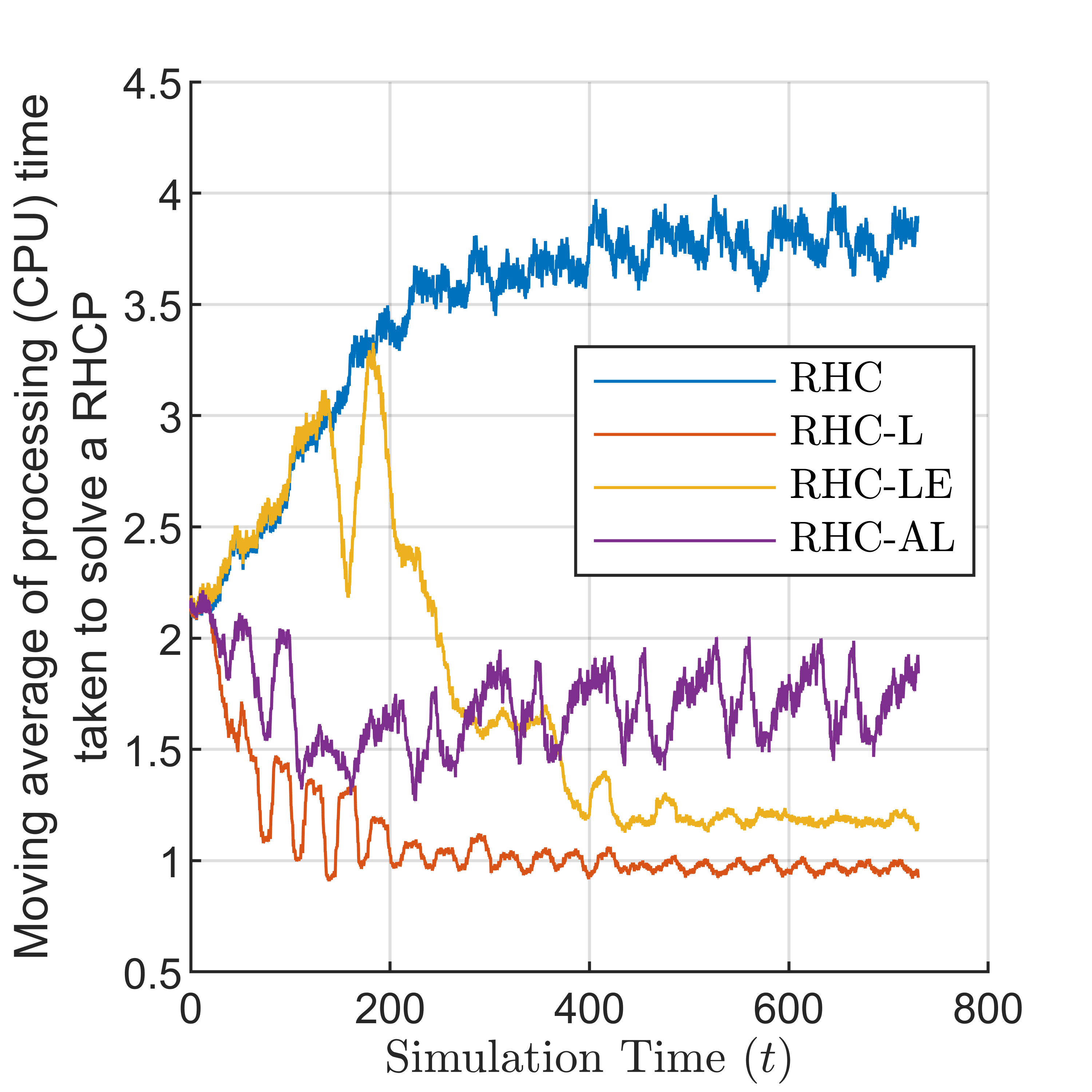}
         \caption{}
         \label{SubFig:1}
     \end{subfigure}
     \begin{subfigure}[b]{0.49\columnwidth}
         \centering
         \includegraphics[width=\textwidth]{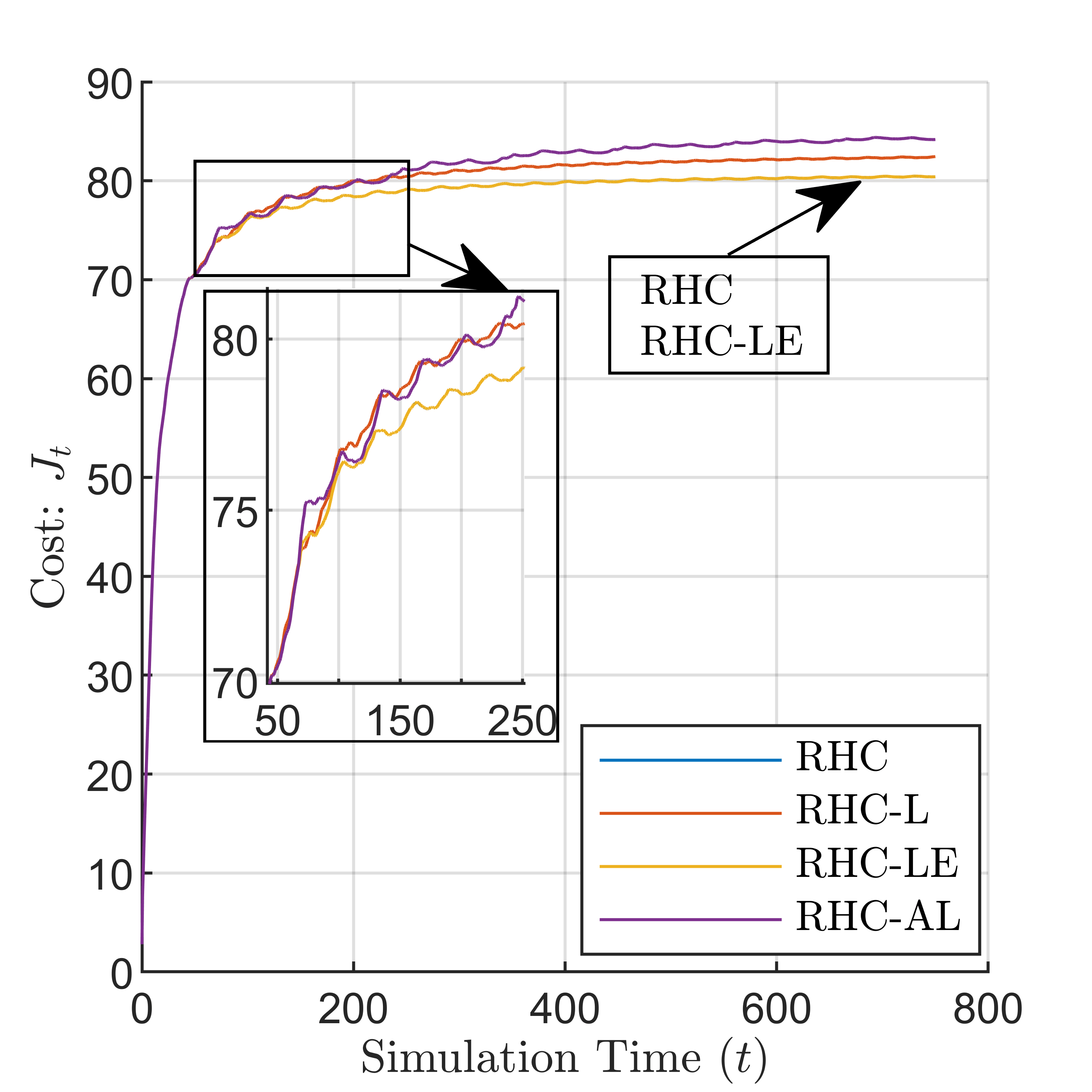}
         \caption{}
         \label{SubFig:2}
     \end{subfigure}
        \caption{Evolution of the average processing time taken to solve a RHCP and the objective function value for PC 2.}
        \label{Fig:ExecutionTimes2}
\end{figure}

As shown in Figs. \ref{Fig:ExecutionTimes1}(a) and \ref{Fig:ExecutionTimes2}(a), the RHC method takes the highest amount of CPU time to solve a RHCP. Its upward trend in the initial stages of the simulations indicates a transient phase of the processor (due to system cache utilization). We point out that this particular transient phase is independent of that of $J_t$ curves shown in respective Figs. \ref{Fig:ExecutionTimes1}(b) and \ref{Fig:ExecutionTimes2}(b). Table \ref{Tab:SteadyStateMetrics} shows that based on the steady-state averages for PC 1 (in Fig. \ref{Fig:ExecutionTimes1}), the RHC-L method spends $86.5\%$ less CPU time compared to the RHC method but at a loss of $4.7\%$ in performance. For the same PC, the RHC-AL method shows a $66.7\%$ reduction in CPU time while having only a $0.1\%$ loss in performance.

In these simulations of RHC-L and RHC-AL methods, for the on-line training of classifiers $f_i(X_i;\mathcal{D}_i)$ (required in \eqref{Eq:RHCLGenStep1}), we have selected the data set size $L=25$ (i.e., $\vert \mathcal{D}_i \vert = 25$). As implied by Figs. \ref{Fig:ExecutionTimes1}(a) and \ref{Fig:ExecutionTimes2}(a), agents have been able to collect that amount of data points well within their transient phase (of the $J_t$ curve). Even though learning based on transient data has a few advantages, it is mostly regarded as ineffective - especially if the learned controller would mostly operate in a steady-state condition. Therefore, we next extend the data set size to be $L = \vert \mathcal{D}_i \vert = 75$ and execute the same RHC-L method, which henceforth is called the RHC-LE method. According to the summarized steady-state averaged data given in Tab \ref{Tab:SteadyStateMetrics}, for PC 1 and 2, the RHC-LE method respectively shows $77.5\%$ and $68.9\%$ reductions in CPU time compared to the RHC method - while having almost no loss in performance ($<0.001\%$) in both cases.



\begin{remark}
The reported simulation results in this section highlight the advantage of the proposed RHC based distributed estimation scheme. In particular, we attribute this superiority of the proposed control solution to its specifically designed agent controllers \eqref{Eq:RHCGenSolStep1}-\eqref{Eq:RHCGenSolStep2}. Note that these agent controllers are non-trivial and operate in a distributed, on-line, asynchronous, and event-driven manner. While such qualities are coveted in real-world applications, unfortunately, they collectively make it difficult to establish theoretical \emph{global} performance guarantees such as asymptotic worst-case and average performances (that have been studied under centralized off-line controllers \cite{Pinto2021}, \cite{Welikala2020J2}). Overcoming this challenge is a subject of future research. 
\end{remark}

\section{Conclusion}
\label{Sec:Conclusion}
The goal of the estimation problem considered in this paper is to observe a distributed set of target states in a network using a mobile fleet of agents so as to minimize an overall measure of estimation error covariance evaluated over a finite period. Compared to existing centralized off-line agent control solutions, a novel computationally efficient distributed on-line solution is proposed based on event-driven receding horizon control. In particular, each agent determines their optimal planning horizon and the immediate sequence of optimal decisions at each event of interest faced in its trajectory. Numerical results show higher performance levels in multiple aspects than existing other centralized and distributed agent control methods. Future work aim to generalize the proposed solution for multidimensional target state dynamics.

\begin{table}[!t]
\centering
\caption{Average over the steady-state period $t\in[500,750]$ of the curves in Figs. \ref{Fig:ExecutionTimes1} and \ref{Fig:ExecutionTimes2}.}
\label{Tab:SteadyStateMetrics}
\resizebox{\columnwidth}{!}{%
\begin{tabular}{|c|c|r|r|r|r|}
\hline
\multicolumn{2}{|c|}{\begin{tabular}[c]{@{}c@{}}Average over steady-state:\\ (Interval: [500, 750])\end{tabular}} &
  \multicolumn{1}{c|}{RHC} &
  \multicolumn{1}{c|}{\begin{tabular}[c]{@{}c@{}}RHC-L\\ ($L=25$)\end{tabular}} &
  \multicolumn{1}{c|}{\begin{tabular}[c]{@{}c@{}}RHC-LE\\ ($L=75$)\end{tabular}} &
  \multicolumn{1}{c|}{\begin{tabular}[c]{@{}c@{}}RHC-AL\\ ($L=25$)\end{tabular}} \\ \hline
\multirow{2}{*}{PC 1} & CPU Time & 10.589  & 1.429   & 2.379  & 3.526  \\ \cline{2-6} 
                        & $J_t$    & 96.132 & 100.611 & 96.132 & 96.269 \\ \hline
\multirow{2}{*}{PC 2} & CPU Time & 3.791  & 0.967   & 1.181  & 1.733  \\ \cline{2-6} 
                        & $J_t$    & 80.289 & 82.190  & 80.289 & 83.968 \\ \hline
\end{tabular}%
}
\end{table}

\appendix

\subsection{Coefficients of the \textbf{RHCP1} objective function \eqref{Eq:RHCP1ObjectiveFunction}}\label{SubSec:CoeffsOfRHCP1}

$$a_1 = \frac{1}{G_j}\log\Big[\frac{G_j-2A_jv_{j1}}{2A_j(v_{j2}-v_{j1})}\Big] + \frac{1}{G_i}\log\Big[\frac{-G_i\Omega_i-Q_i}{Q_i(v_{i2}-v_{i1})}\Big],\ 
a_2 = \frac{1}{G_i},$$ 
$$a_3 = -\frac{G_i\Omega_i+Q_iv_{i2}}{G_i\Omega_i+Q_iv_{i1}},\ 
a_4 = \frac{1}{G_j},\ 
a_5 = -\frac{v_{j2}(2A_j\Omega_j+Q_j)e^{2A_j\rho_{ij}}}{Q_jv_{j2}+2A_j},$$  
$$a_6 = -\frac{G_j-2A_jv_{j2}}{G_j-2A_jv_{j1}}, 
a_7 = -a_5,\ 
a_8 = \frac{1}{v_{i1}},\ 
a_9 = \frac{1}{v_{i2}},
$$
\begin{equation*}
\begin{aligned}
    b_1 = -\frac{Q_i}{4A_i^2}(1+2A_i\rho_{ij}) 
    - \frac{Q_j}{4A_j^2}(1+2A_j\rho_{ij}) - \frac{\Omega_j}{2A_j}\\
    -\sum_{k\in\mathcal{N}_i\backslash\{j\}}\Big[
    \frac{Q_k}{4A_k^2}(1+2A_k\rho_{ij})+\frac{\Omega_k}{2A_k}\Big],\ 
    b_2 = -\sum_{k\in\mathcal{N}_i}\frac{Q_k}{2A_k},
\end{aligned}
\end{equation*}
$$b_3 = -\sum_{k\in\bar{\mathcal{N}}_i\backslash\{j\}}\frac{Q_k}{2A_k},\ 
b_4 = \frac{1}{4A_j^2}(Q_j+2A_j\Omega_j)e^{2A_j\rho_{ij}},$$ 
$$b_5 = \frac{Q_i}{4A_i^2}e^{2A_i\rho_{ij}},\ 
b_{6k} = \frac{1}{4A_k^2}(Q_k+2A_k\Omega_k)e^{2A_k\rho_{ij}},$$
$$
c_1 = -\frac{1}{2A_iv_{i1}},\ 
c_2 = -\frac{v_{i1}\Omega_i-1}{v_{i2}\Omega_i-1},\ c_3 = -e^{2A_i\rho_{ij}},$$
$$c_4 = -c_2,\ 
c_5 = -\frac{G_i\Omega_i+Q_iv_{i2}}{G_i\Omega_i+Q_iv_{i1}}.$$
\bibliographystyle{IEEEtran}
\bibliography{References}

\end{document}